\documentclass[10pt,preprint]{aastex}

\usepackage{natbib}
\usepackage{xspace}
\usepackage{filecontents}
\usepackage{graphicx}
\usepackage{tikz}
\usepackage{subfigure}
\usepackage{amsmath}

\usepackage{url}

\usepackage{mathptmx}       
\usepackage{helvet}         
\usepackage{courier}        
\usepackage{type1cm}        
\usepackage{makeidx}         
\usepackage{graphicx}        
\usepackage{multicol}        
\usepackage[bottom]{footmisc}

\newcommand{\latin}[1]{{\it #1}}
\newcommand{\ie}{\latin{i.e.,}\@\xspace}
\newcommand{\eg}{\latin{e.g.,}\@\xspace}

\newcommand{\etc}{\latin{etc.}\@\xspace}

\newcommand{\ave}[1]{\left\langle #1 \right\rangle}

\newcommand{\slabel}[1]{\label{sec:#1}}

\newcommand{\flabel}[1]{\label{fig:#1}}
\newcommand{\fref}[1]{Fig.~\ref{fig:#1}}
\newcommand{\Fref}[1]{Figure~\ref{fig:#1}}

\newcommand{\elabel}[1]{\label{eq:#1}}
\newcommand{\eref}[1]{(\ref{eq:#1})}
\newcommand{\Eref}[1]{Eq.~(\ref{eq:#1})}

\newcommand{\gpvec}[1]{\mathbf{#1}}

\newcommand{\kvec}{\gpvec{k}}
\newcommand{\rvec}{\gpvec{r}}
\newcommand{\xvec}{\gpvec{x}}

\newcommand{\plaind}{\mathrm{d}}
\newcommand{\dint}[1]{\mathchoice{\!\plaind#1\,}{\!\plaind#1\,}{\!\plaind#1\,}{\!\plaind#1\,}}
\newcommand{\ddint}[1]{\mathchoice{\!\plaind^d#1\,}{\!\plaind^d#1\,}{\!\plaind^d#1\,}{\!\plaind^d#1\,}}

\newcommand{\FC}{\mathcal{F}}
\newcommand{\GC}{\mathcal{G}}
\newcommand{\OC}{\mathcal{O}}

\renewcommand{\exp}[1]{\mathchoice{e^{#1}}{\operatorname{exp}\left(#1\right)}{\operatorname{exp}\left(#1\right)}{\operatorname{exp}\left(#1\right)}}
\newcommand{\erfc}{\operatorname{erfc}}

\newenvironment{subeqnarray}[1]{\begin{subequations}#1\begin{eqnarray}}{\end{eqnarray}\end{subequations}\ignorespacesafterend}

\def\ang{\AA}

\def\gapprox{\lower.4ex\hbox{$\;\buildrel >\over{\scriptstyle\sim}\;$}}
\def\lapprox{\lower.4ex\hbox{$\;\buildrel <\over{\scriptstyle\sim}\;$}}

\shortauthors{McAteer et al. 2014}
\shorttitle{25 Years of SOC: Numerical Detection Methods}

\begin{document}
\renewcommand{\topfraction}{0.95}
\renewcommand{\bottomfraction}{0.95}
\renewcommand{\textfraction}{0.05}
\renewcommand{\floatpagefraction}{0.95}
\renewcommand{\dbltopfraction}{0.95}
\renewcommand{\dblfloatpagefraction}{0.95}


\title{25 Years of Self-Organized Criticality: Numerical Detection Methods}

\author{James McAteer}
\affil{Solar Physics and Space Weather, Department of Astronomy, P.O.Box 30001, New Mexico State University, MSC 4500, Las Cruces, USA;\\
	e-mail: mcateer@nmsu.edu}
\author{Markus J. Aschwanden}
\affil{Lockheed Martin, Solar and Astrophysics Laboratory (LMSAL), STAR Labs, A021S, Bldg.252, 3251 Hanover St., Palo Alto, CA 94304, USA}
\author{Michaila Dimitropoulou}
\affil{Kapodistrian University of Athens, Dept. Physics, 15483 Athens, Greece}
\author{Manolis K. Georgoulis}
\affil{Research Center Astronomy and Applied Mathematics, Academy of Athens, 4 Soranou Efesiou Street, 11527 Athens, Greece}
\author{Gunnar Pruessner}
\affil{Dept. Mathematics, Imperial College London, 180 Queen's Gate, London SW7 2AZ, United Kingdom}
\author{Laura Morales}
\affil{Departamento de F\'isica, Facultad de Ciencias Exactas y Naturales, Universidad de Buenos Aires - Instituto de F\'isica Plasmas (CONICET), Buenos Aires, Argentina}
\author{Jack Ireland}
\affil{ADNET Systems, Inc.,NASA Goddard Space Flight Center, MC. 671.1, Greenbelt, MD 20771, USA.}
\author{Valentyna Abramenko}
\affil{Space Weather Prediction Laboratory, Department of Solar Physics, Celtral Astronomical Observatory of Russian Academy of Science at Pulkovo, 65 Pulkovskoe Sh. St.Petersburg, Russia, 196140\\
Big Bear Solar Observatory of NJIT, 40386 N Shore Dr, Big Bear city, CA, 92314, USA}

\begin{abstract}
The detection and characterization of self-organized criticality (SOC), in both real and simulated data, has undergone many significant revisions over the past 25 years. The explosive advances in the many numerical methods available for detecting, discriminating, and ultimately testing, SOC have played a critical role in developing our understanding of how systems experience and exhibit SOC. In this article, methods of detecting SOC are reviewed; from correlations to complexity to critical quantities. A description of the basic autocorrelation method leads into a detailed analysis of application-oriented methods developed in the last 25 years. In the second half of this manuscript space-based, time-based and spatial-temporal methods are reviewed and the prevalence of power laws in nature is described, with an emphasis on event detection and characterization. The search for numerical methods to clearly and unambiguously detect SOC in data often leads us outside the comfort zone of our own disciplines - the answers to these questions are often obtained by studying the advances made in other fields of study. In addition, numerical detection methods often provide the optimum link between simulations and experiments in scientific research. We seek to explore this boundary where the rubber meets the road, to review this expanding field of research of numerical detection of SOC systems over the past 25 years, and to  iterate forwards so as to provide some foresight and guidance into developing breakthroughs in this subject over the next quarter of a century.
\end{abstract}


\clearpage
\newpage
\section{INTRODUCTION}
Self-Organized Criticality (SOC) is a statistical property of many time-varying systems. \cite{Asch14} (this volume of SSR) present a detailed description of SOC in solar and astrophysical settings; for the purposes of this current paper, SOC is considered in the wider aspect of any physical system that displays the scale invariance in both time and space leading to a critical point. It is often observed in slowly driven, but non-equilibrium, systems and, perhaps most importantly, complexity naturally arises in the system without any fine-tuned parameters as input. Although well-known earlier work \citep[\eg][]{VonN, man75} had shown that complexity could arise from simply-governed, slowly driven systems, the seminal paper of \citet{Baketal87} provided the breakthrough in this subject by showing that all the so-called SOC features (\eg fractal geometry, scale-invariance, power laws) arise from simple systems and lead to a critical point with no fine tuning of the input. Hence the system is both self-organized and critical. The large volume of research resulting from \citet{Baketal87} includes many articles on how to recognize SOC in a system. It is the 25 years of these numerical detection methods that we review in this paper.

The power of SOC lies in the ability to both describe and explain a large variety of physical systems in a quantitative and physically-motivated manner . From sand piles \citep{Baketal87} to solar flares \citep{LuHamilton91}, from fractures \citep{Tur} to forest fires \citep{Dro}; from asteroids \citep{Ive} to accretion disks \citep{Den88}, SOC provides a mathematically tractable and understandable route to study complex systems. The scale-free, dimensionless, nature of SOC conveniently encompasses much of the universe. The concept of simple beginnings - assuming a starting grid and apply a few rules regarding distribution of excess amongst nearest neighbors - is an attractive model to many scientists, spanning subjects from physics and chemistry to economics and sociology. However, every SOC researcher ultimately reverts back to the same set of unanswered questions - How can I tell whether my system is truly SOC, or if it is just displaying SOC-like behavior? How can I detect SOC in such a way that I can confidently distinguish it from other potential physical sources? The route to answering these questions begins in Section~\ref{af} with the seemingly-simple studies of autocorrelations, described in terms of symmetries leading to diffusion models, and correlations functions leading to surface growth models. We end this discussion with a detailed look at the methods of measuring correlation functions, with a emphasis on the Manna model. The models introduced in this section are all guided by simple sets of rules of particle interaction governing how particles spread apart (\ie diffusion), how particles clump together (\ie growth), and the redistribution of particles upon reaching a threshold value. In Section~\ref{strfunc} we move from a discussion of products of field values (\ie correlation functions) to a discussion of increments (\ie structure functions). The value of the structure function as a complementary approach is highlighted with respect to determining linear ranges in log-log plots, with an application to solar magnetic fields. Application-oriented methods (Section~\ref{aom}) provide a third approach to numerical detection of SOC. We end our discussion of numerical methods in Section 2, by studying the advantages of block-scaling as a sub-sampling method to be used when little data is available to the scientist.

With this toolkit in hand, Section~\ref{det} contains a review of the many approaches developed over the last 25 years to identify individual SOC features and events. We split these studies into the three areas of spatial, temporal, and combined spatio-temporal. By performing this three-way split we merely seek a convenient route to provide some narrative to the reader; we do not suggest that these techniques differ in some fundamental way. When studying images in Section~\ref{det1} we usually require thresholding, and considerations of 3D volume. As we typically only have 2D images, this consideration leads us to discuss the potential 2D fingerprint of a 3D SOC system. As a follow-on from this type of thought process, one need only look at that most common feature of SOC detections of power laws in Section~\ref{det2}. It is clearly trivial to plot data on a log-log set of axis and find a straight line fit. The real purpose of this scientific endeavor should be research performing a set of logical deductive steps showing that such data are truly described by a power law, and that this power law can only be the result of an SOC system. The discussion in Section~\ref{powerlaws} shows how rarely we achieve such a scientific nirvana. Only when we fully comprehend issues such as the detection of power laws, and issues of data sampling and  pulse pile-up can we then move to discuss waiting-time distributions as a possible signature of SOC. We conclude in Section~\ref{det3} by showing how spreading and avalanche exponents provide vital tools to study spatio-temporal structures, with an emphasis on examples from magnetospheric and solar physics. 

\clearpage
\newpage
\section{Methods of numerical detections of SOC}
\label{basics}
The basic approach to test for the existence of SOC in numerical or observational data is to extract a series of events and test if these features are in some way connected. Events can are often called features, clusters, storms, objects, explosions, instabilities - the nomenclature is often different but the principle is the same. In Section~\ref{det} we will proceed to perform a synthesis on methods of extraction of these events, however here in Section~\ref{basics} we first review existing methods of testing for connections between events, starting with the autocorrelation function and its modern extensions (Section~\ref{af}), moving onto structure functions (Section~\ref{strfunc}) and then focusing on application-oriented methods developed in the last 25 years (Section~\ref{aom}).

\subsection{Autocorrelation functions}
\label{af}
Autocorrelation functions have a long history in the study of critical systems \citep{Stanley:1971}. While they are defined on the microscopic scale, they bridge the gap to the large scale and typically display scaling on these larger scales in both space and time. As such, correlation functions are at the heart of the theoretical description of scaling phenomena in systems with many interacting degrees of freedom, yet numerically and experimentally they are often inaccessible. The provision of numerical detection methods for the study of SOC systems hinges critically on a fundamental understanding of correlations functions in the study of traditional systems. In the following section, correlation functions are introduced in broad terms, highlighting some basic features and symmetries that are important for a later discussion of SOC. Readers familiar with these two topics may wish to skip to Section~\ref{bdeanm} where we discuss some basic null models in order to motivate the focus on some characteristics of correlations often found in non-trivial systems exhibiting SOC. Some parallels are drawn from the study of surface growth and interfaces and then the basic measurement methods are exemplified using the Manna Model \citep{Manna:1991a}.

SOC systems evolve in time and extend in space due to the interaction of their local degrees of freedom (local activity of avalanching, energy, particle density, height \etc). The propagation of this interaction in time and space can be captured by autocorrelation functions. As SOC systems demand evolution to a critical point, it is expected that every part of a system interacts with every other part of a system, as well as with their history, in such a way that does not allow for degrees of freedom to be dropped on the basis that they are {\em too remote} in space or time. Even the most local features cannot be studied in an isolated fashion, as local degrees of freedom self-interact, mediated by their environment. Correlation functions are therefore used to both measure and quantify these effective interactions at the most basic level.

\subsubsection{Basic features}
The most basic autocorrelation function of local degrees of freedom $\phi(\rvec,t)$, such as the local particle density, energy, magnetization \etc, at position $\rvec$ and time $t$ is
\begin{equation}
\elabel{def_C}
C(\rvec_2,t_2,\rvec_1,t_1) = 
\ave{\phi(\rvec_2,t_2) \phi(\rvec_1,t_1)} - \ave{\phi(\rvec_2,t_2)}
\ave{\phi(\rvec_1,t_1)}
\end{equation}
where $\ave{\cdot}$ takes the expectation value, \ie it is the ensemble average. If $\phi(\rvec_2,t_2)$ and $\phi(\rvec_1,t_1)$ are uncorrelated, in particular when they are independent, the joint probability density of $\phi(\rvec_2,t_2)$ and $\phi(\rvec_1,t_1)$
factorizes and therefore $\ave{\phi(\rvec_2,t_2) \phi(\rvec_1,t_1)}=\ave{\phi(\rvec_2,t_2)}\ave{\phi(\rvec_1,t_1)}$  (\ie the correlation function vanishes) $C(\rvec_2,t2,\rvec_1,t_1)=0$. This is obviously a rather trivial situation - correlations do not matter for these types of  degrees of freedom and the behavior of one is not influenced by the behavior of any other. When $\rvec_1=\rvec_2$ and $t_1=t_2$ the correlation function $C(\rvec_2,t_2,\rvec_1,t_1)$ in fact describes the variance of the local $\phi$. Alternatively $C(\rvec_2,t_2,\rvec_1,t_1)$ may be thought of as a measure of fluctuations relative to the background as \Eref{def_C} can be  re-written as 
\begin{equation}
C(\rvec_2,t_2,\rvec_1,t_1) = \Big\langle \big( \phi(\rvec_2,t_2) - \ave{\phi(\rvec_2,t_2)} \big) \big( \phi(\rvec_1,t_1) - \ave{\phi(\rvec_1,t_1)} \big) \Big\rangle \ .
\end{equation}
The result is large when large fluctuations at $\rvec_1,t_1$ match large fluctuations at $\rvec_2,t_2$, and it is small when they typically miss each other. The correlation function might be negative, signalling \emph{anti-correlations} if positive fluctuations at $\rvec_1,t_1$ typically occur when they are negative, $\phi(\rvec_2,t_2) - \ave{\phi(\rvec_2,t_2)}<0$, at $\rvec_2,t_2$.

\subsubsection{Symmetries}
Symmetries may simplify the dependence of $C(\rvec_2,t_2,\rvec_1,t_1)$ on the two points in both space and time. If the system is translationally invariant, then $C(\rvec_2,t_2,\rvec_1,t_1)$ is a function only of the difference $\rvec_2-\rvec_1$, \ie $C(\rvec_2,t_2,\rvec_1,t_1)=C(\rvec_2-\rvec_1,t_2,0,t_1)$. If it is, in addition, invariant under rotations, then it is only a function of the distance $|\rvec_2-\rvec_1|$. When estimating $C(\rvec_2,t_2,\rvec_1,t_1)$ from numerical or observational data, these invariances can be used to improve the estimates, for example in the form
\begin{equation}
C'(\rvec,t_2,t_1)=V^{-1} \int_V \ddint{r'} C(\rvec',\rvec'+\rvec,t_2,t_1)
\end{equation}
where the integration runs over the entire $d$-dimensional volume $V$ of the system. A system with boundaries cannot be expected to be truly translational invariant, so this is often used as a suitable approximation only in relatively small localizations deep inside the system. Most SOC systems require boundaries in order to dissipate energy or particles driven into it, and they are often not translational or rotational invariant, although some basic symmetries, (\eg due to the shape of the system) remain. A typical example is an inversion symmetry about the origin, so that $C(\rvec_2,t_2,\rvec_1,t_1)=C(-\rvec_2,t_2,-\rvec_1,t_1)$.

Similar simplifications apply in the time domain. If correlation functions are translationally invariant in time the system is said to be stationary, 
\ie $C(\rvec_2,t_2,\rvec_1,t_1)=C(\rvec_2,t_2-t_1,\rvec_1,0)=C(\rvec_2,0,\rvec_1,t_1-t_2)$. By construction of \Eref{def_C}, $C$ is invariant under permutations of the  indices, $C(\rvec_2,t_2,\rvec_1,t_1)=C(\rvec_1,\rvec_2,t_1,t_2)$. If $C$ is additionally invariant under rotation and translation, $C(\rvec_2,t_2,\rvec_1,t_1)=C(\rvec_1,t_2,\rvec_2,t_1)$, then, by definition, 
\Eref{def_C} implies invariance under a change of sign of $t_2-t_1$, 
\begin{equation}
\elabel{cinvar}
C(\rvec_2,t_2-t_1,\rvec_1,0)=C(\rvec_2,t_2,\rvec_1,t_1)=C(\rvec_1,t_1,\rvec_2,t_2)=C(\rvec_2,t_1,\rvec_1,t_2)=C(\rvec_2,t_1-t_2,\rvec_1,0)\ .
\end{equation}
However, correlation functions are often of the form
\begin{equation}\elabel{def_G}
G(\rvec_2,t_2,\rvec_1,t_1) = 
\ave{\phi(\rvec_2,t_2) \psi(\rvec_1,t_1)} - \ave{\phi(\rvec_2,t_2)}
\ave{\psi(\rvec_1,t_1)}
\end{equation}
where $\psi(\rvec_1,t_1)$ denotes a \emph{perturbation} of the system at time $t_1$ and position $\rvec_1$ and $\phi(\rvec_2,t_2)$ is the \emph{response} at time $t_2$ and position $\rvec_2$. In this case a change in the sign of $t_2-t_1$ reverses the causal order and therefore the correlation function $G(\rvec_2,t_2,\rvec_1,t_1)$ is not invariant under that change, as it is not invariant under an exchange of indices - $G(\rvec_2,t_2,\rvec_1,t_1)\ne G(\rvec_1,t_1,\rvec_2,t_2)$, as they refer to different entities. Initial conditions generally play the same role as perturbations or boundary conditions - the presence of initial conditions undermines stationarity and time reversal symmetry, just as the presence of boundary conditions undermines translational invariance and inversion across arbitrary points. In order to distinguish \Eref{def_C} from \Eref{def_G} in the context of SOC, the former is often referred to as the activity-activity autocorrelation function and the latter, less common, is referred to as the propagator or response (correlation) function.

Two-point correlation functions are simply correlation functions evaluated at two sets of coordinates (or, if suitable symmetries are found, differences of two sets of coordinates). In most applications, two-point correlation functions are either evaluated at the same time $t_1=t_2$, known as equal time correlation functions, or at the same point in space $\rvec_1=\rvec_2$, and known as temporal correlation functions or two-time correlation function. The behavior captured by an equal time correlation function is thought to be due to a common source, like the simultaneous ripples on the surface of a pond at two points are caused by a stone dropped at the origin (\eg it is very instructive to study correlations in a deterministic system as simple as $\phi(\rvec,t) = \sin(k_0 |\rvec| - \omega_0 t) / |\rvec|$ for some fixed $k_0$ and $\omega_0$). If the correlation function is intended to measure causal relationships, such as in \Eref{def_G}, it must necessarily vanish at equal times for $\rvec_1\ne\rvec_2$, as a perturbation is expected to require time to propagate from $\rvec_1$ to $\rvec_2$. To stay in the same picture, the response function in \Eref{def_G} would measure the response at $\rvec_2, t_2$ to a stone dropped at $\rvec_1, t_1$.

\subsubsection{Basic diffusion examples and null models}
\label{bdeanm}
In many cases, the field $\phi$ denotes a particle density and the null-models of correlations in time and space are Poisson and Gaussian processes. The former refers to processes where events occur completely independently with constant rate, the latter to the random and interaction-free spreading of a quantity subject to conservation and continuity. In the former case, all connected correlation functions vanish. In the latter case, plain diffusion with constant Brownian diffusion coefficient $D$ introduces correlations between different points in time and space. If a single, freely-diffusing particle is created at time $t_1$ and position $\rvec_1$, the relevant correlation function in $d$ Euclidian dimensions is \citep{vanKampen:1992,Strs07}, 
\begin{equation}\elabel{bare_propagator}
G(\rvec_2,t_2,\rvec_1,t_1) = \theta(t_2-t_1) 
\left(\frac{1}{\sqrt{4 \pi D (t_2-t_1)}}\right)^{d/2}
\exp{-\frac{(\rvec_2-\rvec)^2}{4 D (t_2-t_1)}} \ .
\end{equation}
It describes the expected particle density at $\rvec_2,t_2$ following the creation of the particle at $\rvec_1,t_1$. Equivalently, it is the probability density of finding that particle at $\rvec_2,t_2$ after it has been created at $\rvec_1,t_1$. \Eref{bare_propagator} is also the solution of the deterministic diffusion equation. 

\cite{HwaKardar:1989a} proposed a model more relevant to SOC by introducing a source $\eta(\rvec,t)$, so that  $\phi(\rvec,t)=\int\ddint{r'} \int_0^t\dint{t'} G(\rvec,t,\rvec',t')\eta(\rvec',t')$. If $\eta$ describes Gaussian white noise with some amplitude $2\Gamma^2$, then $\phi(\rvec,t)$ is the height of an interface subject to Edwards-Wilkinson dynamics \citep{EdwardsWilkinson:1982,Krug:1997}. It can be thought of as a surface, or a diffusive field, relaxing under the influence of surface tension $\nu=D$, while being exposed to random addition and removal of material (parameterized by $\Gamma^2$). In one dimension the equal time correlation function becomes
\begin{equation}\elabel{EW_corrfct}
C(\rvec_2,\rvec_1,t,t) =2\Gamma^2
\sqrt{\frac{t}{2 \pi \nu}}
\left(
\exp{-\frac{(\rvec_2-\rvec_1)^2}{8\nu t}}
- |\rvec_2-\rvec_1| \sqrt{\frac{\pi}{8\nu t}} 
\erfc\left(\frac{|\rvec_2-\rvec_1|}{\sqrt{8\nu t}}\right)
\right) \ ,
\end{equation}
and the temporal correlation function starting from a flat interface is then
\begin{equation}
C(\rvec,\rvec,t,t) = 2 \Gamma^2
\frac{\sqrt{t_2+t_1}-\sqrt{|t_2-t_1|}}{\sqrt{4\pi\nu}} \ .
\end{equation}
In terms of observables, this is what is typically studied in SOC systems - namely the correlation of the local height or the particle numbers between sites. 

It is important to note that the distinction between the response function, $G$, and the correlation function, $C$, is more than a technicality. The former is the correlation function for the propagation of a perturbation within the degrees of freedom - it addresses the question of how the degrees of freedom, the field $\phi$, {\em reacts} to a perturbation. The latter, on the other hand, describes the correlations seen in the degrees of freedom as the system evolves. These are mediated by the propagator that communicates events, in particular  any external driving, to other sites in the system. To draw a rough parallel to seismic events: $G$ is the seismic signal measured $\rvec_2,t_2$ throughout the Earth's crust as a bomb detonates at $\rvec_1,t_1$, whereas $C$ are the correlations between the signal at $\rvec_2,t_2$ and $\rvec_1,t_1$ as the earth crust evolves under its natural dynamics.

\subsubsection{Temporal and spatial correlations}
Long-range temporal correlations are frequently found in non-equilibrium systems, even when the microscopic interaction is trivial in the technical sense discussed below \citep{Grinstein:1995}. Even directed models display scaling in  temporal correlation functions \citep{Pruessner_exactTAOM:2003}. Non-trivial spatial, correlations are generally regarded as the signature of interactions that dominates the large scale. Temporal correlations are often quantified by the correlation time $\tau$ (see also the correlation length $\xi$ introduced below). The correlation time is defined by the asymptotic decay of the correlation function $C(\rvec,t_2,\rvec,t_1)\propto\exp{-|t_2-t_1|/\tau}$ for large $|t_2-t_1|$. It can be  defined in a correspondingly similar fashion for the propagator, or response function, $G(\rvec, t_2, \rvec, t_1)\propto\exp{-|t_2-t_1/\tau})$. This structure follows necessarily if the observable $\phi(\rvec,t)$ is subject to Markovian dynamics,  so that $\tau$ is in fact determined by the negative inverse logarithm of the second largest eigenvalue of the Markov matrix \citep{vanKampen:1992}. 

An equation very similar to the Edwards-Wilkinson equation was suggested by \cite{HwaKardar:1989a} as a description of SOC phenomena with a possible {\em mass term}, $\epsilon$, that parameterizes an attenuation of the signal. The resulting equal-time correlation functions in $d=1$ and $d=3$ dimensions are, in the limit of large times,
\begin{subeqnarray}{\elabel{G_n}}
\lim_{t\to\infty} 
C_1(\rvec_2,t,\rvec_1,t) & = & \frac{\Gamma^2 \pi}{\sqrt{\epsilon \nu}}
\exp{-|\rvec_2-\rvec_1|\sqrt{\epsilon/\nu}} \\
\lim_{t\to\infty} 
C_3(\rvec_2,t,\rvec_1,t) & = & \frac{\Gamma^2}{2 \nu |\rvec_2-\rvec_1|}
\exp{-|\rvec_2-\rvec_1|\sqrt{\epsilon/\nu}} .
\end{subeqnarray}
These are also known as Ornstein-Zernike-type correlation functions - namely Fourier transforms of $\Gamma^2/(\nu\kvec^2+\epsilon)$ obtained in the Ornstein-Zernike approximation \citep[][Chap.~7.4.2, \citealp{BarratHansen:2003}, Chap.~5]{Stanley:1971} for the structure function in liquids. In some settings, studying the Fourier transform in space, essentially produces the structure factor whereas studying the Fourier transform in time, essentially produces the power spectrum \citep{AbramenkoETAL:2003}.

The examples above are instances of \emph{trivial} correlations in different disguises. Apart from the fact that the only scale mentioned is that of the diffusion constant $D$ or the surface tension $\nu$, which imposes the typical relation between time and space $t\propto\rvec^2$, it is the triviality in the technical sense that makes them proper null-models. \emph{Trivial} here means that the correlations are produced in the absence of interaction, which, in turn, is absent because the processes considered above are linear, \ie the stochastic partial differential equations of motion are linear in the field $\phi$. The equivalence of linearity and lack of interaction can be understood by noticing that solutions can be superimposed - adding one solution to another produces a new solution. In other words, the solution to an initial condition with two particles initially deposited is just the sum of the solutions for each particle individually - the particles do not {\em see} each other. Therein lies the reason for the interest of statistical mechanics in \emph{non-trivial, spatial} correlations. Their space-dependence is normally quantified by matching correlation functions to the scaling form,
\begin{equation}\elabel{def_eta}
C(\rvec,t,0,t) = a |\rvec|^{-(d-2+\eta)} \GC\left(\frac{|\rvec|}{\xi}\right) \ ,
\end{equation}
with a so-called metric factor $a$ \citep[independent of $\xi$, see][]{ChristensenETAL:2008}, Euclidean dimension $d$, universal exponent $\eta$, also known as the anomalous dimension, and a scaling, or cutoff, function $\GC$ that describes how correlations eventually decay on a scale beyond the correlation length, $\xi$. The divergence of the correlation length at the critical point is probably the most direct signal of criticality. In SOC, where systems are expected to organize themselves to the critical point, the correlation length is naturally limited by the system size $L$ and all scaling of global, system-wide observables in SOC is therefore finite size scaling \citep{Barber:1983}. As such, one of the most direct tests of the system being at criticality is to demonstrate that $\xi\propto L$. 

\Eref{def_eta} is not normally expected to hold on short scales, where lattice effects become important. Rather, it describes an asymptotic behavior in large distances $|\rvec|$ and for large correlation lengths $\xi$. In particular, it is not expected to capture the degeneration of $C(\rvec,t,0,t)$ into the variance at $\rvec=0$. Even when the exponent becomes negative, $-(d-2+\eta)<0$, the scaling function $\GC(|\rvec|/\xi)$ may prevent $C(\rvec,t,0,t)$ from diverging in small distances. In order to illustrate \Eref{def_eta}, the Ornstein-Zernike type correlation functions \Eref{G_n} can be matched against it with
\begin{subeqnarray}{}
\elabel{def_C1}
C_1(\rvec,t,0,t) & = & a_1 |\rvec|
\GC_1\left(\frac{|\rvec|}{\xi}\right)
\text{ with }
a_1=\frac{\Gamma^2\pi}{\nu}\text{ and }\GC_1(x)=\frac{e^{-x}}{x} \\
C_3(\rvec,t,0,t) & = & \frac{a_3}{|\rvec|}
\GC_3\left(\frac{|\rvec|}{\xi}\right)
\text{ with }
a_3=\frac{\Gamma^2}{2 \nu}\text{ and }\GC_3(x)=\exp{-x} \ ,
\end{subeqnarray}
and $\xi= \sqrt{\nu/\epsilon}$. All quantities are determined up to a $\xi$-independent pre-factor, as one demands that all $\xi$-dependence is contained in the scaling function $\mathcal{G}_i$. In both cases $\eta=0$, as expected for the null-models studied. A non-vanishing exponent $\eta$ is a clear signal for non-trivial long-range behavior, (\ie when correlations on the large scale carry the signature of the interaction) which can therefore be considered as shaping the large scale. However, the inverse is not true as $\eta=0$ does not necessarily mean triviality \citep[as found in the response function for the Manna Model, ][]{Pruessner_aves:2013}, as other correlation functions and other observables might still carry the signal of an effective long-range interaction even when the response function does not. The exponent $\eta$ is normally positive, (\ie interaction) and therefore fluctuations make correlations decay quicker. Beyond $\eta=2$ the correlations decay so quickly that coarse grained local degrees of freedom display Gaussian correlations (Section~\ref{BlockScal} and \cite{Pruessner:2012:Book}). In almost all traditional models of equilibrium phase transitions, $\eta$ is a small, positive quantity, with $\eta=1/4$ in the 2D-Ising Model \citep{Stanley:1971} being the \emph{large} exception \citep[\eg][]{BergesTetradisWetterich:2000}.


\subsubsection{Surface growth}
As an example of the use of correlation functions in the numerical detection of SOC, it is instructive to apply them to the study of growth phenomena closely related to SOC, such as the Edwards-Wilkinson equation mentioned above \citep{BarabasiStanley:1995}. Traditionally, exponents in the two areas have been named differently. The roughness of an interface $\phi(\rvec,t)$ above a $d$-dimensional substrate of volume $V=L^d$ and linear extent $L$ is 
\begin{equation}
w^2(L,t)=\frac{1}{2V^2}
\int\ddint{r_1}\ddint{r_2}\ave{(\phi(\rvec_1,t)-\phi(\rvec_2,t))^2} \ .
\end{equation}
Provided $\ave{\phi(\rvec_1,t)}=0$ and assuming translational invariance, this is
\begin{equation}
w^2(L,t)=C(0,t,0,t)-\frac{1}{V^2}
\int\ddint{r_1}\ddint{r_2}
C(\rvec_1-\rvec_2,t,0,t) \ ,
\end{equation}
an example of a sum-rule. According to \citet{FamilyVicsek:1985} the roughness is expected to scale like
\begin{equation}
w^2(L,t)=a L^{2\alpha} \GC\left(\frac{L}{b t^{1/z}}\right) \ ,
\end{equation}
with metric factors $a$ and $b$, roughness exponent $\alpha$, dynamical exponent $z$ and universal scaling function, $\GC$. It is natural to trace the scaling of the roughness to that of the correlation function,
\begin{equation}\elabel{roughness}
C(\rvec,t,0,t) = \tilde{a} |\rvec|^{2\alpha}
\tilde{\GC}\left(\frac{|\rvec|}{\tilde{b} t^{1/z}}\right) \ ,
\end{equation}
even when a number of caveats apply \citep{Lopez:1999} in particular in the presence of boundaries or generally in finite systems \citep{Pruessner_growthdrift:2004}. The language of interface dynamics has a long-standing tradition in SOC and a number of deep-running links between SOC and well understood models of surface growth have been established \citep{PaczuskiBoettcher:1996,Pruessner:2003,Pruessner:2012:Book}.

Comparing \Eref{roughness} to the generalized form of an Ornstein-Zernike correlation function \Eref{def_eta} implies $\alpha=(2-d-\eta)/2$, which for $\eta=0$ reproduces the known results for the Edwards-Wilkinson equation \citep{Krug:1997}. Correspondingly, the correlation length is set by the growth time $\xi\propto t^{1/z}$. \Eref{roughness} remains valid up to a time scale set by the system size, $t\ll L^z$. After that, the correlation length is curbed by the system size, \ie $\tilde{\GC}$ in \Eref{roughness} is replaced by $\FC(|\rvec|/(\tilde{b} t^{1/z}), L/(\tilde{c} t^{1/z}))$. The same exponents characterizing \Eref{roughness} are expected to govern the two-time, two-point correlation function at stationarity,
\begin{equation}\elabel{OZcorr}
C(\rvec,t,0,0) = a |\rvec|^{-(d-2+\eta)}
\GC\left(\frac{|\rvec|}{b t^{1/z}}\right) \ ,
\end{equation}
in an extension of \Eref{def_eta}. An equivalent relation is expected to hold for the response function \Eref{def_G}.

In the presence of a cutoff, set by the system size or other limitations, the decay of correlations on the large scale is characterized by the scaling function, whose typical form is that of an exponential, \ie $\tilde{\GC}$ in \Eref{roughness} and $\GC$ in \Eref{OZcorr} are essentially exponentials. It is common practice to fit $C(\rvec,t,0,t)$ against $A |\rvec|^{\mu} \exp{-\rvec/\xi}$ with some amplitude $A$, exponent $\mu$ and correlation length $\xi$. The latter can be extracted very elegantly, up to the amplitude, by noticing that for $\eta=0$ in \Eref{OZcorr} gives $\sum_{\rvec} C(|\rvec|,t,0,t) \propto \xi^2$ to leading order in $\xi$. On a one-dimensional lattice (where $\xi$ is dimensionless) this is easily verified explicitly using \Eref{def_C1}, as 
\begin{equation}\elabel{xi_square_calc}
\sum_{i=-\infty}^\infty i \exp{-i/\xi} 
= \frac{\exp{-1/\xi}}{(1-\exp{-1/\xi})^2}
= \xi^2 -\frac{1}{6} + \OC(\xi^{-2}) \ ,
\end{equation}
but the same proportionality holds for higher dimensions. As mentioned above, the paradigmatic form of the correlation function (or the propagator) in Fourier space is 
\begin{equation}\elabel{C_Fourier_transform}
\frac{1}{\nu |\kvec|^{2-\eta} + \xi^{-2}} \ ,
\end{equation}
which, for small $\kvec$, converges to $\xi^2$, as expected since
$\sum_{\rvec}C(|\rvec|,t,0,t)$ is the 0-mode of the Fourier transform. Complicated boundary condition either spoil the structure of \Eref{C_Fourier_transform} or require orthogonal functions different from $\exp{-i \kvec \xvec}$. As such, the time separation is exemplified via the {\em time step} and the {\em iteration}, and the slower timescale moves with the number of external perturbations received by the system. Although all correlation functions discussed so far are defined on the microscopic, fast moving time scale, SOC systems normally provide a second, slow time-scale, whose time moves with the number of avalanches generated. Although theoretically less relevant, correlations have also been studied on this coarser time scale \citep{SokolovETAL:2014} which can be linked back to the microscopic dynamics \citep{PickeringPruessnerChristensen:2012,Pruessner:2012:Book}.

\subsubsection{Measuring correlation functions}
There are three main reasons why correlation functions have not received much attention in experimental, numerical, and observational work on SOC: they require high resolution data to start with; they can be technically difficult to determine \eg \citep{Anderson:1971}; they are notoriously noisy or prohibitively expensive in terms of computational effort. The reason for the latter point is not least that the correlation functions have to be determined for a range of different coordinates $\rvec_1,t_1$ and $\rvec_2,t_2$ to reveal the full functional dependence on these parameters. In the presence of boundaries, barely any of the symmetries mentioned above can be exploited to ease the computational effort. In the presence of translational invariance the discrete Fourier transform on a hyper-cubic lattice gives \citep{NewmanBarkema:1999}
\begin{equation}
\tilde{C}(\kvec,t,0,t) = \sum_{\rvec} \exp{i\kvec\rvec}
\tilde{C}(\kvec,t,0,t) = \frac{1}{N} \ave{|\tilde{\phi}(\kvec,t)|^2} \ ,
\end{equation}
where $N=\sum_{\rvec}$ denotes the number of sites and $\tilde{\phi}(\kvec,t)$ is the Fourier transform of $\phi(\rvec,t)-\ave{\phi(\rvec,t)}$, which in the presence of translational invariance equals that of just $\phi(\rvec,t)$ except for $\kvec=0$. In numerical applications, the Fourier transform is available as a  Fast Fourier Transform \citep{PressETAL:1992}. 

Where this is computationally too expensive approximative schemes can be employed \citep{HolmJanke:1993} determining the correlation length from $1/\tilde{C}(\kvec,t,0,t) \propto (\kvec^2+1/\xi^2)$ for a few $\kvec$,  \Eref{C_Fourier_transform}, at least for small $\eta$. Similarly, taking $\nabla^2$ of the correlation function numerically can produce good estimates of the correlation length, assuming the generalised Ornstein-Zernike form, \Eref{def_eta}, provided $\eta$ can be assumed to be small and in particular when $d-2+\eta=0$. Up to some prefecture,the square of the correlation length is also given by the \emph{gap} of the $0$-mode $\tilde{C}(\kvec=0,t,0,t) = \xi^2$. A direct measurement of the correlations, is often hindered by the lack of symmetry. In the presence of conservation, SOC systems have boundaries to dissipate the energy (or particles or whatever is entering the system via the driving) which means that translational invariance is broken. In that case, many of the standard techniques fail when they rely on a standard Fourier transform.


\begin{figure*}
\centering
\hfill
\subfigure[Activity correlations in a linear-linear plot.\flabel{Manna_corr_activity_linear}]{\includegraphics[width=7cm]{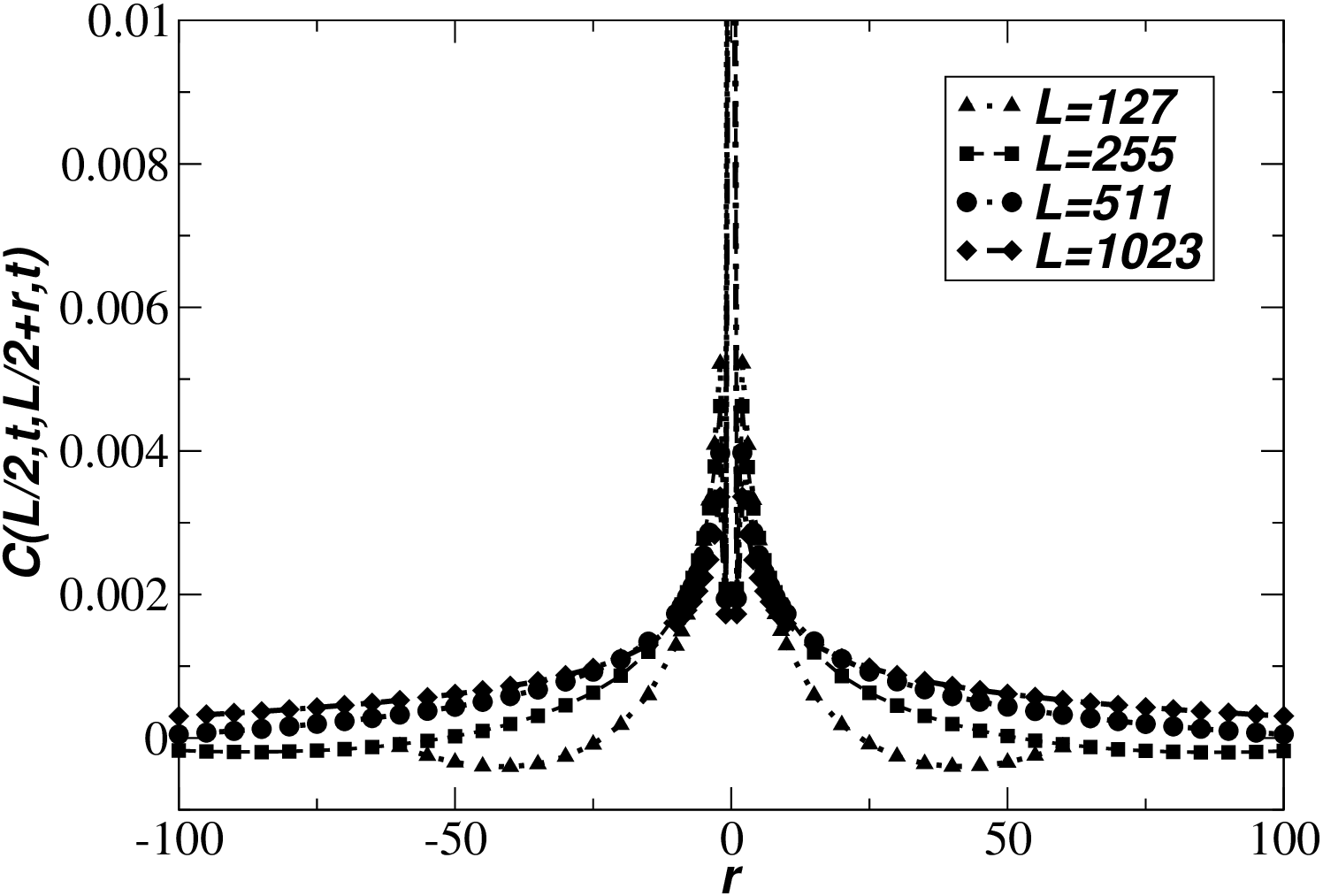}}
\hfill
\subfigure[Collapse of activity correlations in a double-logarithmic plot.\flabel{Manna_corr_activity_collapse}]{\includegraphics[width=7cm]{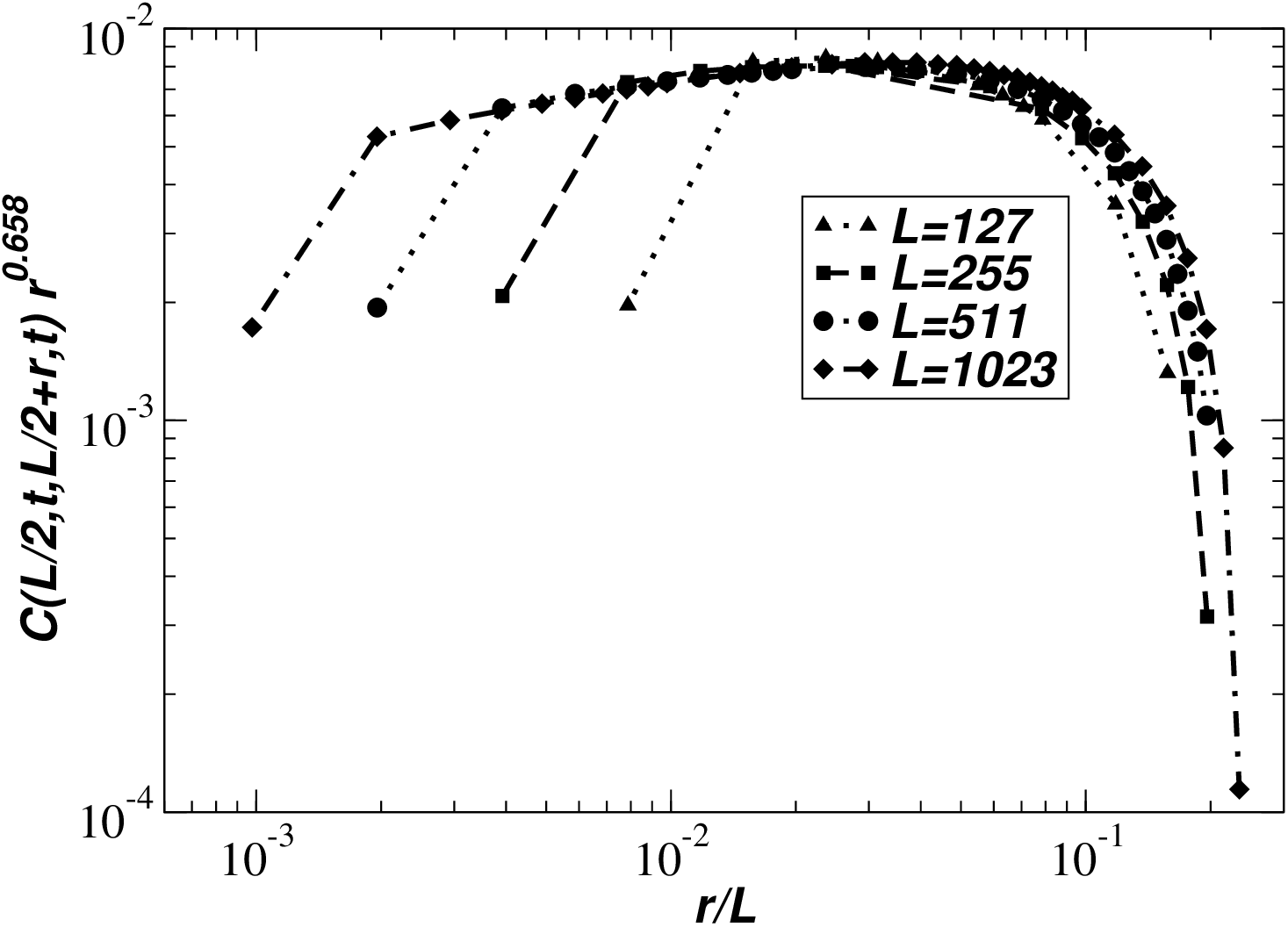}}
\hfill
\caption{\flabel{Manna_corr_activity}The two-point correlation function $C(L/2,t,L/2+r,t)$ of the activity in the Abelian version of the one-dimensional Manna Model \citep{Manna:1991a,Dhar:1999c}. In the language adopted in \Eref{def_C}, $\phi(\rvec,t)$ is the level of activity (\ie at a certain point in space $\rvec$ and a certain microscopic time $t$ the level of avalanching is a Poisson process with unit rate times the number of pairs ready to topple) measured in the middle, $r_2=L/2$, and across the lattice, $r_1=L/2+r$. \subref{fig:Manna_corr_activity_linear} shows the data on a linear scale. That they collapse nicely according to \Eref{def_eta} can be seen in \subref{fig:Manna_corr_activity_collapse}, where the scaling of the abscissa is shown to be compatible with the assumption that the correlation length scales linearly in the system size.}
\end{figure*}

\subsubsection{Example: The Manna Model}
\slabel{Manna_Model}

\begin{figure*}
\centering
\hfill
\subfigure[Substrate correlations in a linear-linear plot.\flabel{Manna_substrate_linear}]{\includegraphics[width=7cm]{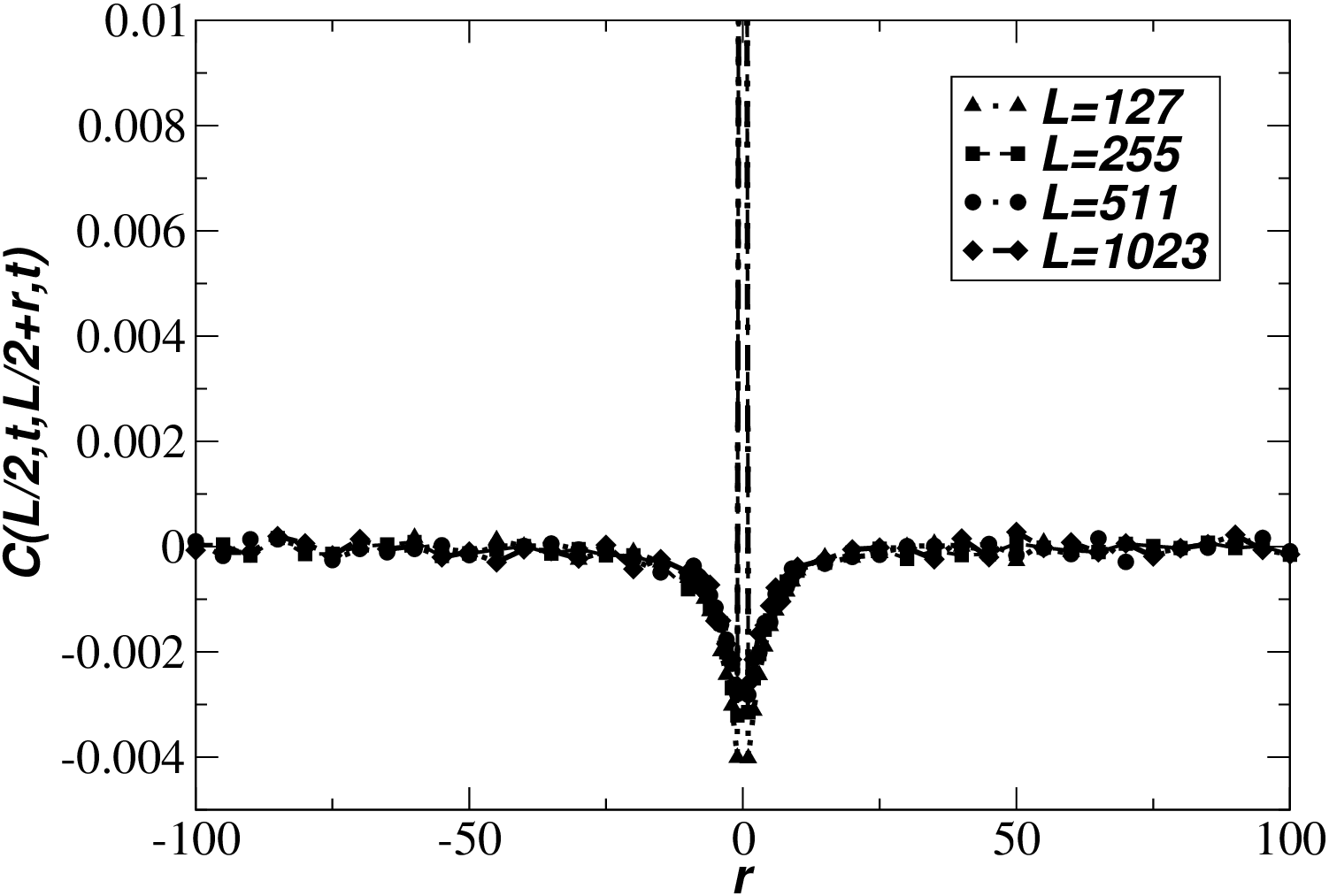}}
\hfill
\subfigure[Attempted collapse of substrate correlations.\flabel{Manna_substrate_collapse}]{\includegraphics[width=7cm]{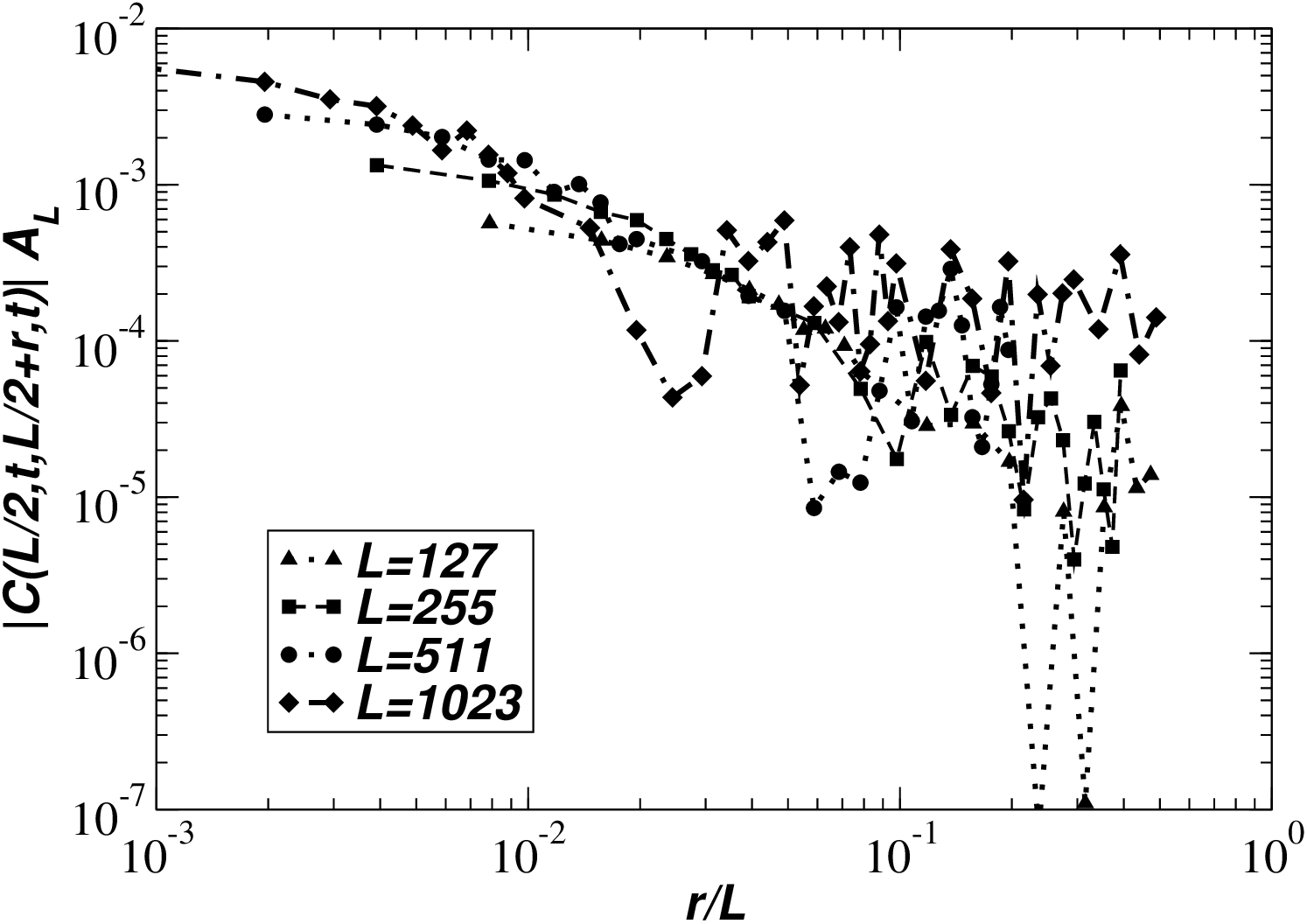}}
\hfill
\caption{\flabel{Manna_substrate}Similar to \Fref{Manna_corr_activity} these plots show the correlations in the inactive particles of the Manna Model (the {\em substrate}) measured during periods of quiescence. In the language of \Eref{def_C} $\phi(\rvec,t)$ is the number of particles resting on $\rvec$ and time $t$, measured between avalanches. \subref{fig:Manna_substrate_linear} suggests some short-ranged correlations, but it also indicates no discernible difference of these correlations for different system sizes. This is confirmed by the failure of the attempted collapse in \subref{fig:Manna_substrate_collapse}, where the amplitudes $A_L$ to rescale the data along the ordinate have been chosen as to facilitate the best collapse.}
\end{figure*}

\Fref{Manna_corr_activity} shows data of $C(L/2,t,L/2+r,t)$ for the Abelian, one-dimensional Manna Model \citep{Manna:1991a,Dhar:1999c} whose correlation function can be determined comparatively easily. In the Manna Model each site is occupied by a non-negative number of particles. As long as any site carries more than one particle, that site redistributes two of them among independently chosen nearest neighbors, potentially making them exceed the threshold and thereby giving rise to an avalanche. While the particle number is conserved in the bulk, sites ÒtopplingÓ along the open boundary can lose one or two particles by moving them outside the lattice. The Manna Model is normally started from an empty lattice, and driven whenever the system is quiescent by depositing particles at randomly, uniformly-chosen sites. The activity for this model is defined in the following as the number of pairs on a site about to be re-distributed. The activity-activity correlation function in \Fref{Manna_corr_activity} displays a long-ranged decay, whose scaling behaviour, however, becomes apparent only when plotted double logarithmically. In fact, the data can be collapsed acceptably well according to \Eref{def_eta} with $\xi=L$ and $d-2+\eta\approx0.658$, \ie $\eta\approx1.658$, rather large compared to, say, $\eta=1/4$ in the Ising Model. Further, the scaling of the two-point activity (\ie activity-activity) correlation function in the Manna model thus differs significantly from that of the propagator $G$, which is known to remain classical, $\eta=0$ (\ie of the form \Eref{G_n}), in the stationary state \citep{Pruessner_aves:2013}.

Various identities exist relating exponents of the activity to exponents of the avalanches \citep[][in particular p.~340]{LuebeckHeger:2003a,Pruessner:2012:Book}. The variance of the activity density, $\Delta \rho_a/L^d$, is expected to scale like $L^{\gamma'/\nu_{\perp}-d}$ \citep{Luebeck:2004}, which is related to $C(\rvec_1,t,\rvec_2,t)$ by the sum rule,
\begin{equation}
\elabel{drho}
\frac{\Delta \rho_a}{L^d}=
\frac{1}{L^{2d}}\int\ddint{r_1}\ddint{r_2}
C(\rvec_1,t,\rvec_2,t) \propto L^{-(d-2+\eta)} \ ,
\end{equation}
which reproduces the well known Fisher scaling law \citep{Stanley:1971} $\nu_{\perp}(2-\eta)=\gamma'$. In the present case $\gamma'/\nu_{\perp}=0.41\pm0.04$ and therefore $\eta=1.59\pm0.04$ \citep{Luebeck:2004} and $2-\eta\approx0.342$ measured above suggests a slight mismatch, which might be explained by finite size effects.


In contrast, \Fref{Manna_substrate} shows the correlations in the inactive particles in the Manna Model (\ie particles that are not moving around) measured during times of quiescence when no avalanche is running. While correlations do exist over a small number of lattice sites, the correlation length does not change with system size. This is clearly visible in \Fref{Manna_substrate_linear} as the data collapses without the need of any rescaling. In fact, the attempted collapse in \Fref{Manna_substrate_collapse} is very poor and does in fact show no sign of scaling. This finding agrees with recent field-theoretical work \citep{Pruessner:2014:FT} which suggests that correlations in the substrate (\ie the background of inactive particles) are either irrelevant or enter only in a very subtle way that is insignificant at large temporal and spatial scales. In other words, the substrate is an unsuitable place to look for correlations and SOC takes place during avalanching, not during quiescence. However, this finding disagrees with the traditional view that the SOC state is one of subtle correlations stored in the substrate \citep[\eg][]{ChristensenOlami:1992b,Lise:2002}. Finally, we note that correlations in the substrate are mostly anti-correlations, \ie fluctuations above the mean are repelling each other. In other words, wherever unusually many particles are found at one point, the environment is depleted, suggesting that the dynamics has led to a pile-up. Again, that ties in well with the self-organization maintaining a particular density of particles, with fluctuations only due to some local re-shuffling.



\subsection{Structure Functions}
\label{strfunc}
\begin{figure*}
\centering
\includegraphics[width=13cm]{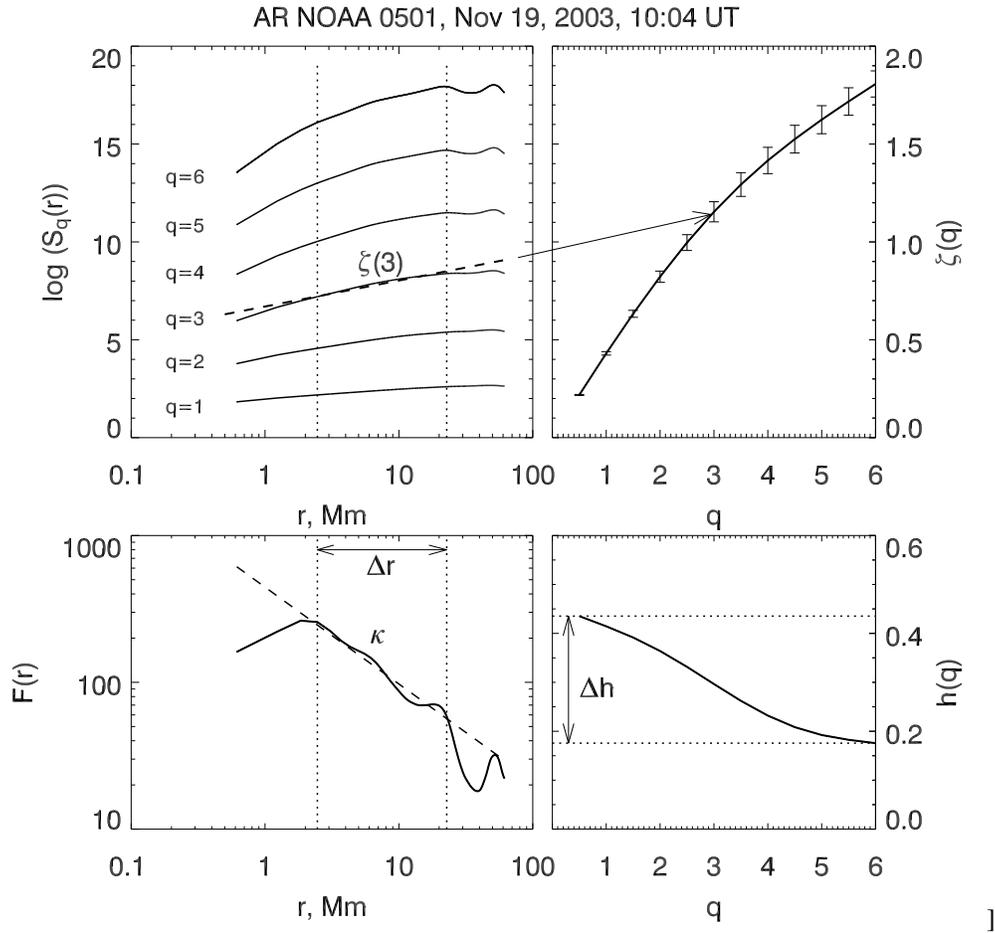}]
\caption{Structure functions $S_q(r)$ ({\it upper left}) calculated from a magnetogram of active region NOAA AR 10501 by Equation (\ref{Sqr}). {\it Lower left:} - flatness function $F(r)$  calculated from the structure functions by Equation (\ref{Fr}) where $S_q(r)$ use the longitudinal component of the magnetic field for {\bf u}. Vertical dotted lines mark the interval of multifractality, $\Delta r$,where flatness grows as power law when $r$ decreases. The interval $\Delta r$ is also marked in upper left frame. The power index $\kappa$ is determined within $\Delta r$. The slope of $S_q(r)$, defined for each $q$ within $\Delta r$, is $\zeta(q)$ function ({\it upper right}), which is a concave for a multifractal and straight line for a monofractal. {\it Lower right: - } function $h(q)$ is a derivative of $\zeta(q)$. The interval between the maximum and minimum values of $h(q)$ is defined as a degree of multifractality, $\Delta h$. }
\label{figu1}%
\end{figure*}

The structure function provides another widely-used, two-points, statistical moment of a random variable in a critical system that can be used to study scaling behavior and inter-scale connections. A phenomenological analogy with the autocorrelation function shows that the {\it product} of field values in two points in the autocorrelation function is replaced by the absolute value of the {\it increment} in the definition of the structure function. The replacement offers an opportunity to consider various powers, $q$, of the increment, and thus to explore the high-order statistical moments, which, in turn, uncover the multifractality and intermittency properties of a system under study. The structure functions were first introduced by \citet{kol41} (hereafter K41) in developing his turbulence theory. Note that the solar photospheric plasma - the medium to which a bulk of our further discussion is applied - is in a state of highly developed turbulence. Structure functions are defined as statistical moments of the increments of a turbulent field ${\bf u }({\bf x})$ as 
\begin{equation} 
S_q(r) = \langle |{\bf u }({\bf x}  + {\bf r}) - {\bf u}({\bf x})|^q
\rangle, 
\label{Sqr}
\end{equation} 
where  ${\bf r}$ is a separation vector, and $q$ is a real number. In the original K41 theory, ${\bf u }({\bf x})$ is assumed to be a fluctuating velocity field, however the structure functions technique is applicable for any random variable, in both temporal and spatial domains,  \citep[\eg][]{sto92, con99, buc06, uri02}. For example, in Figures~\ref{figu1}--\ref{figu5} the structure function technique is applied for the longitudinal component of the photospheric magnetic field. Structure functions, calculated within the inertial range of scales, $r$, ($\eta \leq r \leq L$, where $\eta$ is a spatial scale where the influence of viscosity becomes significant and $L$ is a scaling factor for the whole system) are described by a power law \citep{kol41, mon75, fri95}, 
\begin{equation}  
S_q(r) \sim  (\varepsilon_r({\bf x}) \cdot r)^{q/3} \sim (r)^{\zeta(q)}.
\label{sq3} 
\end{equation} 
where $\varepsilon_r({\bf x})$ is the energy dissipation, averaged over a sphere of size $r$. 

The function $\zeta(q)$ describes one of the most important characteristics of a turbulent field. In order to estimate this function, Kolmogorov assumed that for fully developed turbulence (\ie turbulence at high Reynolds number when the inertial force vastly exceeds the viscous force), the probability distribution laws of velocity increments depend only on the first moment, $\bar \varepsilon$, of the function $\varepsilon_r({\bf x})$. Replacing $\varepsilon_r({\bf x})$ in equation (\ref{sq3}) by $\bar \varepsilon$ we have
\begin{equation}  
S_q(r) \sim  (\bar \varepsilon \cdot r)^{q/3} = C \cdot r^{q/3},  
\label{sq31} 
\end{equation}
where $C$ is a constant. As a result, function $\zeta(q)$ is defined as a straight line with a slope of $1/3$ 
\begin{equation} 
\zeta(q) = q/3. \label{q_by_3}.
\end{equation}
Kolmogorov further realized (see also formulation of Landau's objection concerning the original K41 theory in Frisch 1995) that such an assumption is very rigid and turbulent state is not homogeneous across spatial scales. There is a greater spatial concentration of turbulent activity at smaller scales than at larger scales. This indicates that the energy flow and dissipation do not occur everywhere, and that the energy dissipation field should be highly inhomogeneous, intermittent, and follows a power law,
\begin{equation} 
\langle(\varepsilon_r({\bf x})^{p}\rangle \sim r^{\tau(p)},  
\label{erq} 
\end{equation} 
where $p$ is a real number. Then equation (\ref{sq3}) may be rewritten as
\begin{equation}
S_q(r) \sim  (\varepsilon_r({\bf x}) \cdot r)^{q/3} =
(\varepsilon_r({\bf x}))^{q/3} \cdot r^{q/3} = r^{\tau(q/3)} \cdot
r^{q/3} 
\label{sq33} 
\end{equation} 
or
\begin{equation} \zeta(q) = \tau(q/3) + q/3. 
\label{zetaq}
\end{equation}
Equation (\ref{zetaq}) is referred to as the refined Kolmogorov's theory of fully developed turbulence \citep{kol62a, kol62b, mon75, fri95}. One can see from equation (\ref{zetaq}) that the function $\zeta(q)$ deviates from the straight $q/3$ line - the deviation is caused by the scaling properties of a field of energy dissipation.

Important information on a turbulent field can be derived from the functions $\zeta(q)$ that can be obtained from experimental data. The value of the function at $q=6$ deserves special attention because it defines a power index
\begin{equation}
\beta \equiv 1 - \zeta(6)  
\label{eqn:beta} 
\end{equation}
of a spectrum $E^{(\varepsilon)}$ of energy dissipation $\varepsilon({\bf x})$:
\begin{equation} 
E^{(\varepsilon)} (k) \sim k^{\beta}, 
\label{Ek}
\end{equation}
where $k$ is a wave number as discussed in Section~\ref{turb} below. By measuring $\zeta(q)$ from experimental data and using equation (\ref {zetaq}) one can calculate the scaling exponent $\tau (q/3)$  in equation (\ref {erq}) for the energy dissipation field. The derivative of $\zeta(q)$, 
\begin{equation}  
h(q) \equiv \frac {{\rm d} \zeta(q)}{{\rm d} q}, 
\label{h} 
\end{equation} 
can also be obtained by using the $\zeta(q)$ function (Figure \ref{figu1}, {\it right bottom}). The deviation of $h(q)$ from a constant value is a direct manifestation of {\it intermittency} in a turbulence field, which is equivalent to the term {\it multifractality} in fractal terminology (see further discussion in Section~\ref{turb}).

\subsubsection{The flatness function as an output of two structure functions}
The weakest point in the above technique is to determine the scale range, $\Delta r$, where the slope $\zeta(q)$ is to be calculated (see Figure \ref{figu1}). \cite{Abramenko05b} used the flatness function, defined as a ratio of the fourth statistical moment to the square of the second statistical moment, to visualize the range of multifractality, $\Delta r$. Another option is to use higher statistical moments to calculate the (hyper-)flatness, namely, the ratio of the sixth moment to the cube of the second:
\begin{equation} 
F(r)=S_6(r)/(S_2(r))^3.
\label{Fr}
\end{equation} 
For monofractal structures, the flatness, $F(r)$ is not dependent on the scale, $r$. On the contrary, for a multifractal structure, the flatness grows as a power law, when the scale $r$ decreases: $F(r)\sim k^{\kappa}$. The interval $\Delta r$ of the power law is well defined between the two cutoffs of the spectrum (see Figure \ref{figu1}, bottom left). The power index of the flatness function, $\kappa$, can be used as a measure of multifractality - more complex structures have steeper $F(r)$ spectra. Moreover, the interval $\Delta r$ outlines the range of scales where the property of multifractality and intermittency is met.
 
\subsubsection {Connection to the multifractality spectrum, $f(\alpha)$ }
\begin{figure*}
\centering
\hfill
\subfigure[high-flaring NOAA AR 9077]{\includegraphics[width=8cm]{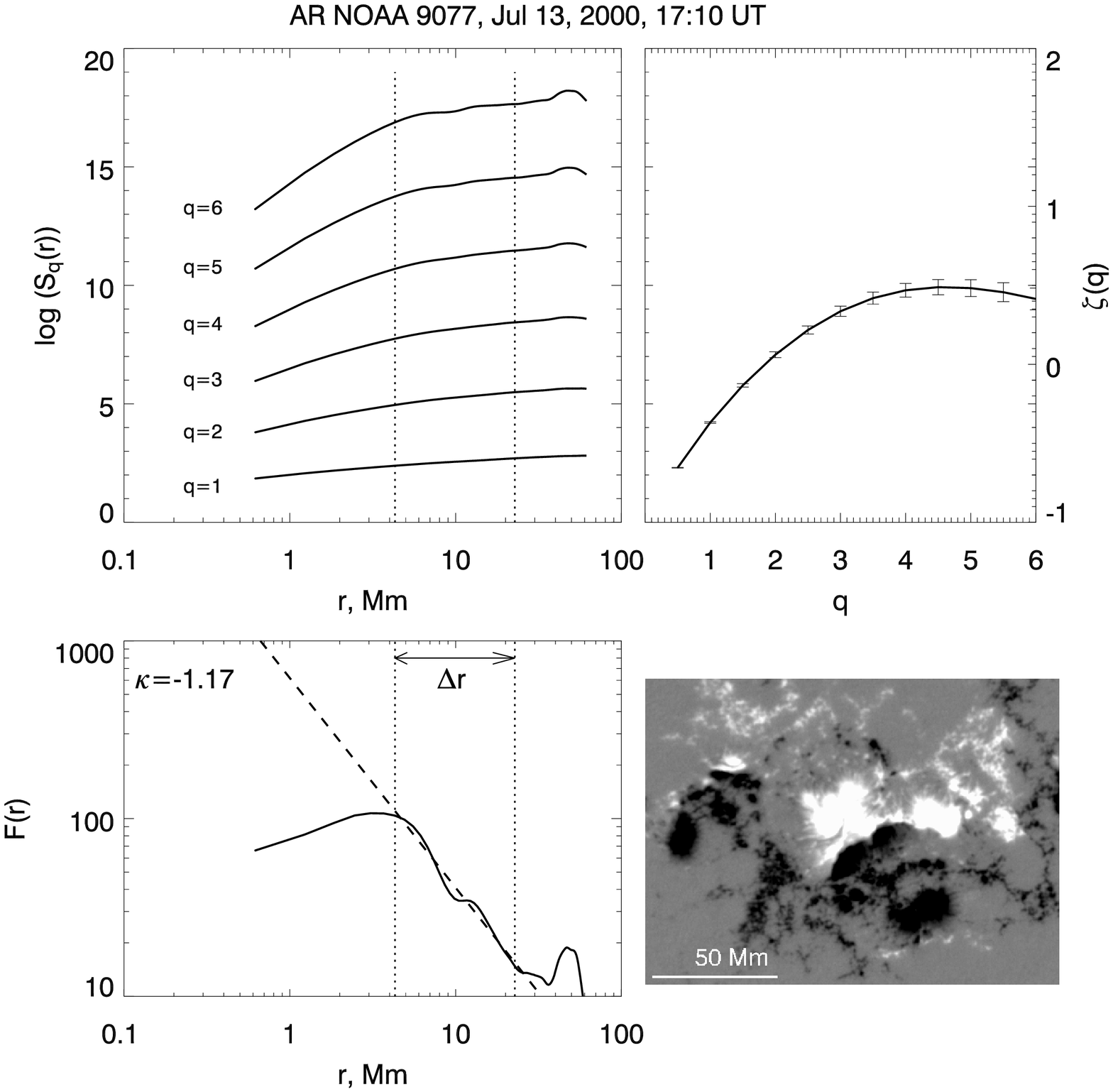}}
\hfill
\subfigure[low-flaring NOAA AR 10061]{\includegraphics[width=8cm]{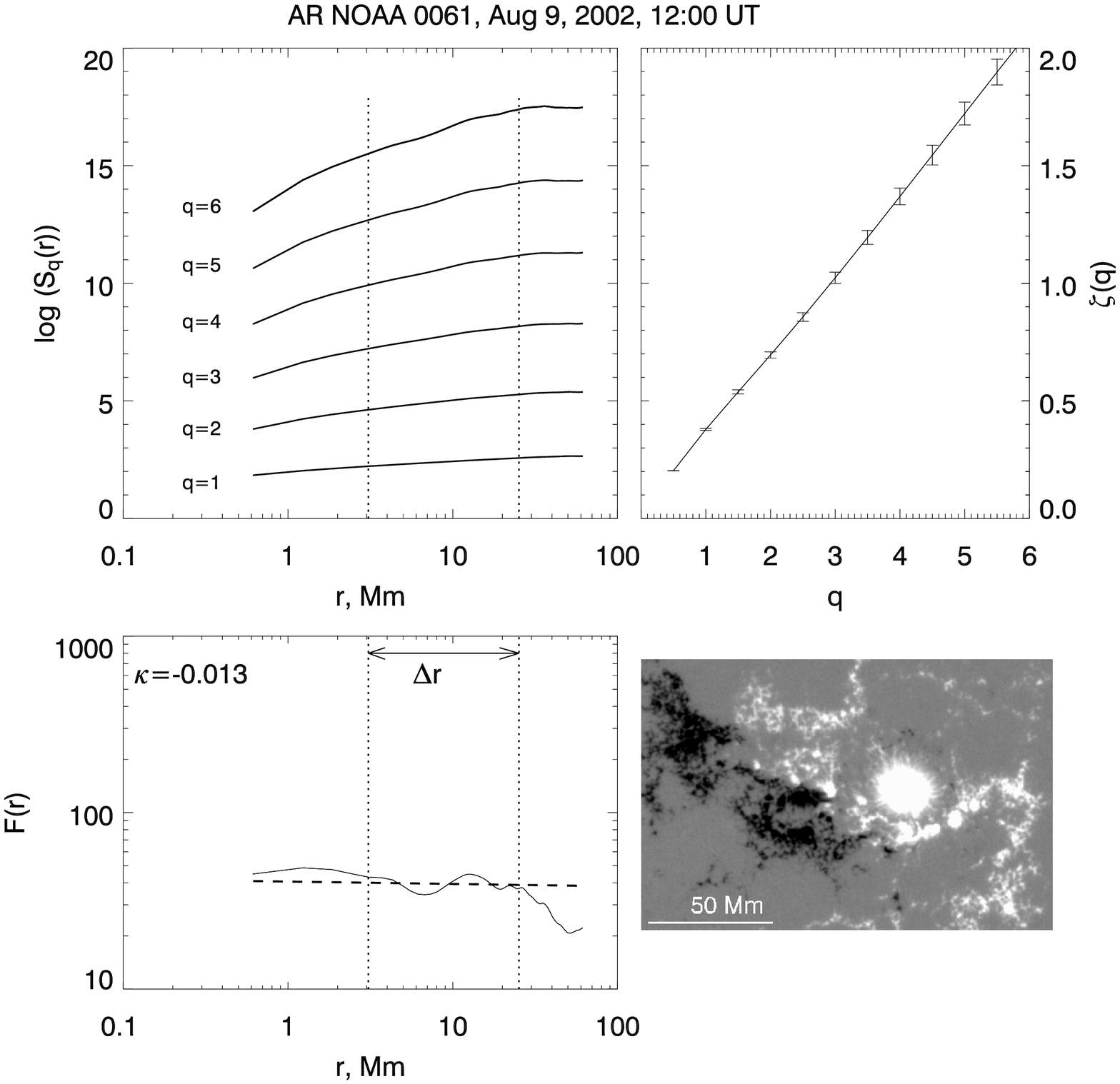}}
\hfill
\caption{Structure functions $S_q(r)$, flatness function $F(r)$ and $\zeta(q)$ function from a magnetogram of high-flaring NOAA AR 9077 ({\it left}, $\Delta h = 0.48$), and from a magnetogram of low-flaring NOAA AR 10061 ({\it right}, $\Delta h = 0.06$). The multifractality index  $\kappa$ is the slope of $F(r)$ calculated inside $\Delta r$. Other notations are the same as in Figure \ref{figu1}. }
\label{figu2}
\end{figure*}

\begin{figure*}
\centering
\includegraphics[width=13cm]{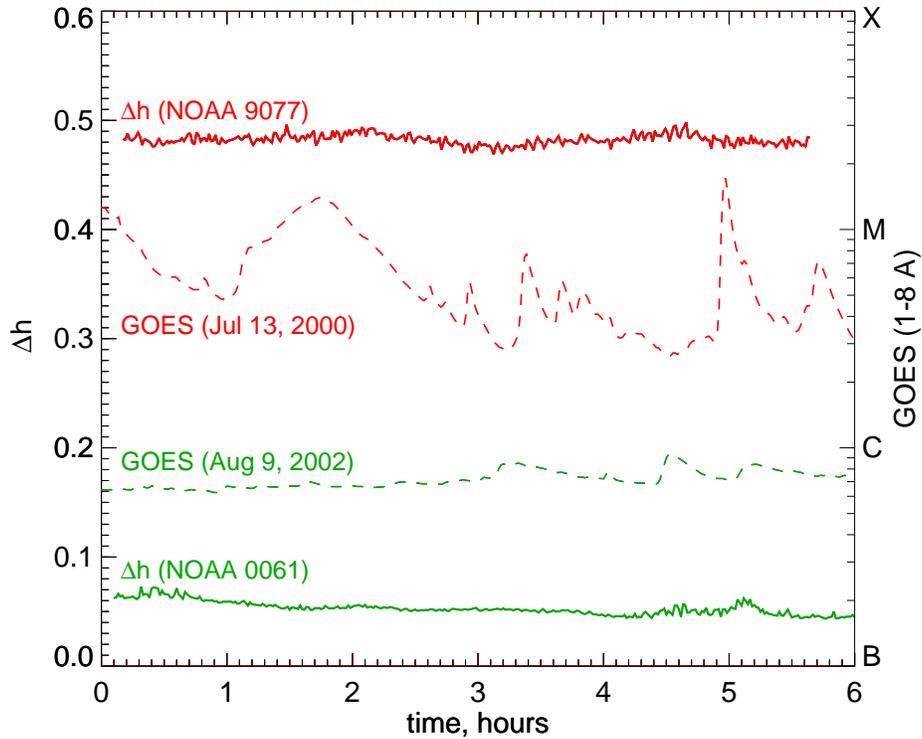}
\caption{Time variations of the measure of multifractality,  $\Delta h$ ({\it left axis}), and GOES soft X-ray flux ({\it right axis, dashed lines}) plotted for six-hour time intervals for the two active regions. Data for NOAA AR 9077 ({\it red lines}) were obtained between 17:00 and 23:00 UT on July 13, 2000 and data for NOAA AR 10061 ({\it green lines}) refer to an interval between 11:00 and 17:00 UT on August 9, 2002. }
\label{figu3}
\end{figure*}

The function $\zeta(q)$ is a straight line for a monofractal, due to a global scale-invariance, whereas it has a concave shape in case of a multifractal. The degree of concavity is usually measured by function $h(q) =\zeta(q) /dq$. All values of $h$ within some range are permitted for a multifractal. For each value of $h$ there is a monofractal with an $h$-dependent dimension $D(h)$ at which the scaling holds with exponent $h$. This representation of multifractality is based on the increments of the field and has its roots in the K41 theory of turbulence. A second representation is based on the dissipation, $\varepsilon$, of the field energy, which relies on the K41 result stating that field increments over a distance $r$ scale as $(\varepsilon r)^{1/3}$, known as the refined similarity hypothesis \cite{mon75}. In multifractal terminology, the refined scaling hypothesis means that for any singularity of exponent $\alpha$ of $\varepsilon r$, there exists an associated singularity of exponent $h=\alpha/3$ for the field of the same set, which has the same dimension $D(h)$. Usually, it is very difficult to measure the local dissipation in the 3D space, and so one-dimensional space averages of the dissipation are typically used. The corresponding dimension $f(\alpha) = D(h) - (d-1)$ is lowered by two units (for the space dimension $d=3$) where one-dimensional cuts of a 3D structure are taken. In the literature $f(\alpha)$ is often referred as the multifractality spectrum \citep[\eg][]{fed88, Lawrenceetal93, fri95, sch00, Conlonetal08, McAteer10}. The values of $D(h)$, in turn, can be  calculated as a Legendre transform of $\zeta(q)$ \cite{fri95},
\begin{equation}
D(h(q)) = inf_{q}(d + q h(q) - \zeta(q)). 
\label{Dh} 
\end{equation}  
When $\zeta(q)$ is concave, then for a given real value of $q$ the extremum in Eq. \ref{Dh} is attained at the unique value $h_{o}(q)$, and 
\begin{equation}
D(h_{o}(q)) = d + q h_{o}(q) - \zeta(q)).
\label{Dh2} 
\end{equation}  

\begin{figure*}
\centering
\includegraphics[width=13cm]{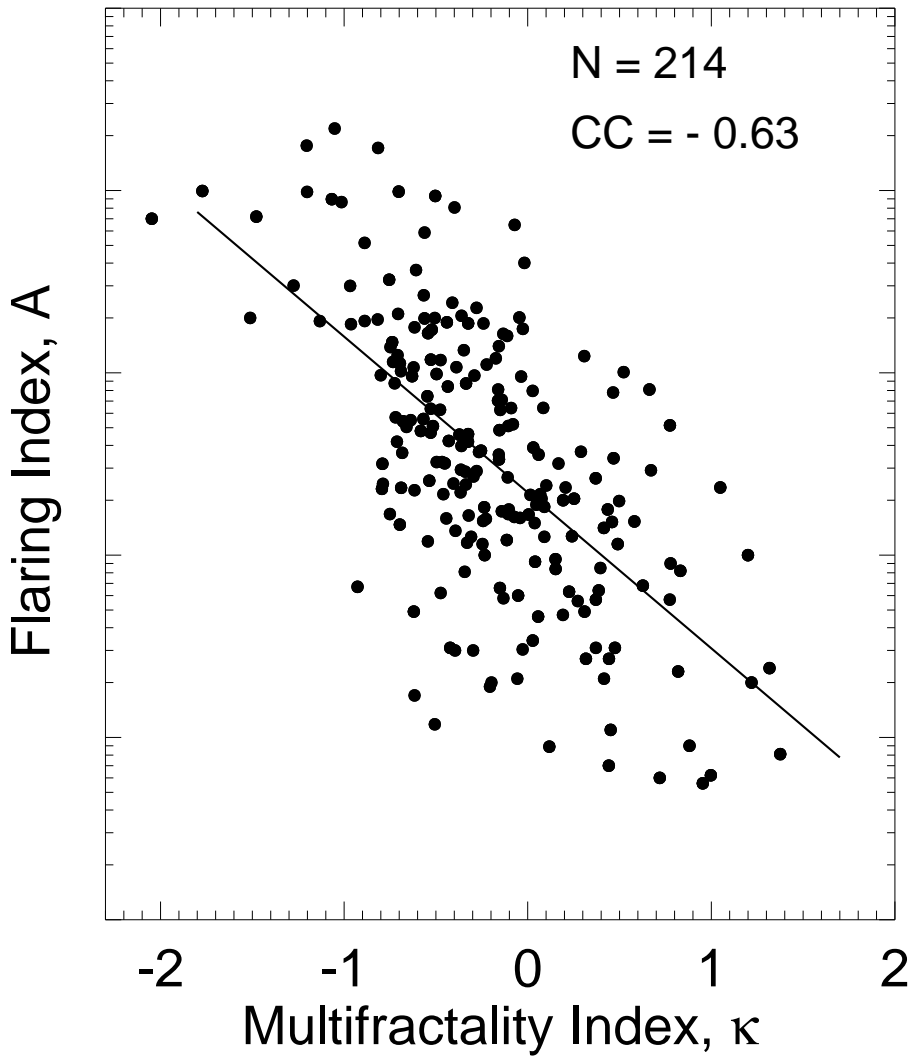}
\caption{Flaring index, $A$, plotted versus the multifractality index, $\kappa$, for 214 regions. The Pearson correlation coefficient is -0.63. From \cite{Abr10}.)}
\label{figu4}
\end{figure*}

\begin{figure*}
\centering
\hfill
\subfigure[high-flare productivity]{\includegraphics[width=8cm]{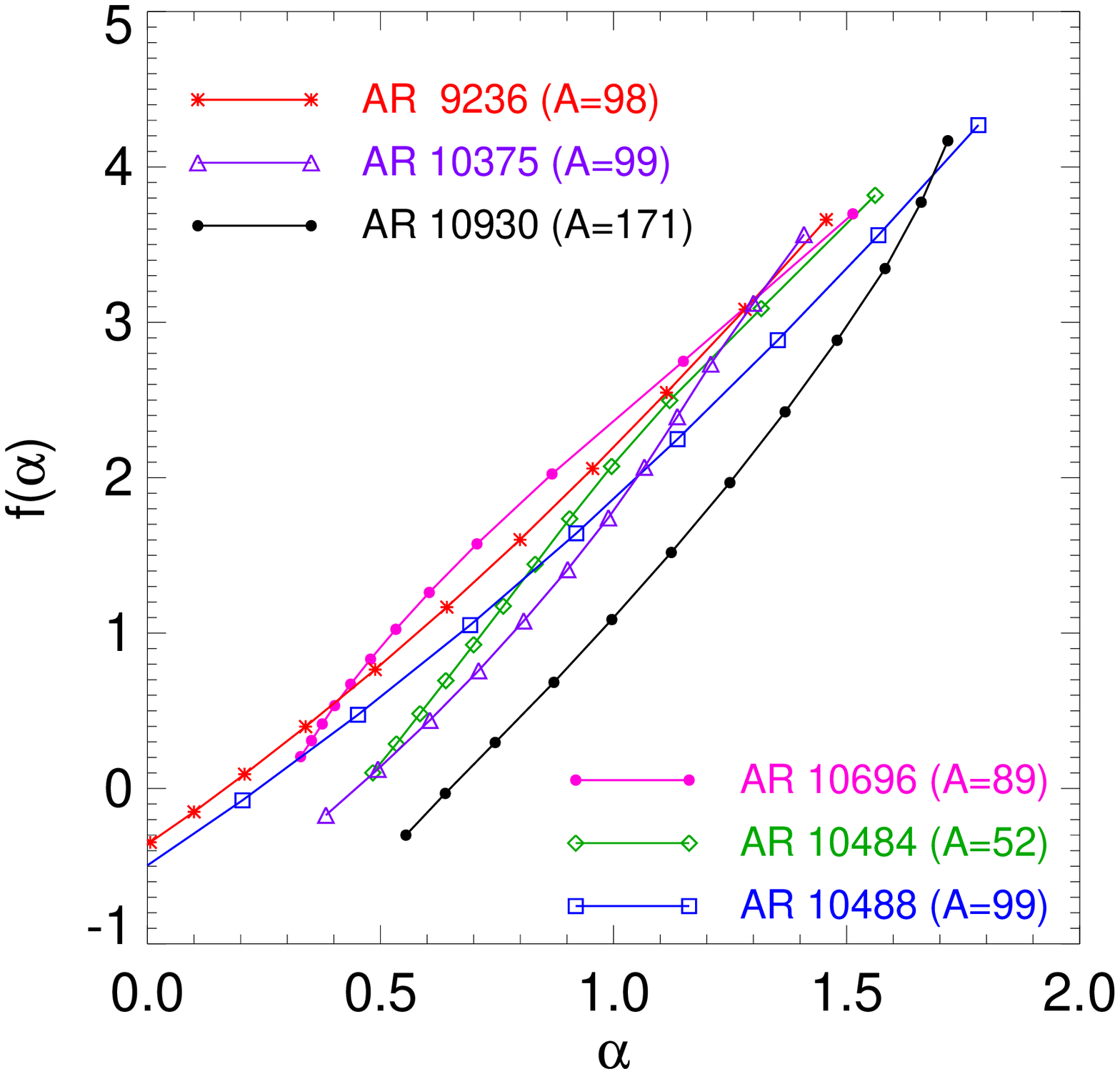}}
\hfill
\subfigure[low-flare productivity]{\includegraphics[width=8cm]{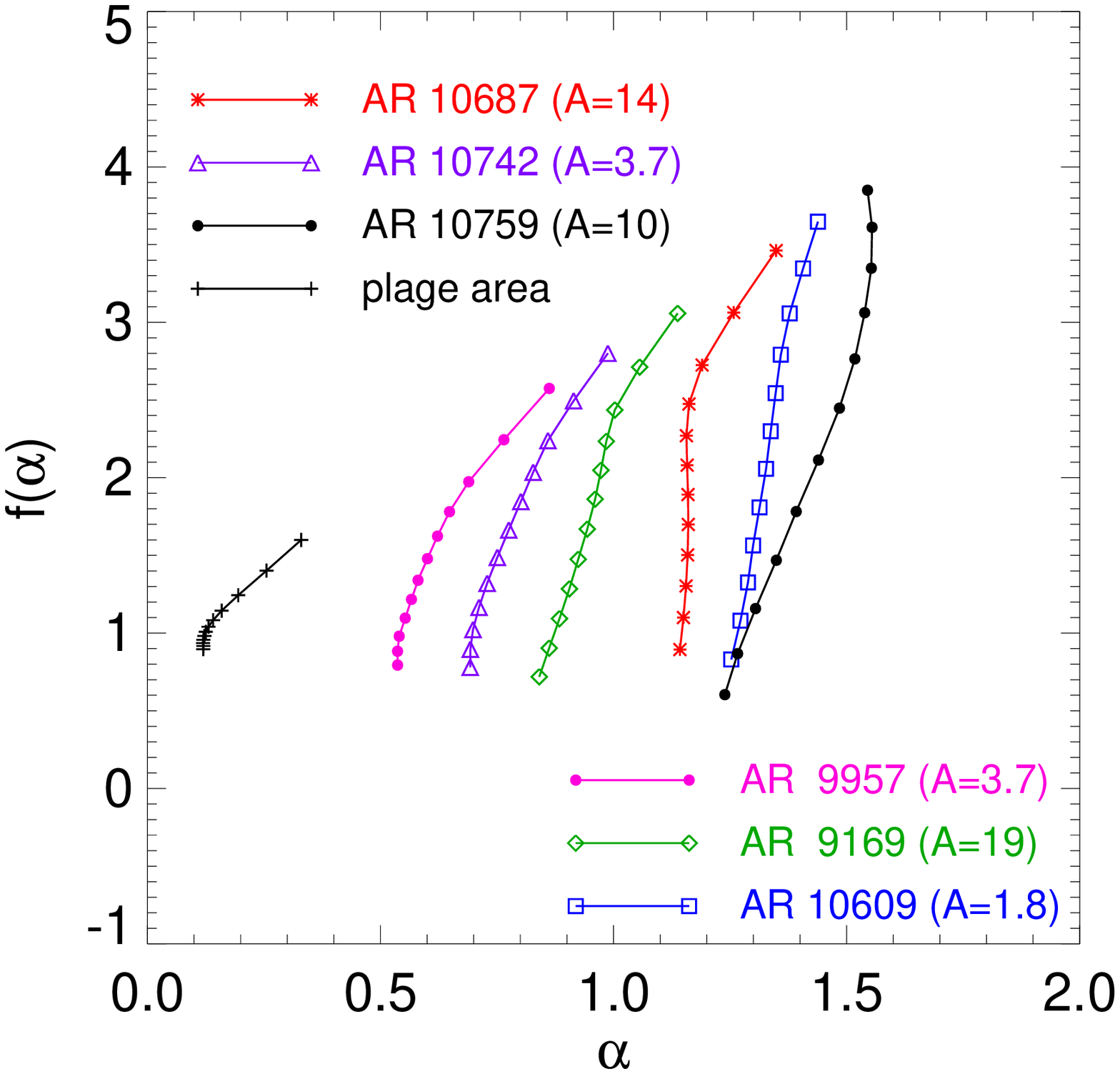}}
\hfill
\caption{Multifractality spectra, $f(\alpha)$, plotted: {\it left - } for regions of high flare productivity; {\it right -} for regions of low flare productivity and for a plage area. Spectra on the left frame are more broad than that on the right frame.}
\label{figu5} 
\end{figure*}

The result of the structure function method as applied to solar active region magnetograms are presented in Figure \ref{figu2} \citep{abr02, Abramenko05b, Abramenko05a, Abr10} . The scaling behavior of the structure functions is different for each region. For the complex and flare-productive NOAA AR 9077 there is a well-defined range of scales, $\Delta r = (4-23)$ Mm where flatness $F(r)$ grows with the power index $\kappa =-1.17$ as $r$ decreases. Function $\zeta(q)$ is concave and the corresponding $\Delta h \approx 0.5$. This implies a multifractal structure of the magnetic field in this active region. To the contrary, the simple non-flaring NOAA AR 10061 (Figure \ref{figu2}, right) exhibits a flatness function that undulates around a horizontal line, which implies a monofractal character of the magnetic field. The function $\zeta(q)$ is nearly a straight line with a vanishing value of $\Delta h \approx 0.05$. Time profiles of $\Delta h$ for the two active regions are compared in Figure \ref{figu3}. The non-flaring NOAA AR 10061 persistently displays lower degree of multifractality, as well as lower X-ray flux, than the flaring  NOAA AR 9077 does. Figure \ref{figu4} demonstrates the statistical relationship between the multifractality index, $\kappa$ and a flaring index, $A$ for 214 regions \citep{Abr10}, from which it is clear that the higher degree of multifractality of the magnetic field may be associated with stronger flare productivity of an active region. Here the flare index $A$ characterizes the flare productivity of an active region per day, being equal to 1 (100) when the specific flare productivity is one C1.0 (X1.0) flare per day. More examples of multifractality spectra $f(\alpha)$ are shown in Figure \ref{figu5} \citep{Abr10}. One can see that the most complex and flare-productive regions (left frame in Figure \ref{figu5}) exhibit broader spectra as compared to that of non-flaring regions (right frame). This means that a set of monofractals that form an observed multifractal, is much more broad in flare-productive regions as compared to non-flaring regions. 

\clearpage
\subsection{Application Oriented Methods}
\label{aom}



As discussed in Section~\ref{af}, the classical autocorrelation SOC detection methods are explicit in theory, but are often challenging in terms of practical application to physical systems, such as the solar atmosphere or the tectonic environment. Over the previous 25 years, and through the evolution of several numerical SOC models created to explain existing physical systems, a variety of application-oriented methods have been developed that together comprise a useful toolkit for the detection of the SOC state. In principle, when the SOC state is reached the system experiences instabilities of all sizes, clustered in cascades of elementary events, or avalanches, all triggered by fixed and small (with respect to the critical threshold) or variable, but statistically small, perturbations (for the latter set of SOC models, see \cite{GeorgoulisVlahos96, GeorgoulisVlahos98}). The main feature of this marginally stable SOC state, where a given small perturbation can cause avalanches of all sizes \citep[\eg][]{kad91, new96} is precisely the absence of a preferred scale for avalanche size. This leads to robust power laws if one examines the distribution function of the event sizes (Section~\ref{powerlaws}). In this sense, a nonlinear dynamical system realizes the SOC state as a statistically-stationary state far from equilibrium. We review these two attributes of marginal stability and statistical stationarity as practical detection methods for an SOC state. We then present a recent non-evolutionary diagnostic SOC-state test and finally discuss block-scaling methodology is detail.

\subsubsection{Marginal stability: a spatially averaged critical quantity}
The diagnostic SOC detection method of marginal stability is based on the stabilization of a spatially averaged system parameter, \ie the parameter compared with the critical threshold. Applying this method to the classical 2D cellular automaton sandpile model of \cite{Baketal87}, it is assumed that each point $i=(x,y)$ of the square grid corresponds to the space occupied by a sand grain. The field variables in this model are the height $h(x,y,t)$ and the slope $G(x,y,t)$ of the accumulated sand at every point, $i=(x,y)$ of the system and in every time step $t$, of its evolution. Referring to the classical cellular automaton, both space and time are discretized: the automaton consists of a discrete grid , e.g., (x,y) in 2D, where each grid site has a position vector $i$ with integer components. The automaton also has two discrete time-scales, namely an integer time step, t, that increases by one with each application of the automaton rules, and an integer iteration that increases by one each time the system is perturbed. The slope $G(x,y,t)$ at a specific point of this automaton's sandpile and for the specific time $t$ is defined as the height difference between the height $h(x,y,t)$ at the point $i=(x,y)$ and the average height of the adjacent grid points $\bar{h}(t)$,
\begin{equation}
\elabel{c3e15} 
\bar{h}(t)=\frac{1}{4}[h(x+1,y,t)+h(x-1,y,t)+h(x,y+1,t)+h(x,y-1,t)] \ .
\end{equation}
Therefore, the slope $G(x,y,t)$ is defined as $G(x,y,t)=h(x,y,t)-\bar{h}(t)$.
The {\it transition} rules describing the evolution of the system when a sand grain is added at a random point $i=(x,y)$ of the grid at time $t$ are defined as $h(x,y,t)\rightarrow h(x,y,t)+1$.
The instability criterion embedded in the transition rules of the system reflects a critical value of the slope $G_{c}$. A point $i=(x,y)$ of the system is considered unstable when the inequality $G(x,y,t)>G_{c}$ is fulfilled.
When such an instability occurs at the point $i=(x,y)$ and at the time $t$, then the dynamical system responds at the time $t+1$ according to the following evolution or redistribution rules, 
\begin{eqnarray}\label{c3e19}
h(x,y,t+1)=h(x,y,t)-4  \ , \\ 
h(x\pm 1,y,t+1)= h(x\pm 1,y,t)+1 \ , \\
h(x,y\pm 1,t+1)= h(x,y\pm 1,t)+1 \ .
\end{eqnarray}
Transition and evolution rules comprise the driving and relaxation mechanisms, respectively, that inexorably lead the system to marginal stability. A practical SOC-state detection mechanism based on this marginal stability reached by system in such a state was presented by \cite{GeorgoulisThesis00}. This mechanism monitored the temporal evolution of the mean value of the field variable(s) that determine(s) the instability threshold for the system. For the \cite{Baketal87} model described above, \cite{GeorgoulisThesis00} monitored the temporal evolution of the mean height $\bar{H}(t)$ of the sandpile throughout the grid, where $\bar{H}(t)=\frac{\int h(\textbf{i},t)d\textbf{i}}{\int d\textbf{i}}$, with $\textbf{i}$ being the position vector. Equivalently, one can monitor the temporal evolution of the mean slope $\bar{G}(t)$ throughout the grid, where $\bar{G}(t)=\frac{\int G(\textbf{i},t)d\textbf{i}}{\int d\textbf{i}}$, as SOC can be reached in both critical-slope and critical-height cellular automata models \citep{Kadanoffetal89}.

\begin{figure*}
\centering
\includegraphics[width=15cm]{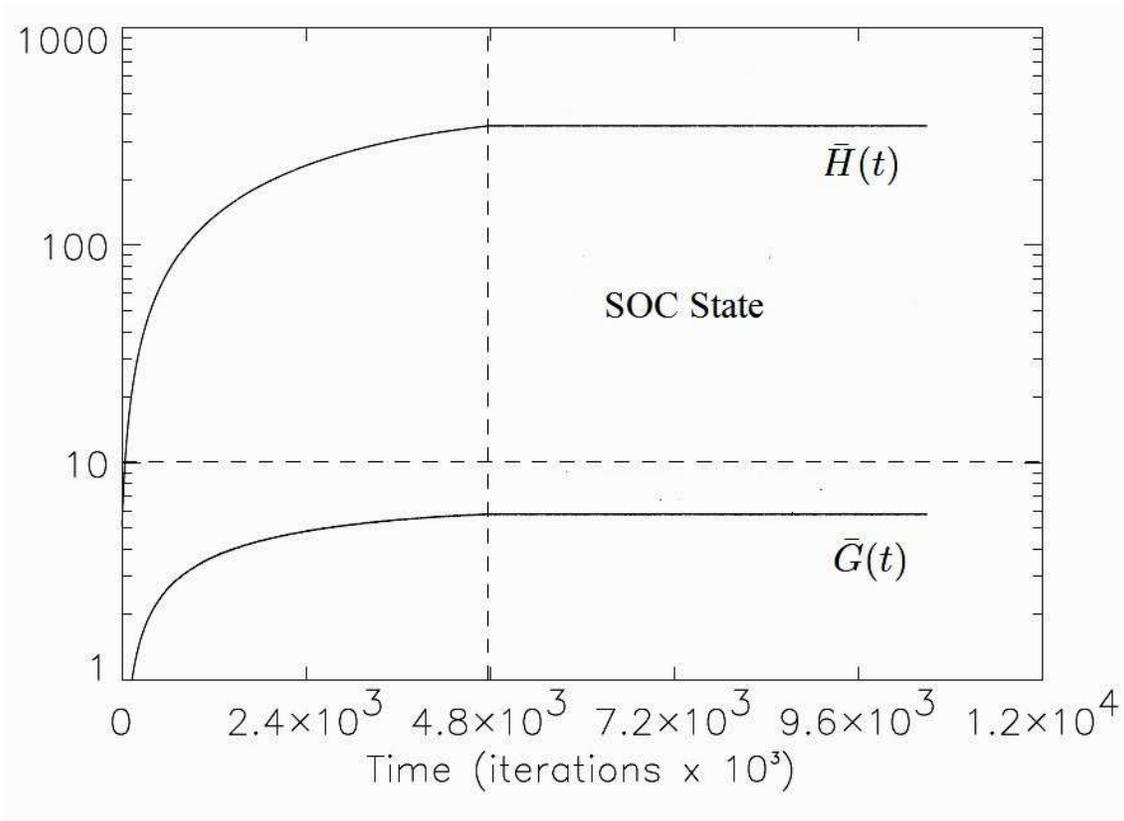}
\caption{Time evolution of the mean height $\bar{H}(t)$ and the mean slope $\bar{G}(t)$ for a 3D statistical Flare cellular automaton sandpile with dimensions $20\times 20\times 20$. The dashed vertical line corresponds to the time at which the system enters the SOC state. The horizontal dashed line corresponds to the critical threshold value of the slope, which defines the instability criterion of the system. From \cite{GeorgoulisThesis00}).}
\label{Fig0304}%
\end{figure*}
Figure~\ref{Fig0304} presents the temporal evolution of the mean height $\bar{H}(t)$ and the mean slope $\bar{G}(t)$ for a 3D sandpile cellular automaton model with dimensions  $20\times 20\times20$. Initially both the mean height  $\bar{H}(t)$ and the mean slope $\bar{G}(t)$ are increasing. This ascending course corresponds to the sequence of the metastable states, through which the system evolves towards the SOC state. This marginally stable state is reflected in the stabilization of both variables after the dashed vertical line. This line determines the time, in system iterations, after which the system enters the SOC state, generating avalanches lacking a characteristic scale in size or duration. Figure \ref{Fig0304} also shows that after the SOC state is reached, the mean slope $\bar{G}(t)$ stabilizes around a value slightly lower than that of the critical threshold $G_{c}$. In the cellular automaton model used in this example, the critical threshold (horizontal dashed line) is $G_{c}=10$, in arbitrary system units. In addition, the SOC state is reached after $\sim 4.8 \times 10^6$ iterations, which corresponds to $\sim 0.6 N^3$ iterations, where $N=8\times10^3$ is the number of nodes, or grid sites, in the SOC system used here. This number of iterations is in order-of-magnitude agreement with the prediction of \citet{Char01} regarding the number of iterations needed to reach SOC ($\sim N^d$, where $d$ is the Euclidean dimension of the system), although the proportionality factor here is $\sim1$, where in the prediction of \citet{Char01} it is typically $\gg1$. Possibly this is due to the fact that the statistical flare model of \cite{GeorgoulisVlahos96, GeorgoulisVlahos98}, which is the one used in Figure~\ref{Fig0304}, does not apply a fixed, infinitesimal driving, but rather uses a perturbation of variable amplitude that is small on average as compared to the critical threshold. This appears to shorten the driving time needed for the system to reach the SOC state.

\begin{figure*}
\centering
\includegraphics[width=13cm]{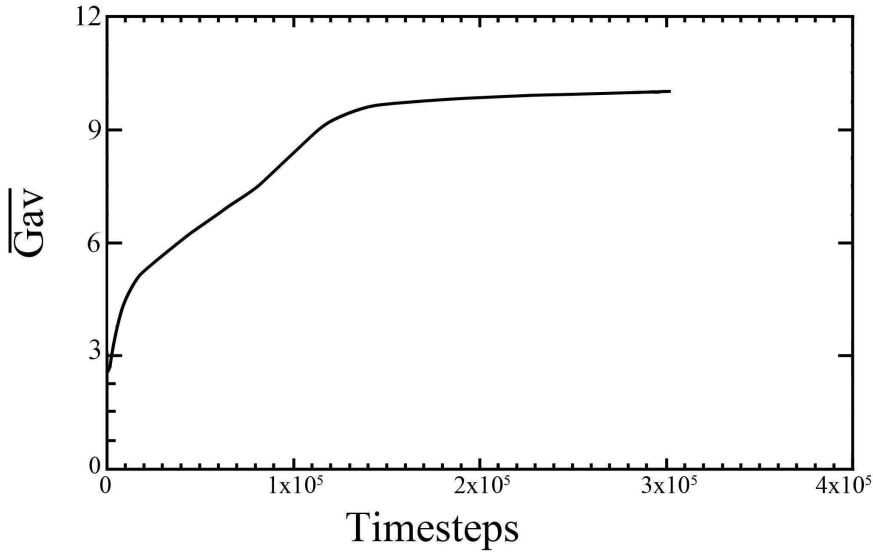}]
\caption{Average Laplacian {\bf G}$_{av}$ over the grid for $3 \times 10^{5}$ timesteps for NOAA 10570. $\bar{G}_{av}$ increases gradually for $1.4 \times 10^{5}$ timesteps, after which the SOC state is reached, with $\bar{G}_{av} \lesssim G_{cr} = 10G$. From \cite{Dimitropoulou11}.}
\label{Fig0302}%
\end{figure*}

\begin{figure*}
\centering
\includegraphics[width=13cm]{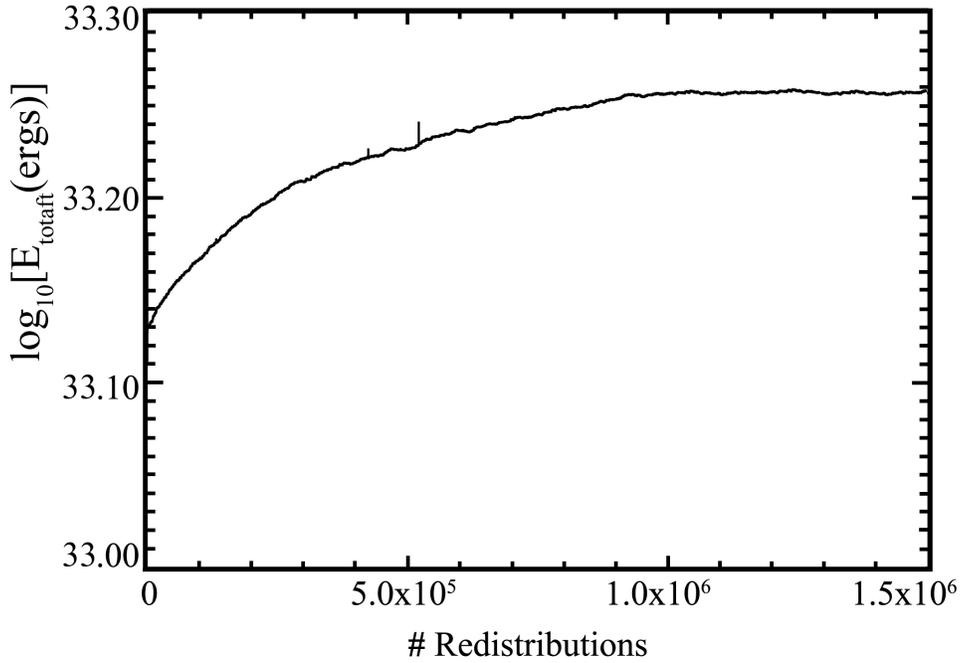}
\caption{Total Volume Energy, $\log_{10}(E_{totaft})$, after each redistribution for NOAA AR 10570. As in Figure~\ref{Fig0302}, $E_{totaft}$ increases gradually until an asymptotic stable state is reached. From \cite{Dimitropoulou11}.}
\label{Fig0303}%
\end{figure*}

The same method was adopted by \cite{Dimitropoulou11} for the detection of the SOC state in a 3D cellular automaton that included vector, rather than scalar, magnetic fields such as the seminal models of \citet{LuHamilton91} and \citet{Luetal93}. The novel element of this work, however, is that the magnetic field vector is {\it data-driven}, i.e., relying on actual solar active regions. The model uses an observed photospheric vector magnetogram of a given active region and extrapolates it via a nonlinear force-free extrapolation \citep{Wiegelmann08} into the overlaying corona, thus obtaining the initial 3D vector field. The configuration is subsequently evolved into the SOC state using conventional cellular-automata rules. This model has been coined the static integrated flare model (S-IFM) by \cite{Dimitropoulou11} because it refers to a single, simultaneous magnetogram. In this model it is assumed that instabilities occur if the magnetic field stress exceeds a critical threshold. For every site $\rvec$ within a cubic grid with dimensions $32\times 32\times 32$, the magnetic field stress $G_{av}(\rvec)$ is calculated as $G_{av}(\rvec)=|\bf{G}_{av}(\rvec)|$ where
%
\begin{equation}\label{c3e21} 
{\bf G}_{av}(\rvec)={\bf B}(\rvec)-\frac{1}{nn}\sum_{nn}{\bf B}_{nn}(\rvec) \ ,
\end{equation}
where $nn$ is the number of nearest neighbors for each site $\rvec$ and ${\bf B}_{nn}(\rvec)$ is the magnetic field vector of these neighbors. Depending on the location of each site within the volume, the number of nearest neighbors $nn$ can be 3, 4, 5, or 6 in 3D, for an edge, vertex, boundary or interior location of the examined grid site, respectively. As $G_{av}$ is related to the diffusive term of the induction equation, it was selected by \cite{Dimitropoulou11} to be compared against the critical quantity of the system such that every site $\rvec=(i,j,k)$ for which the inequality $G_{av_{i,j,k}}\geq{G_{cr}=10 G}$ is satisfied is considered unstable and undergoes magnetic field restructuring according to specific evolution rules. By monitoring the volume average $\bar{G}_{av}$ of the critical quantity $G_{av}$, it was shown that $\bar{G}_{av}$ increases gradually during the continuous driving of the system. When the system reaches the SOC state, $\bar{G}_{av}$ stabilizes around a value slightly lower than the threshold value $G_{cr}$. Figure \ref{Fig0302} shows $\bar{G}_{av}$ value over $3 \times 10^{5}$ time steps for a solar active region (NOAA AR 10570). $\bar{G}_{av}$ increases up to time step $\sim 1.4 \times 10^{5}$, thereafter asymptotically tending to the critical threshold at $G_{cr}=10G$. A second indication that the system has reached the SOC state is that the total volume energy attains an asymptotic value stemming from the competing tendencies of injecting energy in the system via driving and dissipating it via relaxation events. Figure \ref{Fig0303} shows the logarithm of the volume magnetic energy $E_{totaft}$ after each scan of the grid for possible re-distributions. $E_{totaft}$ shows when the system appears to reach the SOC state, namely at $\sim10^6$ iterations, or $\sim(1/32) \times N^3$, where $N=32^3$ is the number of system nodes in this case. This is again dimensionally consistent with the prediction of \cite{Char01}, although the proportionality factor is much smaller than the one predicted in that study, even though the driving perturbations in \cite{Dimitropoulou11} have a fixed amplitude.

\subsubsection{Statistical stationarity: number of avalanches per fixed time interval}

\begin{figure*}
\centering
\includegraphics[width=13cm, height=7cm]{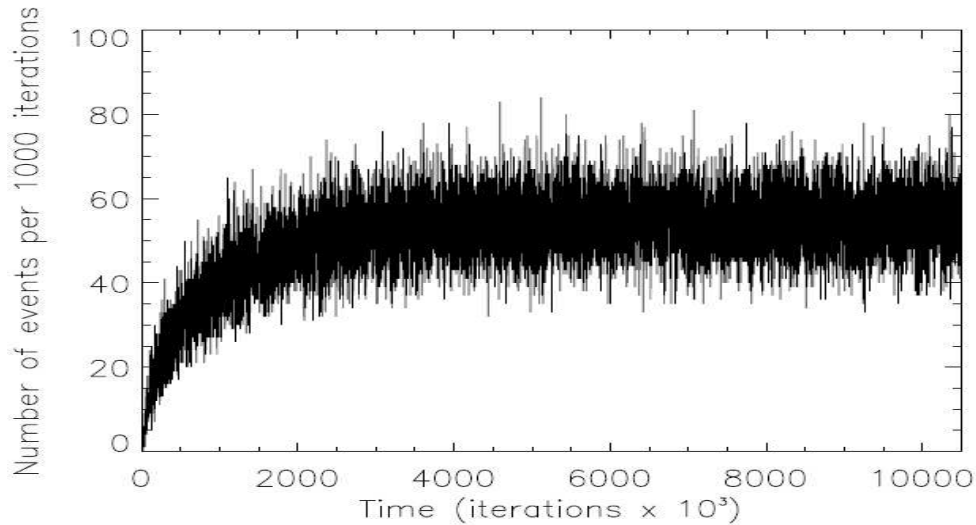}
\caption{Time series of the number of avalanches produced per 1000 iterations for the same 3D statistical flare cellular automaton model discussed in Figure \ref{Fig0304}. A statistical stabilization of the average number of events is shown, after the system has reached the SOC state, beyond the first $2 \times 10^6$ iterations. From \cite{GeorgoulisThesis00}.}
\label{Fig0305}%
\end{figure*}
 
 \begin{figure*}
\centering
\includegraphics[width=14cm]{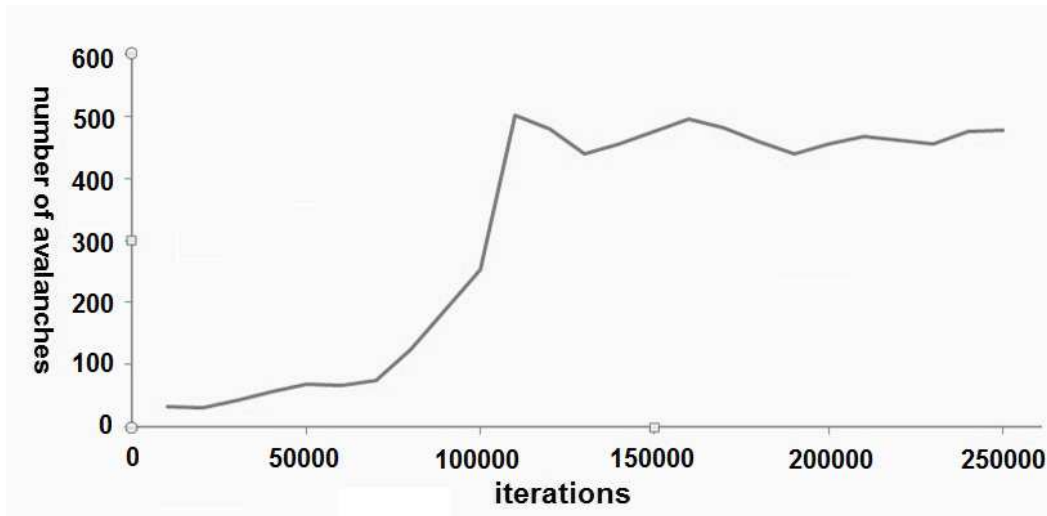}
\caption{Time series of the average number of avalanches produced per 1000 iterations for the static, data-driven cellular automaton of NOAA AR 11158. A statistical stabilization of the average number of events is shown, after the system has reached the SOC state, beyond the first 130,000 iterations. From \cite{Dimitropoulou11}.}
\label{Fig0306}%
\end{figure*}

Statistical stationarity can also be used as an applied diagnostic method towards the detection of the SOC state. This is based on the premise that after a dynamical system has entered the SOC state, the number of avalanches produced within a fixed time interval will vary around a well defined average value \cite{GeorgoulisThesis00}. Figure~\ref{Fig0305} shows an example of this variance that corresponds to the same 3D cellular automaton model of \cite{GeorgoulisThesis00} described in Figure~\ref{Fig0304}. In particular, Figure~\ref{Fig0305} shows a time series of the number of avalanches produced in fixed time intervals consisting of 1000 model iterations. A new iteration is triggered when a sand grain is added to the modeled sandpile at one specific, randomly chosen, grid point (\ie $h(x,y,t)\rightarrow h(x,y,t)+1$, as above). In accordance to conventional SOC models, the driving of the system is not continuous, with each new iteration requiring the complete relaxation of all avalanches in the system. As a result of the statistical stationarity embedded in the SOC state dynamics, the number of avalanches per 1000 iterations varies around a well defined average value of $\sim$50 events, regardless of event size.

The same method was applied to the static, data-driven, integrated flare model \citep{Dimitropoulou11}, as described in the previous paragraph. Figure \ref{Fig0306} shows the average number of avalanches, this time for a single vector magnetogram of the observed NOAA AR 11158, as a function of the simulation iterations. The driving of the system is also not continuous and is applied to a single, random grid point as long as there are no ongoing avalanches. It is shown that after approximately the first 130,000 iterations the average number of the produced avalanches stabilizes around $\sim$450 events per 1000 iterations, which attests to the statistically stationary SOC state reached by the system.

\subsubsection{Non-evolutionary diagnostic SOC-state test}
A third SOC-state test is made possible from the coupling between two data-driven solar flare cellular automata models: the static (S-IFM) model  and the dynamic (D-IFM) model. Rather than detecting the SOC state in line with the previous tests (i.e., on an evolution time series of a possible  SOC system), this non-evolutionary diagnostic aims to determine whether a given 3D snapshot magnetic configuration could be in the SOC state. Both the classical (e.g., autocorrelation test of Section~\ref{af}) and the applied methods of marginal stability and statistical stationarity tests rely on an SOC-state detection based on a continuous monitoring of the evolution of a potential SOC system. This non-evolutionary test instead offers an indication of whether an instantaneously observed system is {\it possibly} in an SOC state, among other possible physical mechanisms that may have led it to the observed configuration.

\begin{figure*}
\centering
\includegraphics[width=14cm]{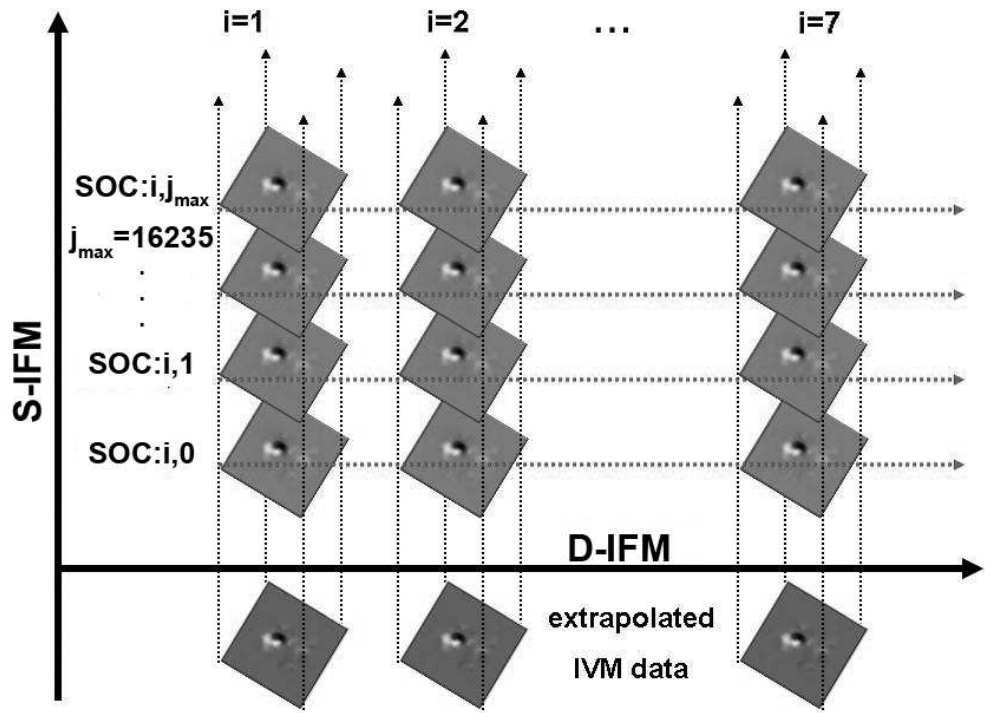}
\caption{Graphical description of the D-IFM, applied to 7 vector magnetograms of NOAA AR 8210: each vertical sequence indicates a separate application of the S-IFM to a single IVM vector magnetogram. This leads to 7 3D SOC-state magnetic field configurations that can be evolved indefinitely. For each horizontal group of 7 3D configurations, a spline interpolation progresses the magnetic field vector from the one to the next configuration, collecting avalanches and their properties. This action corresponds to a single application of the D-IFM. In this test, 16,235 events (i.e., septuplet groups) have been collected in order to attain sufficient statistics. From \cite{Dimitropoulou12}.}
\label{Fig0307}%
\end{figure*}

A brief description of the D-IFM method is attempted here for context: in D-IFM, the single vector magnetogram of S-IFM is replaced by a time series of vector magnetograms of a given active region. Each magnetogram of the time series is subjected to the S-IFM methodology, i.e., an initial nonlinear force-free extrapolation to obtain the 3D coronal magnetic field and a randomly driven evolution into the SOC state. Each magnetic configuration is confirmed to have reached the SOC state through the marginal stability and statistical stationarity tests. The D-IFM then proceeds by slowly driving the magnetic configuration from the one 3D SOC snapshot to the next via a spline interpolation of the magnetic field components. The number of iterations is typically $>> 1$ for observational cadence of the order tens of minutes and depends on the Alfv{\' e}n time required to cross a distance equal to the line element (pixel size) assuming a constant, typical coronal Alfven speed of $~10^{8}$ cm/s (Dimitropoulou {\em et al.} 2013, Table 3). In this course, avalanches occur and are relaxed, giving rise to a sequence of SOC-state events with properties that are studied statistically. Figure~\ref{Fig0307} depicts this basic D-IFM concept applied to a time series of 7 vector magnetograms of the observed NOAA AR 8210. Avalanches occur when the critical threshold of the magnetic field Laplacian is exceeded. Moreover, numerous sequences, or groups, of 3D configurations can be obtained, for each of which one may independently apply the D-IFM and collect the statistics jointly.

\begin{figure*}
\centering
\includegraphics[width=15cm]{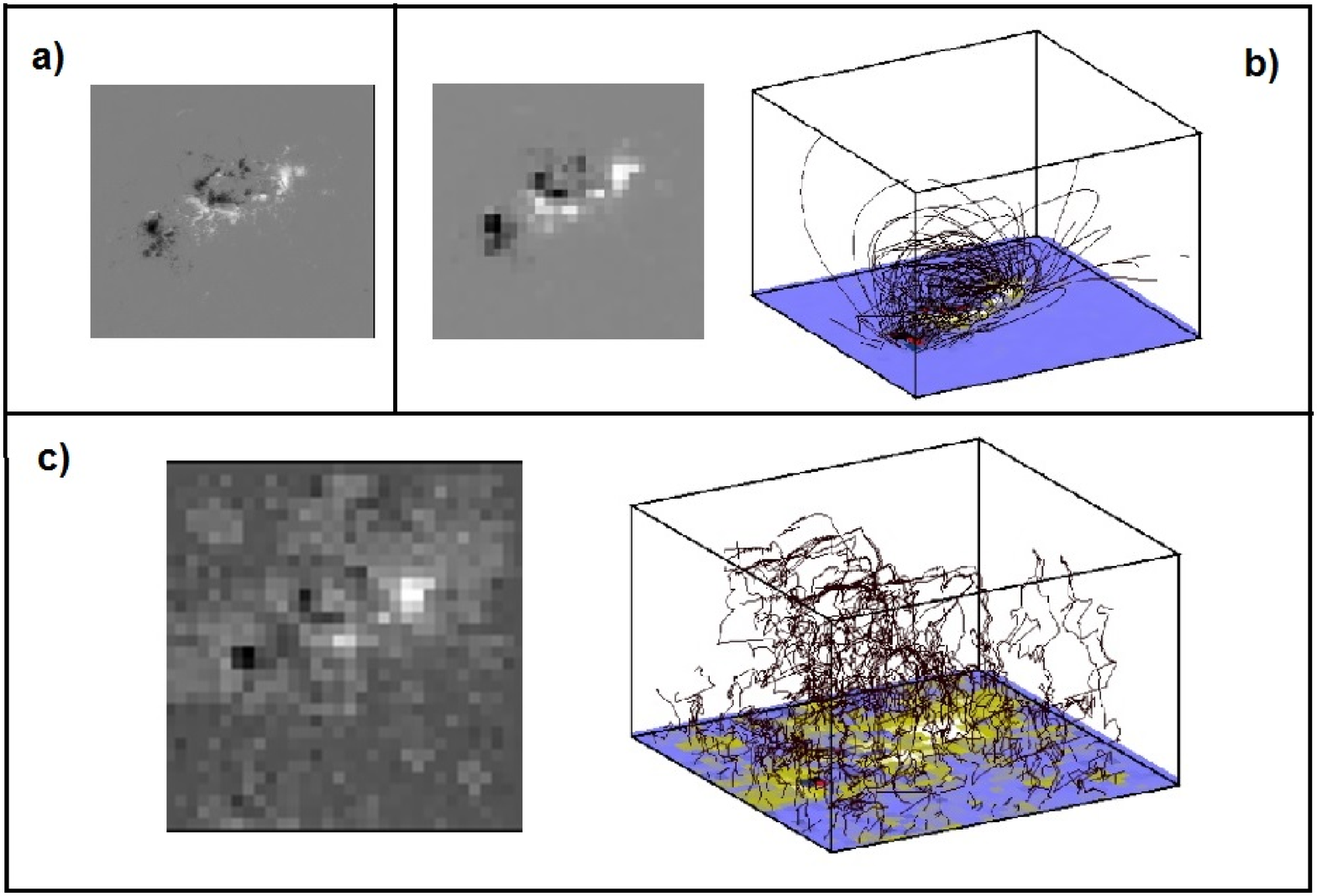}
\caption{a) Observed vertical component of NOAA AR 11158 (13  February 2011, 15:58:12 UT). b) Left: re-binned photospheric magnetic field to grid dimensions  $32\times 32$. Right: extrapolated coronal magnetic field with grid dimensions $32\times 32\times 32$. c) Similar to {\it b}, but after the S-IFM application for $2.5\times 10^5$ iterations.}
\label{Fig0308}%
\end{figure*}
It is this coupling between the static and dynamic models that inspires the concept of the following non-evolutionary diagnostic SOC-test. The principal idea is to apply the S-IFM to an observation (vector magnetogram), leading the initial NLFF field solution into a SOC-state magnetic configuration. The random forcing of the S-IFM will give rise to a very different SOC-state configuration, as compared to the initial NLFF field solution. Then, the same instability criterion is used to revert the configuration to the initial NLFF field solution via the D-IFM, \ie through a continuous interpolation. Since the final S-IFM snapshot is proved to have reached the SOC state and the D-IFM demonstrably retains the SOC characteristics, reverting this snapshot to the original NLFF field solution via the D-IFM is a good indication that the initial NLFF field is indeed in a SOC state. This would be impossible to claim otherwise for any given static 3D magnetic field solution.

\begin{figure*}
\centering
\includegraphics[width=15cm]{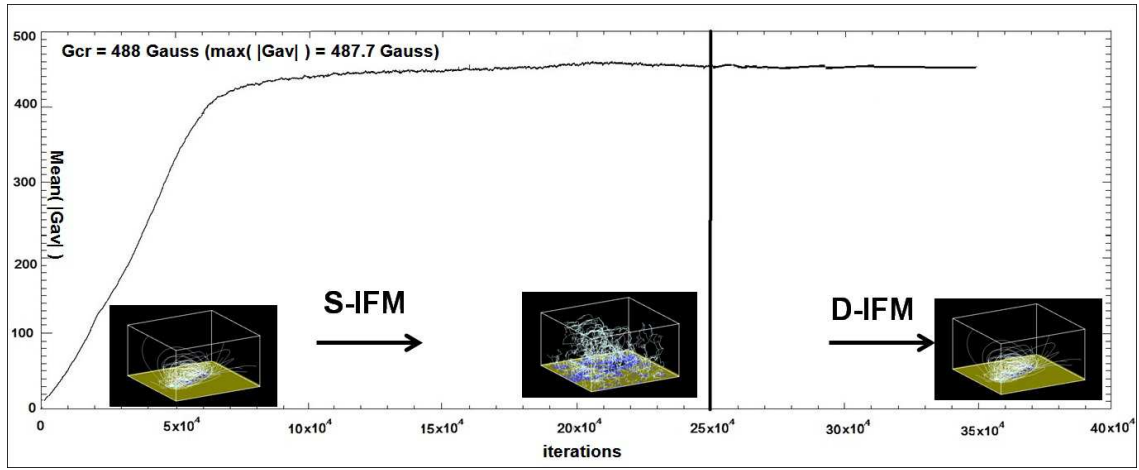}
\caption{Non-evolutionary SOC test run on a snapshot of the observed NOAA AR 11158, shown in Figure~\ref{Fig0308}. The S-IFM has brought the snapshot to an SOC state after $\sim 0.8\times 10^5$  iterations. This is confirmed by the stabilization of the averaged slope in the grid (curve). To ensure the unambiguous evolution to the SOC state, the S-IFM is applied for an additional $2.5\times 10^5$ iterations (vertical line). The system is then reverted back to the initial 3D-extrapolated magnetic configuration via the D-IFM, reaching it after $\sim10^5$ iterations, without exiting the SOC state.}
\label{Fig0309}%
\end{figure*}

Figure \ref{Fig0308} presents the S-IFM part of the non-evolutionary diagnostic SOC-test concept applied to the observed NOAA AR 11158. Figure \ref{Fig0308}a depicts the vertical component of the studied photospheric vector magnetogram, while Figure \ref{Fig0308}b shows the preprocessing necessary in order to apply the S-IFM, namely the re-binning of the magnetogram into a grid of 32x32 (left) and the subsequent 3D nonlinear force-free extrapolation (right). The choice of coarse grid resolution is determined by computational power available for the iterations. Figure \ref{Fig0308}c illustrates the photospheric vertical field component (left) and the corresponding 3D coronal configuration (right) after the S-IFM application for $2.5\times 10^5$ iterations. Notice the severe distortion of the magnetic field vector, caused by the randomness of the S-IFM forcing. This configuration, however, is both a valid (i.e., divergence-free) magnetic field solution and is demonstrably in the SOC state. Retaining the same instability threshold, the D-IFM is then applied, aiming to revert the 3D configuration of Figure \ref{Fig0308}c into that of Figure \ref{Fig0308}b, with the results shown in Figure \ref{Fig0309}. Evidently, the system reverts back to the configuration of Figure \ref{Fig0308}b after $\sim10^5$ iterations. The marginal stability test shows that the system remains in the SOC state until the end of the simulation, and therefore in the course of the continuous interpolation to the initial 3D field. Continuous interpolation would not be possible if the critical threshold $G_{cr}$ for the extrapolated field in Figure \ref{Fig0308} was not the same with the one in the S-IFM.

This simple SOC diagnostic suggests that both observed and force-free extrapolated solar magnetic configurations may already be in an SOC state - at least this is indicated by the successful test on NOAA AR 11158. It should be followed by including an investigation on how a far-from-equilibrium, SOC, state prevails on a force-free equilibrium magnetic field solution. If confirmed this finding may have important ramifications on whether the global solar magnetic field, at least the low-$\beta$ corona, is into an SOC state or whether this feature restricts to (many, most, or all) active regions. This aligns with the discussion on open problems and questions in SOC applicability, detailed in the review of \citet{Asch14}. 
   
\subsubsection{Block scaling}
\label{BlockScal}
A sum rule, similar to equation~\eref{drho} above relating $\Delta\rho_a$ and $C(\rvec_1,t,\rvec_2,t)$, can be used to extract scaling in systems when very little data is available. Although the basic concept also applies to the variance and thus to the two-point correlation function, it can be applied much more directly to one-point functions, \ie to the basic degree of freedom $\phi(\rvec,t)$ (the local activity, energy density, particle density \etc). SOC occurs only right at the critical point, therefore the globally averaged activity (the order parameter) is normally very small. Although there are strong spatio-temporal fluctuations (\ie the activity might flare up locally and even globally on occasion) the local activity (or generally order parameter density) can be averaged spatially over local patches. In the following section, these patches are referred to as {\em blocks}. There are $N=(L/\ell)^d$ such blocks of linear extension $\ell$ in a $d$-dimensional system, $V$, with overall linear extension, $L$. Within each such block $B_i$ (such as illustrated in \Fref{FigBS1}a) the local activity density can be defined as
\begin{equation}
\phi_i(t) = \frac{1}{\ell^d} \int_{B_i} \ddint{r} \phi(\rvec,t) \ ,
\end{equation}
first suggested by \citet{Binder:1981a} for the order parameter in a ferromagnetic phase transition. Obviously, the arithmetic mean over the blocks is invariant under a
change of $\ell$, because 
\begin{equation}\elabel{unconditional_ave}
\frac{1}{N} \sum_i^N \phi_i(t) = \frac{1}{L^d} \int_{V} \ddint{r}
\phi(\rvec,t) \ ,
\end{equation}
independently of $\ell$. One may introduce, however, a level of activity $T$, effectively a threshold, which has to be present somewhere in the patch if the patch is to be considered active, say
\begin{equation}
a_i(t) = \theta(\max\left\{\phi(\rvec,t)| \rvec\in B_i\right\} - T) \ ,
\end{equation}
where $\theta$ denotes the Heaviside theta function and $\max\left\{\phi(\rvec,t)| \rvec\in B_i\right\})$ is the maximum activity $\phi(\rvec,t)$ in the block $B_i$. As a result $a_i(t)$ is unity if $\phi(\rvec,t)$ exceeds $T$ somewhere in an {\em active} block. Otherwise, it vanishes. To facilitate better data analysis, $\phi(\rvec,t)$ may be a function of the original raw data, with a background subtracted and/or the modulus taken to make it non-negative. Conditioning the average to active blocks produces the conditional activity
\begin{equation}\elabel{def_rho}
\rho(t,\ell) = 
\frac{\sum_i^N a_i(t) \phi_i(t)}{\sum_i^N a_i(t)} \ ,
\end{equation}
\ie $\rho$ is the average activity exceeding the threshold. This quantity displays a dependence on $\ell$, as opposed to \Eref{unconditional_ave} (which displays so such dependence). In the presence of correlations, non-vanishing $a_i(t)$ is indicative of large levels of activity in the whole block, such that $\rho(t,\ell)$ should increase as $\ell$ decreases. This is strictly true for $T=0$ and non-negative $\phi(\rvec,t)$, in which case $\sum_i^N a_i(t) \phi_i(t)=\sum_i^N \phi_i(t)$, because $a_i(t)=0$ implies $\phi_i(t)=0$ if $T=0$. In that case $\sum_i^N a_i(t)/N$ cannot increase as $\ell$ decreases and so $\rho(t,\ell)$ increases with decreasing $\ell$: it is a matter of standard finite size scaling that $\rho(t,\ell)\propto\ell^{-\beta/\nu_{\perp}}$ \citep{Pruessner:2008b} with $\beta/\nu_{\perp}=(d-2+\eta)/2$ from  the usual scaling relations \citep{Luebeck:2004,Pruessner:2012:Book}. If $T=0$, the scaling is driven by the dominator in \Eref{def_rho} and amounts to counting the number of blocks containing a certain level of activity $\phi(\rvec,t)$. The procedure is then not dissimilar to the box-counting method used in the study of fractals \citep{Falconer:2003,McAteer05}.

\begin{figure*}
\centering
\hfill
\subfigure[HMI Magnetogram from 11 Feb 2014.]{\includegraphics[width=6cm]{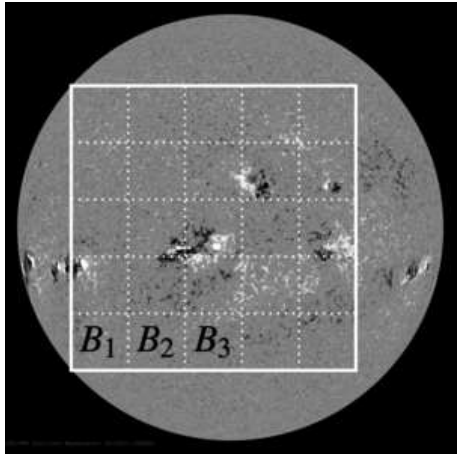}}
\hfill
\subfigure[Block scaling of the conditional activity extracted from the image.]{\includegraphics[width=6cm]{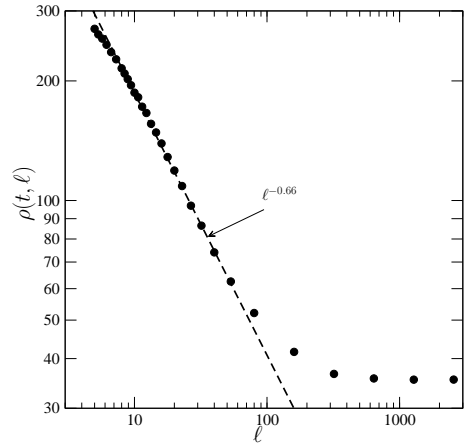}}
\hfill
\caption{A block scaling analysis of a snapshot of an HMI Magnetogram (11 Feb 2014). (a) The large quadratic patch covering most of the sun ($2560\times2560$ pixels) is divided into smaller blocks (here $5\times5$ blocks of linear extension $512$, some of which are labelled) as shown by the dotted lines, \ie $L=2560$, $\ell=512$. (b) Processing the data as described in the text produces a narrow scaling region with an approximate exponent $0.66$. Ordinate $\rho(t,\ell)$  denotes the activity at the time $t$ when the snapshot in (a) was taken averaged over those blocks of size $\ell$ which exceed a (high) threshold $T$ somewhere within the block.}
\flabel{FigBS1}%
\end{figure*}

The same behavior, $\rho(t,\ell)\propto \ell^{-(d-2+\eta)/2}$ is expected for $T>0$, as the fraction of blocks with some activity above the threshold decreases with decreasing $\ell$, while those blocks $i$ containing such high levels generally have a higher average activity $\rho_i(t)$, \ie the numerator is expected to increase and the denominator to decrease with decreasing $\ell$. As suggested by the exponent $\eta$, see \Eref{def_eta}, the scaling of $\rho(t,\ell)$ is indicative of correlations. If blocks are large, then most of them will exceed the threshold somewhere (\ie they will be active) and in fact $\rho(t,\ell)$ approaches the unconditional average \Eref{unconditional_ave} as $\ell\to L$ (as long as the threshold is smaller than the global maximum). If $\phi(\rvec,t)$ were completely independent at different $\rvec$ then selecting them according to activity exceeding a threshold amounts to a random, independent selection. Provided only that the blocks are big enough that the single site where the threshold is exceeded does not introduce a significant bias, $\rho(t,\ell)$ will barely increase with decreasing $\ell$, even when working on a lattice. Correlations, however, have the effect that regions with an activity beyond a certain threshold are generally more active, or, in the case of anti-correlations, significantly less active. 

A relation similar to $\rho(t,\ell)\propto\ell^{-(d-2+\eta)/2}$ applies to the variance of the conditional activity, that is the variance of $\phi_i(t)$ conditional to $\phi(\rvec,t)$ exceeding some threshold within the block. In effect, block scaling gives access to finite size scaling, without changing the system size. In block scaling, the cutoff in correlations, avalanche size distributions \etc, is implemented not by the system size, but by the block size. However, the linear extent of the block $\ell$ is an \emph{additional} scale whose upper cutoff is set by the system size. Proper asymptotic scaling can be expected only when $\ell/L\ll1$. On the other hand, $\ell\gg a$ (the lattice spacing or some other microscopic cutoff) must be fulfilled to avoid some smaller scale physics or other effects such as resolution limitations to take over and dominate the behavior of $\rho(t,\ell)$. Block scaling therefore is a form of intermediate scaling \citep{Barenblatt:1996}.

Nevertheless, the block scaling method provides access to a whole range of scales, even when, ultimately, it cannot replace finite size scaling. It is a tool to quantify correlations allowing a possible universality class to be identified. It has the advantage of requiring little data such as a single but highly resolved snapshot. It is in effect a sub-sampling scheme \citep{Efron:1982}, designed to extract as much information as possible from a (comparatively) sparse source. However, although block scaling instantly indicates the presence of correlations and its scaling, it cannot serve as an unique indicator for the presence of SOC.

\Fref{FigBS1} shows the results of a block scaling procedure applied to a full disk solar magnetogram. By design, this process specifically filters the active region patches from the quiet Sun. The data encoded in the grey-level of the magnetogram were processed by taking the modulus of the deviation from the overall average, and considering as active only those regions which are close to the maximum. In other words, the magnetic field in regions that count as active deviate very strongly from the mean magnetic field. \Fref{FigBS1}b shows a narrow region of power law, which may terminate or bend for very small patch sizes, where the analysis gets close to the resolution limit. Correlations of strong active regions are of course expected and \Fref{FigBS1}b shows $\beta/\nu\approx 0.66$ and therefore $\eta\approx1.32$ in the present case, again comparatively large. For comparison, $\eta\approx1.54$ in the Manna Model \citep{Luebeck:2004,Pruessner:2012:Book} in two dimensions. 


\clearpage
\section{Detection of SOC-state events}
\label{det}
With a powerful set of tools designed to study the correlations expected to be present between features in SOC systems, we now turn our focus to the question of what determines a feature. In this context a {\em feature} is considered as collection of density enhancements in space, a variation in time, or a variation of density enhancements in spatio-temporal data. In this section we discuss the relevant problems with each method, and review some method-specific tools that have been determined as useful tools for analyzing SOC systems.

\subsection{Feature Detection in the Spatial Domain}
\label{det1}
\subsubsection{Thresholding}
Feature detection in space usually consists of dealing with a 2-dimensional greyscale image captured on a charge-coupled device (CCD), and often calibrated (\eg simple CCD considerations of flatfielding, dark subtracting, etc. have been removed). However, these data still remain in digital number (DN) space. As such, the scientist usually considers a series of image processing routines, (\eg based on standard procedures available in \cite{Gonzalez:2008}, or \cite{Starck:2006}) that can be used to identify potential SOC features, to separate them from any noise or non-SOC background, and to characterize them for further analysis. One of the simplest approaches is to apply a fixed threshold in DN space, and group contiguous pixels into one feature. One of the earliest uses of this thresholding and grouping was in studies of colloidal dynamics or Brownian motion \citep{perrin1920, cro96}, and the use of such an algorithm extends to diffusion limited aggregation \citep{Havlin}, particles in Saturn's rings \citep{zeb85}, and urban growth \citep{Batty89}. The case study of solar bright points - small scale, short lived brightenings in the solar corona  - provide some insight into the power of such a method. The threshold is usually considered at 2 or 3 standard deviation amplitudes above a background mean  \citep[\eg][]{mcateer02, mcateer03}. By adding on rules regarding feature size and feature lifetime \citep{mcateer03PhD}, this procedure makes it possible to track features over a sequence of images \citep[\eg][]{Def07, Lam08, Lam10, Kirk12, Kirk13}. With such set of extracted features, the final step is a search for correlations and power laws in their distributions \citep{KruckerBenz98, ParnellJupp00, parnell09}. Although thresholding and grouping provides a simple and convenient method of identifying features, it is also prone to problems with sensitivity in the chosen threshold and in differentiating between feature disappearance and feature clumping. 

\subsubsection{A volumetric consideration}
\label{vc}

\begin{figure*}
\centering
\includegraphics[width=13cm]{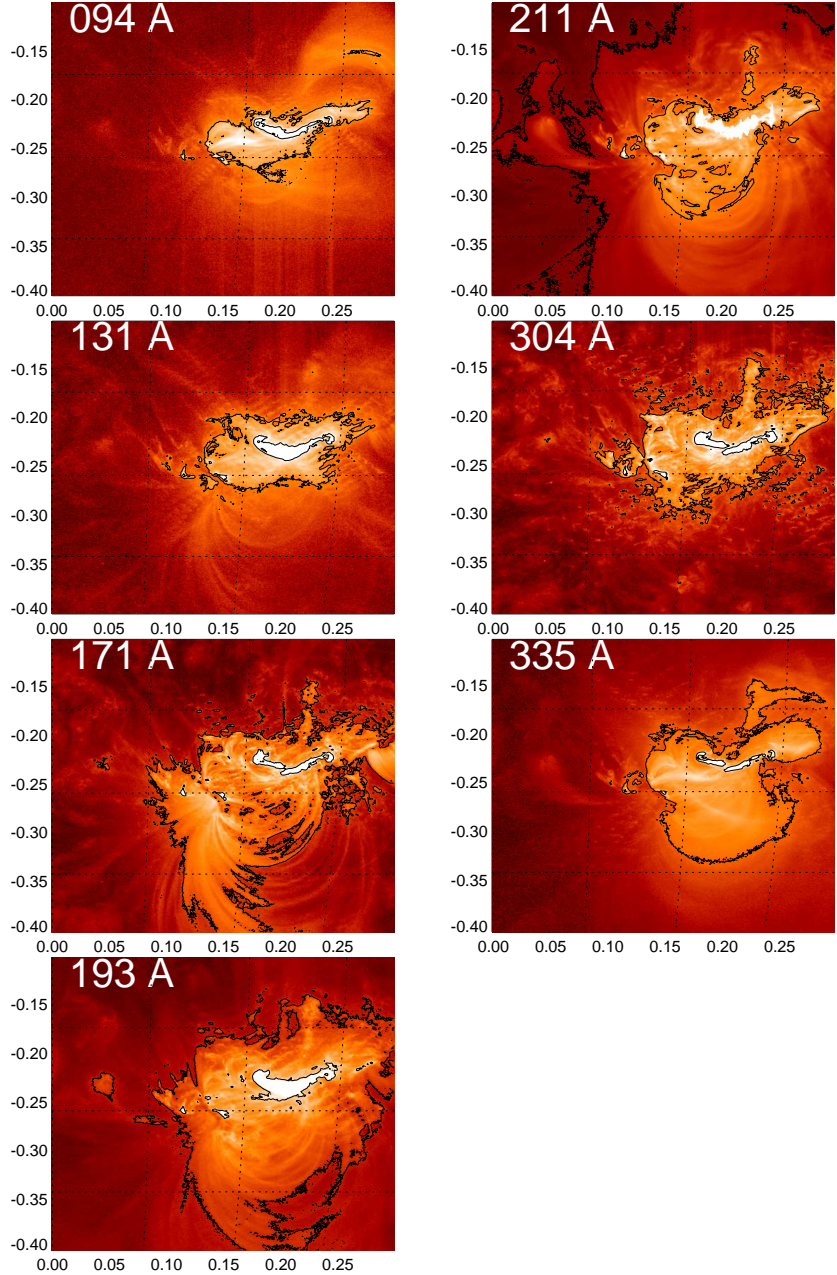}
\caption{Seven multi-wavelength EUV images of the X2.2-class flare observed with AIA/SDO on 2011-Feb-15 01:50:00 UT, in the wavelengths of 94\AA, 131\AA, 171\AA, 193\AA, 211\AA, 304\AA, and 335\AA. The spatial scale of an image side is $\approx 0.3$ solar radius ($\approx 200$ Mm) and the flare area is indicated with black contours at the 50\% and 75\% peak flux level.}
\flabel{FigEUV}%
\end{figure*}

One method to overcome the known problems associated with thresholding and grouping is to use multiple images of the same feature, as observed at different wavelengths. In astrophysical observations, power-law distributions of fluxes or fluences of candidate SOC events have been measured in almost every wavelength, from gamma-rays, hard X-rays, soft X-rays, EUV, visible light, to radio wavelengths. While numerical lattice simulations of SOC models quantify the size of an SOC event simply by the number of active nodes that are unstable and subject to a local re-distribution during any time of an SOC avalanche, the size of an astrophysical SOC avalanche can only be quantified in terms of an observed flux or fluence \ie the time-integrated flux over the duration of an avalanche. However, astrophysical fluxes or intensities, with physical units of energy per time unit, are wavelength-dependent, and thus depend on the instrumental wavelength filter response function, expressed as a function of emission measure per temperature unit, $R(T)$. There are different methods to convert the observed flux into wavelength-independent quantities that can be suitable for the characterization of the size of an SOC avalanche: conversion into radiated energy, \ie $E = n_{phot} h\nu = n_{phot} h c / \lambda$, where $n_{phot}$ is the number of photons that produce a flux $F_\lambda$; conversion into an emission measure by inversion of the flux $F_{\lambda} = \int [dEM/dT]\ R(T) dT$; conversion into thermal energy $E_{th} = 3 n_e k_B T_e V$, which requires a determination of the electron density (\eg from the volumetric emission measure, $n_e = \sqrt{EM/V}$) and the electron temperature $T_e$. Whatever quantity is preferred to characterize the size of an SOC avalanche, this is an extra step that is usually not part of any numerical or mathematical SOC theory.

\begin{figure}[tpbh]
\centerline{\includegraphics[width=0.6\textwidth]{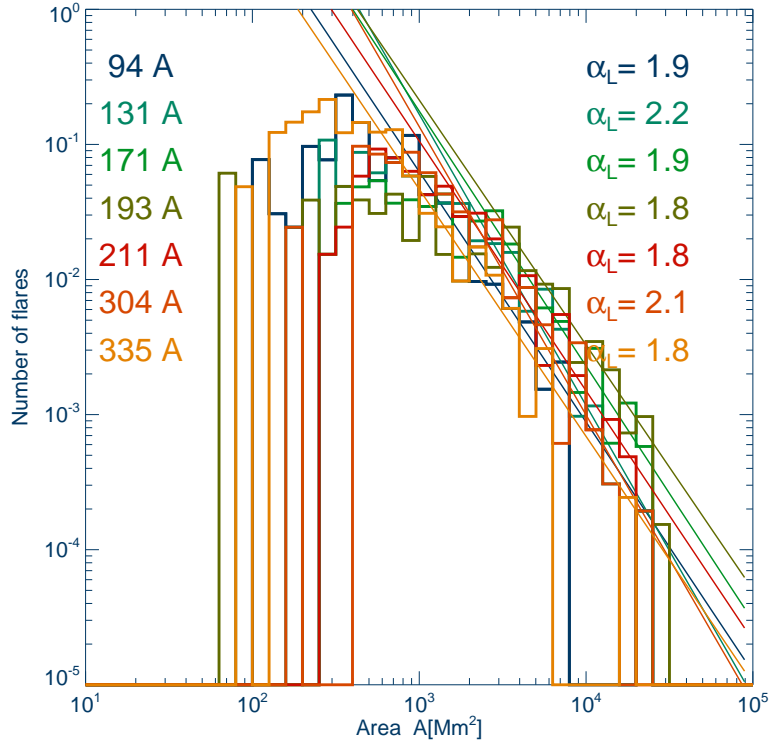}}
\caption{Size distribution of 155 solar flare areas, obtained in 7 different wavelengths (represented in different colors with the wavelengths indicated on the left side, and the index of the power law, $\alpha_A$, on the right side). The 155 flare events include all M- and X-class flares observed with AIA/SDO during 2010 May 13 and 2012 March 31. From \cite{Aschwanden13a}.}
\flabel{FigALPHAS}
\end{figure}

A study of how events appear in different wavelengths provides insight on the spatial structuring of an SOC system. \Fref{FigEUV} shows 7 EUV wavelength images of a large solar flare just at the peak of the emission, observed with AIA/SDO on 2011 February 15, 01:50 UT. A bright sigmoidal white structure is evident in the core of the active region, evidence of a high emission measure and a high-density heated plasma, confined in a helically twisted magnetic filament. Brightness contour levels at 50\% and 75\% of the flux maximum, include somewhat less dense heated plasma loops that surround the core, and make up a substantial fraction of the active region. Using 50\% contours to demarcate the flare area $A(t)$, the relative size varies considerably across different wavelengths, with a minimum size in the 94 \ang\ filter, and a maximum size in the 193 \ang\ filter. To measure the actual flare area $A(t)$, one has to subtract a pre-event background image $A(t_0)$, which will filter out all static emission from the active region. It is usually not possible to know {\em a priori} which wavelength is the best to measure the flare area, or what flux threshold level is most appropriate to define the flare area. Thus, it is advisable to measure the flare area with different thresholds and in different wavelengths, in order to determine any possible nonlinear scaling between different wavelengths, which could in turn affect the slope of the power-law distributions of flare areas, $N(A)$. Such a study has been performed with 5 different threshold levels and 7 wavelength filters for 155 flares \citep{Aschwanden13a}. The resulting flare area distributions are shown in \Fref{FigALPHAS}, after normalizing the flare area to the same flux threshold. Almost identical indices are obtained for the flare areas obtained in the 7 wavelength filters in \Fref{FigALPHAS}, which indicates that the flare areas measured in different wavelengths are statistically either identical or differ only by a fixed proportionality constant. The individual indices are also tabulated in Table~\ref{pls}. This result simplifies future analysis enormously, because it essentially implies that the choice of wavelength does not affect the statistical distributions of geometric parameters, such as the size distribution of lengths $L$, areas $A$, or volumes $V$ of candidate SOC events.

\begin{figure}[tpbh]
\centerline{\includegraphics[width=1.0\textwidth]{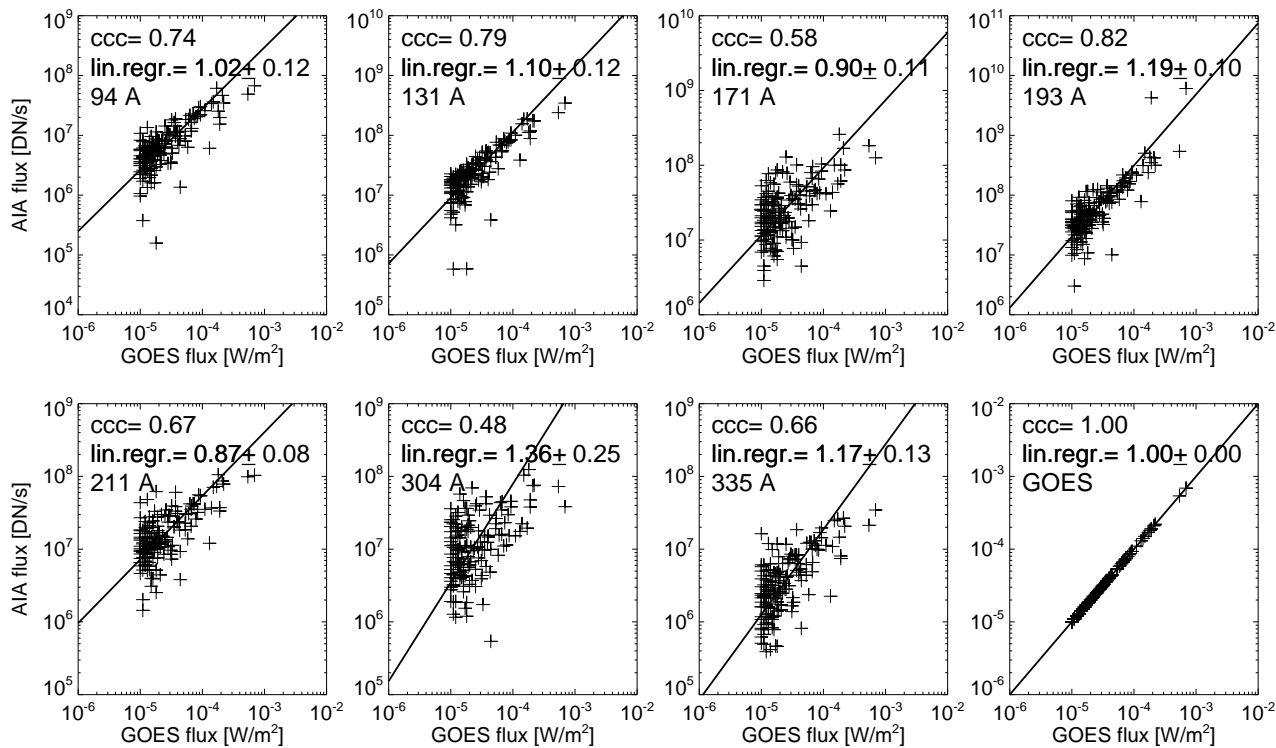}}
\caption{Correlations between the observed fluxes $F_{\lambda}$ in 7 different AIA wavelengths with the GOES flux $F_{GOES}$ for 155 M- and X-class flares observed with AIA/SDO. From \cite{Aschwanden13b}.} 
\flabel{FigCORRS}
\end{figure}

\begin{table}[tpbh]
\begin{center}
\caption{The index of the power law of size distributions of flare areas $a_A$ and fluxes $F_{\lambda}$, and scaling exponents $\gamma$ (discussed in \Fref{FigCORRS}) for 155 flares observed with AIA/SDO observed in 7 wavelengths.}
\medskip
\begin{tabular}{lllll}
\hline
Instrument	&Wavelength 	&Index of       &Index of     &Cross correlation\\
                &		&power law of       &power law     &exponent        \\ 
                &		&area           &flux  	      &AIA vs. GOES     \\
		&$\lambda$ [A]  &$\alpha_A$     &$\alpha_F$   &$\gamma$         \\
\hline
\hline
AIA		&  94 		&2.0$\pm$0.1	&2.2$\pm$0.04 & 1.02$\pm$0.12	\\
AIA		& 131 		&2.2$\pm$0.2	&2.0$\pm$0.02 & 1.10$\pm$0.12	\\
AIA		& 171 		&2.1$\pm$0.5	&2.0$\pm$0.1  & 0.90$\pm$0.11	\\
AIA		& 193 		&2.0$\pm$0.3	&2.0$\pm$0.1  & 1.19$\pm$0.10	\\
AIA 		& 211 		&2.0$\pm$0.4 	&2.1$\pm$0.2  & 0.87$\pm$0.08	\\
AIA		& 304		&2.1$\pm$0.2	&2.1$\pm$0.9  & 1.36$\pm$0.25	\\
AIA		& 335		&1.9$\pm$0.2	&1.9$\pm$0.1  & 1.17$\pm$0.13	\\
GOES		& 1-8 		&		&1.92	      &                 \\
\hline
FD-DOC prediction &		&{\bf 2.00}     &{\bf 2.00}   &{\bf 1.00}       \\
\hline
\label{pls}
\end{tabular}
\end{center}
\end{table}

A complementary study of the wavelength dependence of observed fluxes provides further insight into SOC processes. \Fref{FigCORRS} shows scatterplots of the 7 AIA flare peak EUV fluxes with the higher energy (GOES) soft X-ray flux, for the same set of 155 M- and X-class flares \citep{Aschwanden13b}. Apparently there exists a correlation between each of the EUV fluxes and the soft X-ray flux. The cross-correlation coefficients vary from $CCC=0.82$ for the 193 \ang\ filter, which shows the closest correlation with the GOES 1-8 \ang\ flux due to their overlapping high-temperature response (i.e., the 193 \ang\ filter is sensitive to the Fe XXV line at a temperature of $T_e \approx 20$ MK), down to $CCC=0.48$ for the 304 \ang\ filter, which is most sensitive to cooler chromospheric plasma. Although the proportionalities between the EUV and soft X-ray fluxes have some significant scatter, their size distributions are similar, as the indices $\alpha_F$ listed in Table 1 demonstrate. Consequently, it is reasonable to also expect near-proportionality for linear regression fits between the EUV and soft X-ray fluxes, i.e., $F_{EUV} \propto F_{SXR}^\gamma$, with a scaling exponent of $\gamma \approx 1$. Indeed, Table 1 shows an average exponent of $\gamma = 1.1\pm0.2$ for these 7 wavelengths. This important result of near-proportionality of EUV to SXR fluxes implies the wavelength independence of flux size distributions, which again eases comparisons of SOC statistics in astrophysical objects considerably.

\subsubsection{Turbulence and Fractals: A direct 2D fingerprint of 3D SOC?}
\label{turb}

\begin{figure}[tpbh]
\centerline{\includegraphics[width=0.75\textwidth]{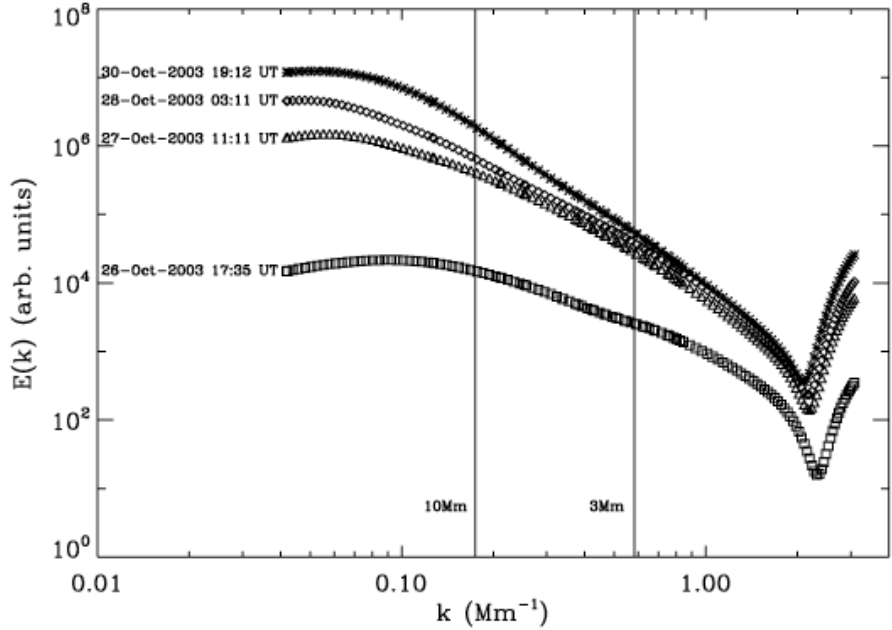}}
\caption{The Fourier spectrum of a 2D slice of the active region magnetic field, plotted in log $E(k$) - log (k) show a clear linear range as a signature of SOC, and changes over shallow (27-Oct) to steep (30-Oct). From \cite{Hewett08}.} 
\label{HewFour}
\end{figure}

Direct imaging has the potential to provide a direct fingerprint of detecting SOC in the spatial domain. Under this paradigm, it is assumed that any SOC system will involve power laws across spatial scales, and that this will manifest in terms of turbulence and fractality \citep{McAteer10, mcateer13b, mcateer15}. Indeed, since \cite{kol41} and \cite{man75} first introduced the ideas of turbulence and fractals, respectively, complex systems have been found to be ubiquitous in many areas of human and natural sciences. Spatial power laws provide the connection between turbulence and SOC as discussed above in Section~\ref{strfunc}. The calculation of the spatial energy spectrum is given as 
\begin{equation}
E(k) \sim k^{-\beta} \ ,
\end{equation}
where the spatial energy, $E$, varies with wavenumber, $k$, risen to a scaling index, $\beta$. (where $\beta$ = 5/3 for fully developed turbulence in fluids). Energy in this terminology strictly refers to the energy in the Fourier spectrum of the data. The scaling index is often calculated from a linear regression of the $E(k)$ plot over a chosen linear range of wave numbers (see Section~\ref{strfunc} for examples applied to solar active regions, where \cite{Abramenko05b} and \cite{Hewett08} use $3-10$~Mm, see Figure~\ref{HewFour}). More power at small $k$ (hence large spatial scales) results in a larger scaling index, and so large $\beta$ is suggestive of increased complexity in the system. \cite{Georgoulis12} studied a sample comprising hundreds of solar active regions and showed many of them follow non-Kolmogorov power-spectrum scaling, with $\beta > 5/3$. Extended to a multi scale approach, this method can be used to eliminate any background non-SOC component \cite{Hewett08}. Fractals are defined in a similar manner as the self-similarity of an image across all scale sizes, or the scaling index of any length, $l$, to area, $A$,
\begin{equation}
A \sim l^{\alpha} \ .
\end{equation}
The fractal dimension, $\alpha$, and various other forms of fractal dimension (see \cite{mcateer13b} for a complete list), is often calculated via a thresholding and contouring approach. The more complex the thresholded contour, the more space it fills, and therefore the larger the fractal dimension. \cite{McAteer05} and \cite{Conlonetal08} use such an approach to study the complexity of solar active regions. \cite{Georgoulis02} adopt a similar approach to show the dust-like nature of small scale brightenings. \cite{kestener10} and \cite{conlon10} extended this to a multifractal approach that can be used to eliminate non-SOC backgrounds from images to show a clear relationship between the remaining multifractal spectrum of an active region and its potential to produce large solar flares. The power of these approaches, as evident in Figure~\ref{figu5} and Figure~\ref{HewFour} is that they may provide a means of linking the clear time-varying nature of SOC avalanches in the emission from an active region \citep{mcateer07, mcateer13} with a 2d spatial slice of the 3D SOC nature of spatial structures. However, it is important to note that although turbulence and fractality may be a signature of an SOC system there may be several other reasons for their occurrence. Therefore, these techniques should be accompanied by studies in time to confirm the existence of SOC \citep{mcateer15}.

\subsection{Feature Detection in the Temporal Domain}
\label{det2}
An SOC system inevitably results in a series of catastrophies or avalanches, detectable in both observational data and simulations a release of energy. In an idealized dataset, each event would be well separated in space and time. A scientist simply needs to only identify each event, and can be secure in the knowledge that there is no overlap. However, such an idealized dataset is rare. Instead data often contains events that overlap significantly. In such a case of pulse-pile up, it may still be possible to separate out the signature of each individual event, and study these to determine if the waiting time distributions are the unique signature of SOC, or otherwise. 

\subsubsection{Power laws}
\label{powerlaws}
The Fourier power spectrum is a useful and simple tool to examine event occurrence in the temporal domain. Many systems exhibit power spectra such that the power spectral density $P(\nu)$ is proportional to a negative power law of frequency $\nu$,
\begin{equation}
\label{eqn:ji:plps}
P(\nu)\propto \nu^{-p}
\end{equation}
where $p\ge0$. A nomenclature for noise spectra has emerged depending on the value of the index $p$, and is described in Table~\ref{tab:ji:nomen}. Flicker, or shot noise, is common in electrical signals, and it was the analysis of this noise that produced a physically based model that 
is highly relevant for SOC models.
\begin{table}[tpbh]
\begin{center}
\caption{Nomenclature of noise spectra \citep{Aschwanden11}}
\medskip
\begin{tabular}{ll}
\hline
Index of power-law 	&Spectrum        \\
$p$                	&Nomenclature \\ 
\hline
\hline
$0$		& white noise				\\
$1$		& pink noise, shot noise, flicker noise, $1/f$ noise	\\
$2$		& red noise, Brown(ian) noise			\\
$3$		& black noise				\\
\hline
\label{tab:ji:nomen}
\end{tabular}
\end{center}
\end{table}
Briefly, we envisage the electrical signal in an RCL-circuit as consisting of the superposition of current spikes, parameterized as Dirac $\delta$-functions having random arrival times $t_{j}$, \ie
\begin{equation}\label{eqn:ji:current}
I(t) = \sum_{j}q\delta(t-t_{j}).
\end{equation}
The general autocorrelation function given in \Eref{def_C} can be rewritten for a such a time series $I(t)$ as
\begin{equation}\label{eqn:ji:ac}
C(t') = \lim_{T\rightarrow\infty}\frac{1}{T}\int_{-T/2}^{+T/2}I(t)I(t+ t')dt.
\end{equation}
The Weiner-Khinchin theorem \citep{chatfieldtimeseries} states that the power spectra density $P(\nu)$ of a stationary random process is the Fourier transform of the corresponding autocorrelation function,
\begin{equation}\label{eqn:ji:wk}
P(\nu) = 2 \int_{-\infty}^{+\infty}C(t')e^{-i2\pi\nu t'}dt'.
\end{equation}
This enables the calculation of the power spectra of models of random processes $I(t)$. \citet{vanderziel} and \citet{Aschwanden11} use the current model above to derive Schottky's result \citep{schottky} for the white noise spectral power distribution in electrical circuits. This general procedure in going from a model of the process to its power spectrum is used below to generate other power-law power spectra.

Power-law power spectra have been observed in solar phenomena. \cite{mcateer07} find power laws in solar flare X-ray data, and they then use this a means of studying the source of these X-rays in \citet{mcateer13}. \cite{2014AA...563A...8A} observe power laws in the integrated emission of small portions of active regions and the quiet Sun as observed in the 195\AA\ passband images from EIT over the frequency range 0.01 - 1 mHz. \cite{ireland2014} observe power laws in power spectra of AIA 171\AA\ and AIA 193\AA\ in active region, moss and quiet Sun areas in the frequency range 0.5 - 10 mHz. \citet{gupta2014} showed power-law power spectra in the intensity at six single points in AIA 171\AA\ coronal plumes extending over the frequency range $0.3\rightarrow 4.0$ mHz. Further out in the solar atmosphere at 2.1 $R_{sun}$, \cite{bem08} show the presence of power-law power spectra in Ultraviolet Coronagraph Spectrometer observations of the intensity of Lyman-$\alpha$ in the frequency range $2.6\times10^{-6}\rightarrow1.3\times10^{-4}$ Hz. Lower in the solar atmosphere, \cite{rea08} show the presence of power-law Fourier power spectra, in the range 7-20 mHz, in the Doppler velocity of the chromospheric Ca II 854.2 nm line.

Many models can generate power spectra that exhibit power laws. One simple model is the autoregressive process,
\begin{equation}\label{eqn:ji:ar}
X_{t} = \alpha X_{t-1} + N(0,\sigma)
\end{equation}
for $t\ge1$, $\alpha>0$ and Gaussian noise $N(0,\sigma)$ with zero mean and standard deviation $\sigma$. This simple process generates a power spectrum with index $p=2$ in the limit of high frequencies \citep{chatfieldtimeseries}. \cite{Aschwanden11} gives the example of a shot noise spectrum of exponentially decaying pulses. Each pulse is modeled as an exponentially decaying function of time $t$
\begin{equation}
\label{eqn:expdecay}
f(t) = \frac{E}{T} \exp{-\frac{t}{T}},
\end{equation}
for some timescale $T$ and energy $E$. The corresponding Fourier power spectrum is (using Equations \ref{eqn:ji:ac} and \ref{eqn:ji:wk})
\begin{equation}
\label{eqn:ftexpdecay}
P(\nu) = \frac{E}{1 + (2\pi \nu T)^{2}}.
\end{equation}
The total Fourier power spectrum of a distribution $N(T)$ of these decaying pulses is 
\begin{equation}
\label{eqn:sumftexpdecay}
P_{total}(\nu) = \sum_{T}N(T)P_{T}(\nu).
\end{equation}
Further, if the number of events of a given energy $E$ is assumed to be
\begin{equation}
\label{eqn:energydistrib}
N(E) \propto E^{-\alpha_{E}}
\end{equation}
and the total energy in each event depends on its time scale $T$ such that
\begin{equation}
\label{eqn:energytime}
E \propto T^{1+\gamma}
\end{equation}
then it can be shown that the observed power spectrum can be approximated by
\begin{equation}
\label{eqn:finalfps}
P_{total}(\nu) \propto \nu^{-(2-\alpha_{E})(1+\gamma)}.
\end{equation}
This derivation shows it is possible to generate a power law using swarms of statistically similar events. 

A power law may seem evident from a simple plot of the data, but the determination that a power law is actually present in the data is a subject that requires much attention \citep{Clauset:2007}. There are essentially two parts to determining the properties of a power law in the data. Firstly, one must determine that a power law is an appropriate representation of the data. This should involve a combination of testing many different parameterizations/models of the data and making some determination as to which one best explains the data. Choosing a model also requires that the researcher think about the physical processes that may be occurring to generate the observations \citep{2002MNRAS.335..389P, 2005ConPh..46..323N, 2010MNRAS.402..307V}. This could be roughly classed as the model selection stage. The second stage is to actually determine the values of the parameters of the power law, properly taking in to account the details of the instrument, observational effects, and the statistics of the measurement. This is the parameter estimation stage. Clearly the two stages are intertwined to some extent.

The identification that a power law is better than other reasonable models for the data is discussed by \cite{Clauset:2007}. A summarized procedure for deciding if a given data set follows a power law is given. This procedure is applied to twenty four real-world datasets, drawn from a broad variety of disciplines, including physics, earth sciences, biology, ecology, paleontology, computer and information sciences, engineering, and the social sciences to test for the presence of power laws. The paper finds that in general, it is extremely difficult to tell the difference in the data between a log-normal behavior and that of a power law.

Estimation of the parameters of a power law, along with an error estimation, is often crucially important as the index of the power law is often used as an indicator of the underlying physical process. Fitting a straight line to binned data is not recommended, as it introduces an arbitrary parameter the histogram binsize. \cite{whi08} show that no histogram binning yields values of the index of the power law consistently close to the true value. However, better methods exist. Let us assume a set of observations $X_{i}$ $1\le i \le N$, drawn from a power-law probability density function of the form $p(X)\propto X^{-\gamma}$. \cite{New:2005} derives that the likelihood function for this set of observations is given by 
\begin{equation}
\label{eqn:ji:plindex:mle}
\hat{\gamma} = 1 + \frac{1}{\frac{1}{N}\sum_{i=1}^{N}\ln Y_{i}}.
\end{equation}
There is an implicit assumption here that the observer has collected all the data perfectly, which is rarely the case. It can be appropriate to more carefully consider how the observation is made, and to include that in the estimation of the index of the power law. For example, \cite{ParnellJupp00} consider the observation process in the determination of the distribution of small heating events in the solar corona. It is assumed that the observed energy $E_{obs}$ of a small heating event is related to its true energy $E$ by
\begin{equation}
\label{eqn:ji:parnell}
E_{obs} = uE
\end{equation}
where $u$ is an under-reporting factor which satisfies $0\le u\le 1$. After some assumptions on the distribution of $u$, the under-reporting of the true energy of the event can be compensated for in the analysis; the final value of the index of the power law fully incorporates the modeling of the under-reporting. The likely physical nature of the energy deposition has also been considered for the same problem of the distribution of the energy in heating events in the corona. \cite{2001ApJ...563L.165M} consider the geometry of the energy deposition event in the corona, which is shown to have a strong influence on the final value of the index of the power law. These two studies show that a careful consideration of the likely physical process, and the way it is observed, is required in order to fully realize the potential of the data.

\subsubsection{Pulse-pileup effects}
One of the tenets of slowly-driven SOC models is the separation of time scales, which means that the waiting time (\ie the time interval between the starting times of two subsequent events) is larger than the event duration of the first event, so that there occurs only one event at a time, while no two events overlap with each other. While this requirement can easily be controlled in numerical cellular automaton simulations, it cannot be taken for granted when an automated pulse detection algorithm is applied to a time series of observations. In principle, numerical detection schemes can be designed to end one event before the next is detected, but this may truncate the duration of the earlier event or ignore a later event that starts during the decay phase of the earlier event. In practice, it is expected that the time separation criterion will be fulfilled during quiescent periods with low event rates, but it is possible that events start to overlap during more active periods, an effect known as {\em pulse pile-up}. This effect can be investigated by considering solar flare statistics during various phases of solar activity.
	
\begin{figure}[tbp]
\centerline{\includegraphics[width=0.9\textwidth]{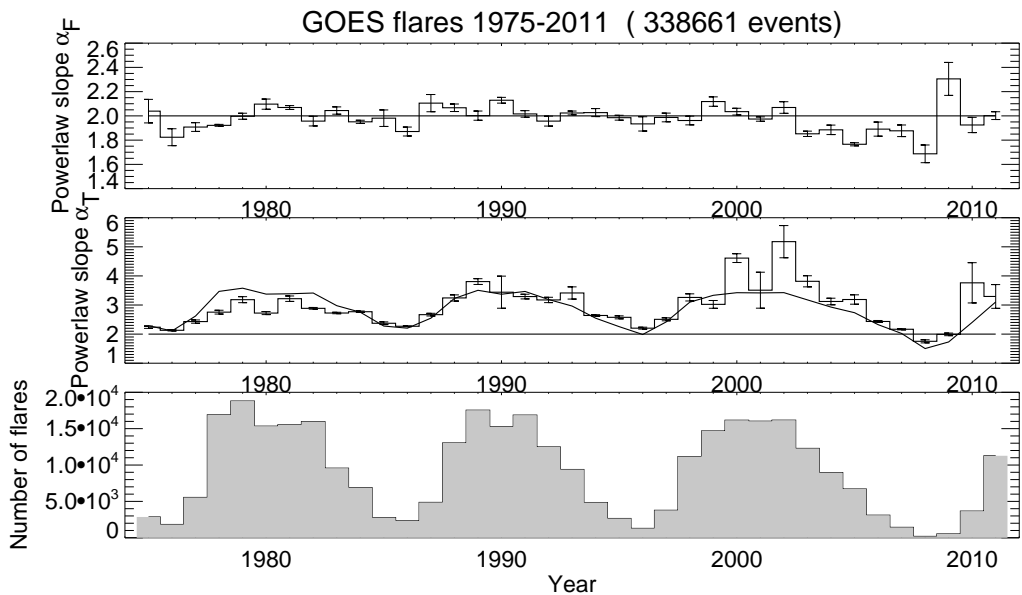}}
\caption{Variation of the index of the power law, $\alpha_P(t)$, of the soft X-ray 1-8 \ang\ peak flux (top panel) and the flare rise time $\alpha_T(t)$, detected with GOES (middle panel), and the annual variation of the number of flares over 3 solar cycles (bottom panel). The flare rate predicts the variation in $\alpha_T(t)$ of the flare time duration (smooth curve in middle panel) as a consequence of the violation of the separation of time scales. From \cite{AschwandenFreeland12}.}
\flabel{FigSunspot}
\end{figure}

\cite{AschwandenFreeland12} studied flare statistics from the GOES satellite sampled over a period of 37 years (1975-2011), covering about three solar cycles. The soft X-ray flux from the Sun varies by about two orders of magnitude during each solar cycle, due to the variation of emerging magnetic fields and the resulting coronal plasma heating rate, which is all driven by the solar magnetic dynamo. This makes the Sun an ideal system to study SOC systems with variable drivers. While the power law of the soft X-ray peak rate is invariant, $\alpha_F=1.98\pm0.11$ (\Fref{FigSunspot} top), during different solar cycles, the time durations do have a variable slope from $\alpha_T\approx 2.0$ during solar minima to $\alpha_T\approx 2-5$ during solar maxima. This is explained in terms of a flare pile-up effect. The variability of the flaring rate is shown in \Fref{FigSunspot} (bottom), from which the steepening of the index of the power law can be estimated by using the ratio of the mean inter-flare time interval to the mean flare duration (\Fref{FigSunspot} middle panel, solid curve), agreeing with the variability of the observed flare rate (\Fref{FigSunspot}, middle panel, histogram). Apparently, the long flare durations are underestimated due to subsequent flares that start during the decay phase. This also affects the statistics of waiting times accordingly. In other words, the separation of time scales (i.e., the waiting times and flare durations) is violated during the busy periods of the solar cycle maximum.

The influence of different pulse detection methods on the shape and index of the power law has also been studied in \cite{buchlin05}, who compares a peak detection method, a threshold method, and a wavelet method. The peak method requires a relatively noise-free smoothed time profile, so that noise fluctuations do not contaminate the statistics with multiple peaks per time structure, leading to an excess of short waiting times. The threshold method requires that the time profiles return to a sub-threshold background level for each event, otherwise events in the decaying tail of a pulse time profile are ignored. The wavelet method has the ability to detect simultaneous pulses with different time scales, which would be impossible with the peak or threshold method. Interestingly, the three methods reveal quite different waiting-time distributions in each case. The threshold-based method seems to produce distributions that resemble power laws, while the peak-based and wavelet-based methods produce exponential-like distributions, at least in the regime of large waiting times. This result imposes some ambiguity in the interpretation of waiting-time distributions. The effect of event definition on the distribution of waiting times has also been numerically simulated with the continuously driven Olami-Feder-Christensen (OFC) model \citep{olami92} by \cite{hamon02}. 

\subsubsection{Waiting-time distributions}
In cases where pulse pile up can be neglected, or at least estimated and removed, it is possible to then study the waiting times between events as a possible signature of SOC. This leads naturally to the following key questions: do waiting-time distributions (WTDs) comprise an indisputable SOC-state feature? Can physical systems exhibiting different WTDs from the ones predicted in the original SOC concept be safely excluded from the long list of potential SOC systems? Since the development of the first avalanche models, it was suggested that the associated exponential-function WTDs should convey a necessary SOC signature. The context of solar flare dynamics provides a useful insight into this debate. Numerous researchers analyzed hard X-ray flare data in an attempt to construct the corresponding WTDs. Their results were initially conflicting. \cite{Biesecker94} used 1 yr of Gamma Ray Observatory (GRO) BATSE data to produce a WTD. The observed distribution was essentially exponential, covering the gaps due to lack of observational data through a simulation representing a Poisson process with a time-varying rate. \cite{Pearceetal93}, however, using 10 yr of Solar Maximum Mission hard X-ray burst spectrometer (HXRBS) data, found a WTD that was closer to a power law than to an exponential. This result suggested that the HXRBS events are interdependent. \cite{Crosby96} reported a distribution over a wide range of waiting times that could be fitted by a power law with an exponential rollover based on hard X-ray events observed in a single active region by the WATCH experiment onboard the GRANAT satellite. The index of the power law was close to that found by \cite{Pearceetal93}. 

Faced with these apparently conflicting results, \cite{Wheatlandetal98} re-examined the WTD of solar flare hard X-ray bursts. The WTD constructed from the ICE/ISEE 3 data showed an overabundance of short waiting times (10 s - 10 min) in comparison to a simulation of the time history of bursts as a Poisson process. This over-clustering with respect to a Poisson process indicates, according to \cite{Wheatlandetal98}, the interdependence of some of the bursts that occurred in temporal proximity. Such a Poisson process would yield an exponential distribution for the waiting times of the solar flares and, according to \cite{Boffettaetal99}, such a distribution would only be expected if the events were completely uncorrelated. Moreover, \cite{Boffettaetal99} suggested that SOC models are expected to display an exponential WTD $P(\tau_L) = <\tau_L>^{-1} exp(−\tau_L/<\tau_L>)$, where $<\tau_L>$, is the average waiting time, which depends on the parameters of the model. This behavior is related to the fact that the avalanche duration is much smaller than the loading time (\ie the time between two successive injections of magnetic field in random positions) and charging place (\ie the random position in which the injection of the magnetic field takes place) is independent from the avalanche position. Then one expects no correlation between successive bursts and thus a trivial, exponential statistics for the waiting times. However, various caveats on this assessment were thereafter voiced: first, Buchlin (2005) suggested that thresholding the event time series may result in WTD resembling power laws in an SOC system. Based on a non-stationary Poisson model as introduced by \cite{Wheatland2000} and further discussed in \cite{WheatlandLitvinenko02}, \citet{aschwanden10} reviewed numerous studies and data sets to conclude that WTD for solar flares can generally be approximated by a non-stationary Poisson distribution of the form $P (\Delta t) \propto \lambda_0 (1 + \lambda_0 \Delta t)^{-2}$, where $\lambda_0 = 1/\Delta t_0$ is the flare rate corresponding to a waiting time $\Delta t_0$, below which there is a high flare rate, or clustering, of small released energies. Above this time, the flare rate decreases with flare magnitude (released energy) giving rise to a WTD that resembles a power law. Evidence that this WTD can in fact correspond to an SOC system also stems from the analytical predictions of the avalanche model of solar flares \cite{Char01} and the fractal-diffusive SOC model described by\cite{aschwanden14} and discussed extensively by \cite{Asch14} (this volume).

\begin{figure*}
\centering
\includegraphics[width=14cm]{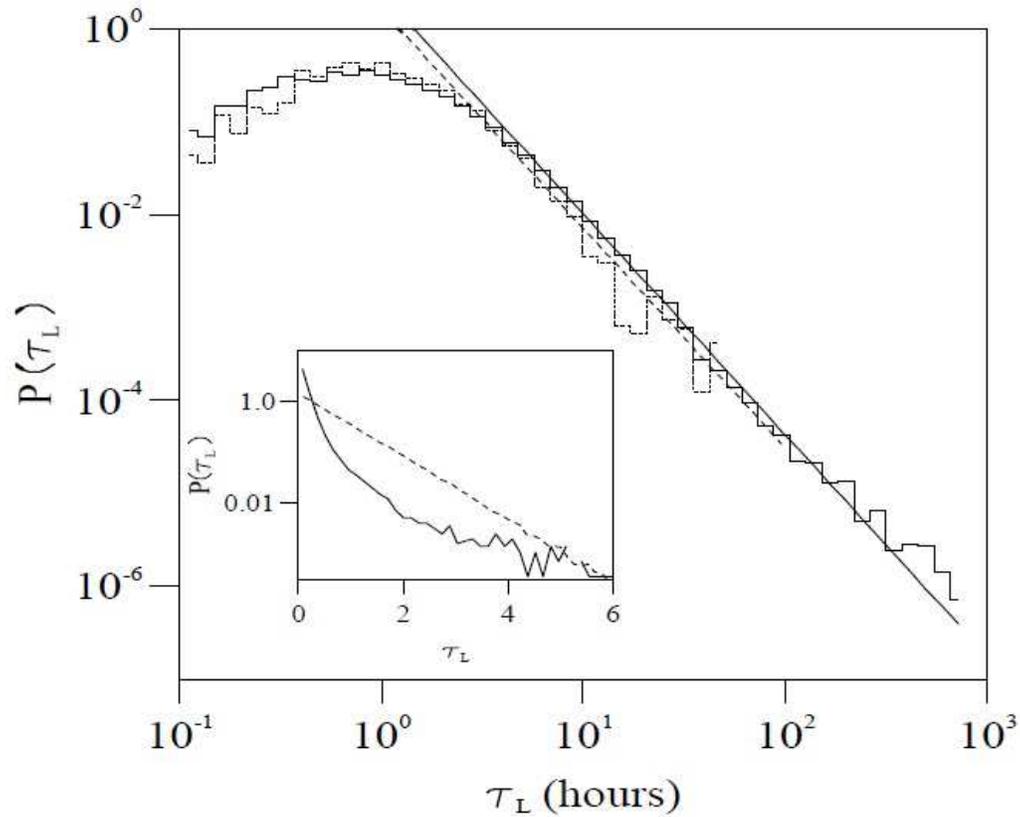}
\caption{Probability distribution function of the waiting time $P(\tau_L)$ between two X-ray flares for two datasets (A, dashed line and B, solid line). The straight lines are the respective least-squares fit of a power law. The inset shows the distribution for dataset B (solid line) and the distribution obtained through a reference SOC model (dashed line) that exhibits an exponential distribution. The variables shown in the inset have been normalized to the respective root-mean-square values. From \cite{Boffettaetal99}.}
\label{Fig0310}%
\end{figure*}

Nonetheless, \cite{Boffettaetal99} calculated the waiting times for flares recorded in hard X-rays during the period 1976-1996. Two different datasets were created: dataset A, by calculating only the differences between the time of occurrence of flares within the same active region and dataset B, by calculating the time differences between two successive maxima of flare intensity regardless of the position of the flare on the Sun's surface. The results presented in Figure \ref{Fig0310} distinctively show a power-law distribution of WTDs for both datasets A and B. In the inset of this figure \cite{Boffettaetal99} show the WTD distribution for dataset B (solid line) derived from observations, compared with the corresponding distribution obtained through a cellular automaton model (dashed line) used by \cite{Boffettaetal99} as reference of the exponential behavior of SOC simulations.

These results have beed used to argue against the relevance of SOC in solar-flare dynamics. It has been also proposed that SOC should be discarded in plasma turbulent transport dynamics in magnetic confinement devices after carrying out the same analysis on edge electrostatic fluctuations from the reversed-field experiment (RFX) pinch \citep{Spadaetal01}. Yet, as suggested by \cite{Sanchezetal02}, such tests must be considered with extreme care. In their work, \cite{Sanchezetal02} stressed that the {\em waiting time} definition is of crucial importance with regards to the resulting WTD of a physical or simulated system. Until then, some authors used the time interval between triggers, others the time interval between two consecutive maxima in burst intensity, and finally others considered the time lapse between the end of a burst and the beginning of the next one. \cite{Sanchezetal02} showed that only the quiet time would yield an exponential WTD for non-correlated triggers in an SOC system. \cite{Sanchezetal02} carried out their simulations on a 1D running sandpile, consisting of $L$ cells, and with a closed and an open boundary, respectively, located at the first and last cells. At each iteration, $U_0$ grains of sand are dropped at each cell with probability $P_0$. Whenever the local sand slope, $Z_j=h_j-h_{j-1}$, exceeds some prescribed critical value $Z_c$, $N_f$ grains of sand are moved to the next cell. The sandpile reaches the critical state after the incoming sand flux is balanced by the flux leaving the system through the open boundary. With these results, \cite{Sanchezetal02} claimed that the lack of an exponential WTD should not be used to discard SOC dynamics when all other signatures (i.e.,  $f^{-k}$ regions in fluctuation power spectra, or Hurst exponents $H>0.5$ from the rescaled range (R/S) analysis) suggest the existence of SOC. They propose that an exponential WTD is not a necessary condition for SOC state in the following cases: 
\begin{enumerate}
\item[a)] When the avalanche durations are longer than the quiet times; then power laws can appear because waiting times become contaminated by the event-duration scaling go the power law.
\item[b)] When the avalanche durations are much shorter than the quiet times; then power laws can still appear if the measurements' maximum resolution lies within the self-similar range, since all detected avalanches then become strongly correlated. This argument was supported by the earlier study of \cite{ChristensenOlami92}, in the context of a spring-block model for earthquakes. The model showed that waiting times could follow power-law distributions in case events larger than a certain size are only considered. 
\item[c)] When experimental resolution is sufficiently high to detect events of all possible sizes; the lack of exponential waiting times in this case might simply imply that the system is driven in a correlated way. The physical origin of the correlated driver in this case is system-dependent and should be determined on a case-by-case basis. 
\end{enumerate}
It is therefore possible that a system governed by SOC dynamics can lack exponential WTD statistics, not only when the experimental resolution lies within the self-similar scale range, but also when the system is slowly driven in a correlated way. Appreciating the long-standing debate at this point, we recommend caution in the interpretation of a given WTD and suggest that waiting-time statistics should not be used as a {\it necessary} test of SOC behavior in physical systems.

\subsection{Feature detection in the Spatial-Temporal Domain}
\label{det3}
The previous two sections have focused on identifying features either in space or in time. This is appropriate as scientists are often relegated to studying such datasets. A time series is often all that is obtained from stellar observations. Although this can reveal time-separable pulses that can be used for testing the statistics of SOC phenomena, all spatial information is concealed in a dot-like point source. More informative from imaging observations can exhibit the detailed fractal spatial structure of SOC phenomenon, but temporal information is commonly lacking or ignored. Combining the two domains of space and time into spatio-temporal event detection methods clearly present a powerful means to analyze SOC phenomena. However, these methods are quite complicated and hence need a sophisticated initial setup in order to work correctly.

\subsubsection{Spreading and Avalanche Exponents}
The relationships between the spreading and avalanche exponents \citep{Munoz99} and spatio-temporal structures provides a useful method to study if the system is in an SOC state. The concepts of spreading and avalanche exponents were put to use in the case of numerical models for magnetospheric \citep{MorCha08b} and solar flare \citep{MorCha08a} phenomena  as well as with observations of auroral emissions (Uritsky et al., 2000) and multi-wavelength data for solar flares \citep{Aschwanden12a}. When an SOC system arrives in the vicinity of criticality the spreading of an active site can be described by a number of scaling laws that characterize its dynamical properties. Generally the measured quantity is a survival probability $P(t)$ that an instability is still active after $t$ iterations and the number of active sites at a given time, $n(t)$ \citep{bona07}. Both quantities are expected to satisfy a power law with $t$,
\begin{equation}
n(t) \sim t^{\eta}~,\qquad P(t) \sim t^{\delta} \ ,
\end{equation}
where $\eta$ and $\delta$ are the so-called spreading exponents \citep{Munoz99}. This implies that the total number of active sites having a lifetime $T$ scales as $n_{s} \sim T^{\eta+\delta}$, and therefore its time integral should be characterized by the exponent $\kappa = 1+\eta+\delta$. Provided that these scaling relations hold, then the total number of avalanching sites - the size of the event - $S$, scales with its lifetime $T$ as:
\begin{equation}
S(T) \sim T^{\kappa} \ .
\label{spread3}
\end{equation}
Another spreading exponent that characterizes the probability distribution of avalanche sizes is therefore found as  $P(S) \sim S^{-\beta}$.

\begin{figure}
\centering
\includegraphics[width=20pc]{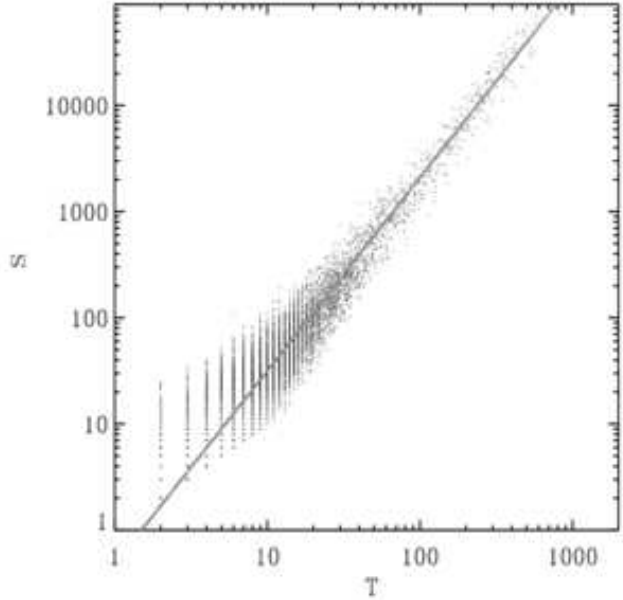}
{\caption{Correlation plot of avalanche sizes $(S)$ vs lifetimes $(T)$ for a simulation on a square lattice of size $N=128$ and angular threshold $\Theta_{c} = 2.25$ rad. The gray line is a least-squares fit, computed using only avalanches with lifetime $T > 40$ iterations. The value of the spreading exponent in this case is $\kappa = 1.82\pm0.3$. As $S = L^3 \sim T^{\kappa}, L \sim T^{\kappa/3}$, which is this case is $L \sim T^{0.61}$, close to classical diffusion ($L \sim T^{0.5}$). From \cite{MorCha08b}.}
\flabel{figure3}}
\end{figure}

As avalanches of size $S$ can have different durations $T$, the probability of an avalanche reaching a size $s$ before dying is 
\begin{equation}
P(s) = \int^{t_{max}}_{t_{min}} P(s|t) \,(1 - t^{-\delta}) {\rm d}t \ ,
\label{ps1}
\end{equation}
where $t_{min}$ and $t_{max}$ are the upper and lower duration bounds of size-$s$ avalanches, and $P(s|t)$ is the conditional probability of an avalanche having reached size $s$ at time $t$ since onset. $P(s|t)$ is bell-shaped and peaks at $t \sim 1/s^{1+\eta+\delta}$ so it can be shown \citep{Munoz99} that $P(s)$ scales as
\begin{equation}
P(s) \propto  s^{-\beta}~,
\qquad \beta = \frac{1 + \eta + 2\delta}{1+\eta+\delta} \ ,
\label{eqn:beta2}
\end{equation}
with the same scaling as expected for $P(S)$.

These redundant relations provided in Equation~\ref{ps1} and \ref{eqn:beta2} can be considered as another way of verifying if an avalanching system is in an SOC state. These relations were confirmed and presented for the case of an anisotropic SOC model for solar flares that used magnetic field lines as a basic dynamical element, and the angle between field lines as the threshold value \citep{MorCha08a}. The typical correlations found between the avalanche sizes and lifetimes are displayed in \Fref{figure3}. The same analysis has also proved useful for the case of an SOC model for the magnetosphere \citep{Liu11}. In the last decades it has been claimed that the solar corona and the Earth's magnetosphere might be in SOC. Several models have been produced in order to prove this assertion and the formalism of spreading exponents indeed provides an excellent venue to test observational data and models. 

\subsubsection{Spatio-Temporal Structures}
Spatio-temporal structures are well defined in classical SOC models, such as a numerical cellular automaton simulation like the BTW model \citep{Baketal87}. Once an SOC avalanche starts at time $t_1$, the evolution of the avalanche size is updated as described in Section~\ref{aom} above. In this section we describe the spatial-temporal evolution that determines the resulting size off the avalanche.The initial size of the avalanche at time $t_1$ has then the size $s_i=1$, which represents the unstable node in the lattice grid. In the next time step, zero to four next neighbors can become unstable (in a 2D lattice grid), after the application of the SOC re-distribution rule, and thus the avalanche has a size of $s_2=1,...,4$ nodes, or dies out ($s_2=0$). If the avalanche is further unstable, the size can grow to $s_i = 1, 2, ...,8 (i>2) $ next neighbors, and so forth. The cumulative avalanche size after time step $t_n$ is the time-integrated instantaneous size of the avalanche, i.e.,
\begin{equation}
S = \int_0^{t_n} s(t) dt = \sum_{i=1}^n s_i \ .
\end{equation}
If the same spatial pixel is active multiple times during an avalanche event, it is counted multiple times correspondingly. Consequently, the so-defined avalanche size $S$ is not a geometric volume, but rather a volume in hyper space (with $d$ geometric dimensions plus one time dimension). Note that the time step and the spatial pixel (or voxel) size are dimensionless in numerical lattice simulations and are set to unity for convenience. Non-imaging astrophysical observations typically record the spatio-temporal information of an SOC phenomenon by a flux or intensity $F_i=F(t=t_i)$ at time $t_i$ with a cadence or time interval $dt$. The summed flux adds up to a  time-integrated fluence or energy $E$ as discussed in Section~\ref{vc}. The flux $F_i$ corresponds to the emission from all active or unstable pixels in a cellular automaton avalanche, and thus represents the instantaneous energy dissipation rate  $dE_i/dt$ at time $t_i$. The total dissipated energy per avalanche, $E$, corresponds then to the time-integrated size $S$, as $E = \Delta E \ S \propto S$, with a constant energy dissipation quantum $\Delta E$ per pixel or voxel.
	
\begin{figure}[t]
\centerline{\includegraphics[width=0.9\textwidth]{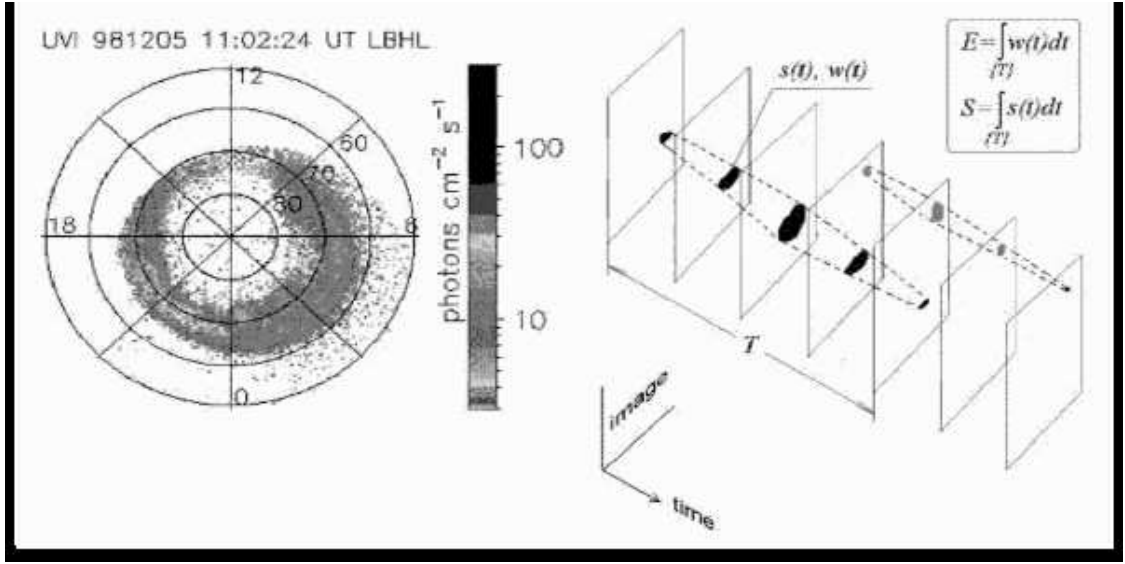}}
\caption{{\sl left:} An example of a POLAR UVI image. {\sl Right:} A schematic drawing illustrating the method of identifying spatio-temporal auroral events from POLAR UVI images. The elliptical spots in the image planes indicate the time evolution of two time-overlapping auroral events with the photon flux exceeding some noise threshold. From \cite{uritsky02}.}
\flabel{FigPOLAR}
\end{figure}

With the luxury of high-resolution imaging when observing a candidate SOC phenomena in astrophysical data, it is possible to additionally measure the (possibly fractal) area $A_i$ at  each time $t_i$. This renders a snapshot of the instantaneous contours of an SOC avalanche, defined by the sum of pixels with a flux in excess of some noise threshold, $F_i > F_{th}$ (similar to Sections~\ref{vc} and~\ref{turb}). The area information $A(t)$ is not sufficient to reconstruct the volume $V(t)$ at a given time $t$, as there is no direct information on the column depth along the line-of-sight. However, the information of the avalanche location is crucial to separate multiple avalanches occurring at the same time, or over-lapping in time, at different spatial locations. The concept of the spatio-temporal tracking of two time-overlapping avalanches is shown in \fref{FigPOLAR} for the case of auroral sizes recorded with POLAR UVI \citep{uritsky02}. The area distribution of auroral sizes was found to have a power-law distribution with a slope of $\alpha_A=1.73\pm0.03$ for the auroral observations during Jan 1997. In contrast, earlier measurements by \cite{lui2000} of the same data yielded a much flatter distribution with a slope of $\alpha_A =1.21\pm0.08$, because multiple time-overlapping auroral events were not spatially separated, and thus led to an over-estimation of large areas. The flatter index is also not consistent with predictions of a theoretical SOC model (see Section 3.3.1 in \cite{Asch14}, this volume). Therefore, the proper spatial separation of time-overlapping events in spatio-temporal detection methods is very important to obtain the correct SOC statistics. 

\begin{figure}[t]
\centerline{\includegraphics[width=0.8\textwidth]{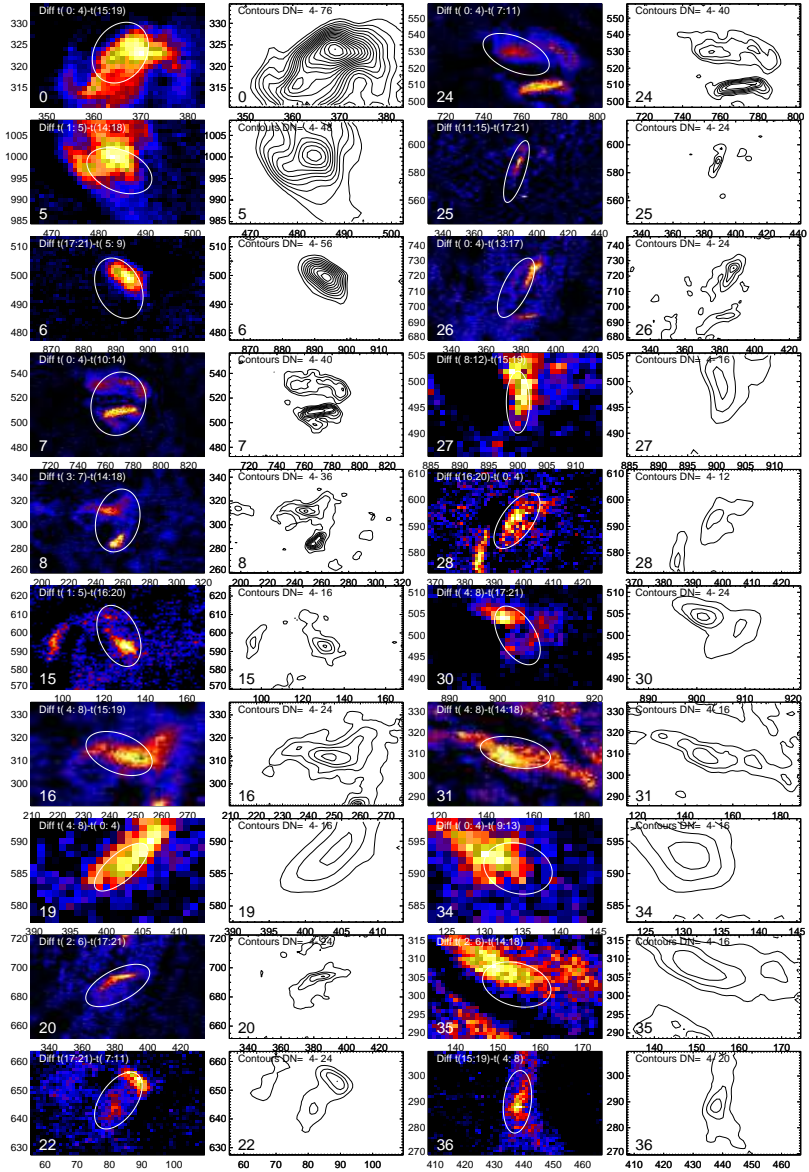}}
\caption{Spatial maps of 20 EUV nanoflare events are shown, observed with TRACE in 195 A on 1999 February 17, 02:16-02:59 UT. The greyscale images (first and third column represent difference images taken at the peak and minimum time of each nanoflare and averaged over five cadences. The contours of these difference images of detected nanoflares (second and fourth column) have a flux increment of 4 DN. From \cite{Aschwanden00b}.} 
\flabel{Fig_EUVNANO}
\end{figure}


\begin{figure}[tpbh]
\centerline{\includegraphics[width=10cm]{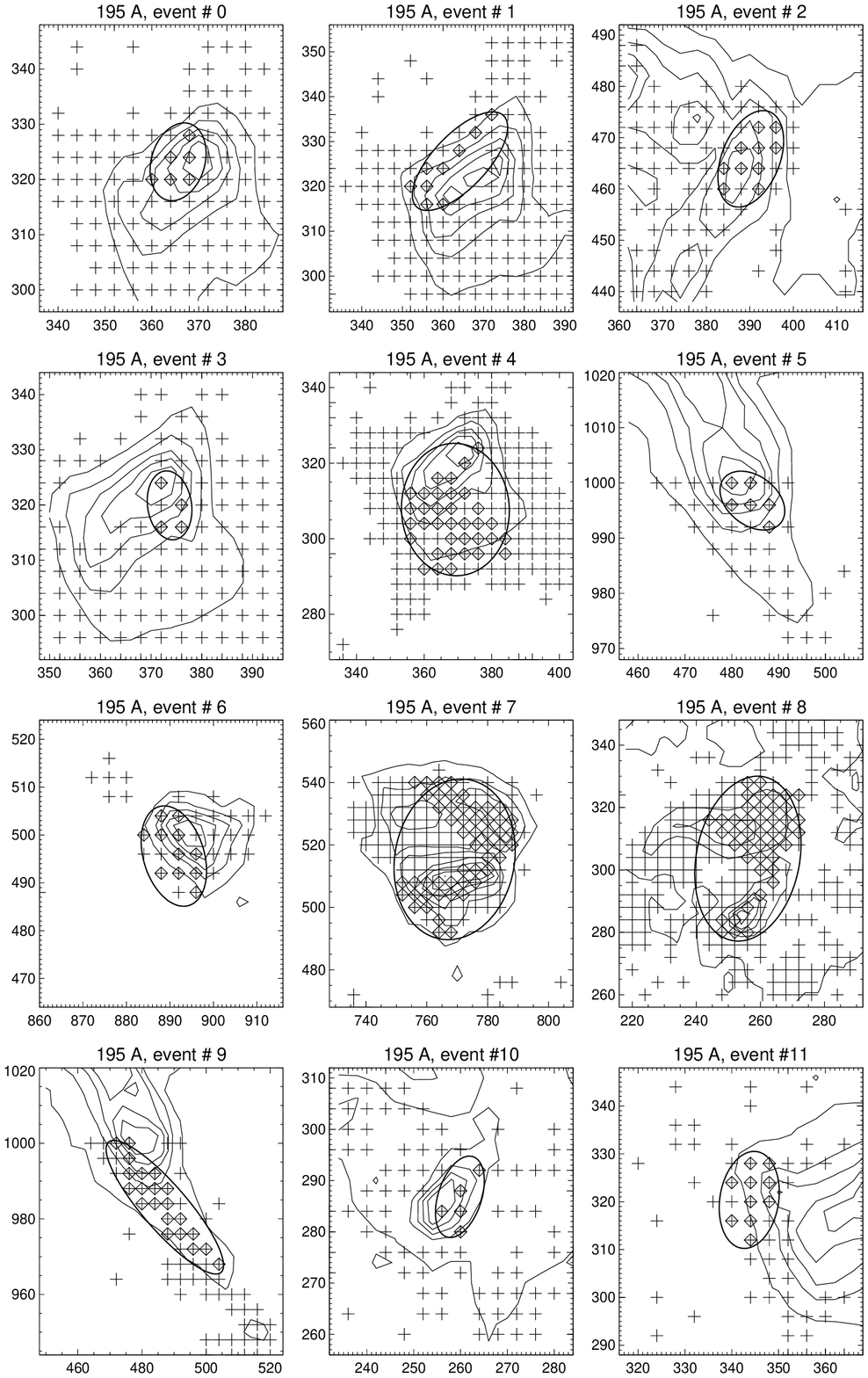}}
\caption{Spatial clustering of the pattern recognition code is illustrated for the 12 largest events on 1999 February 17, 02:15-03:00 UT. The contours outline local EUV 
intensity maps around the detected structures. The crosses mark the positions of macropixels with significant variability ($N < 3 \sigma$). The spatiotemporal pattern algorithm starts at the pixel with the largest variability, which is located at the center of each field of view, and clusters nearest neighbors if they fulfill the time coincidence criterion. These macropixels that fulfill the time coincidence criterion define an event, marked with diamonds, and encircled with an ellipse. Each macropixel that is part of an event, is excluded in subsequent events. From \cite{Aschwanden00a}.}  
\flabel{Fig_EUVTIME}
\end{figure}

Spatio-temporal detection of nanoflares in the solar corona present a good example of the power of this technique. Nanoflares often occur near-simultaneously in different spatial locations, and thus require a sophisticated automated feature detection algorithm. While an absolute flux threshold, i.e., $F_i > F_{th}$, was used in the foregoing description of detecting auroral events, solar nanoflares cannot be detected by an absolute flux threshold, because they are associated with much weaker and fainter local brightness enhancements than the variation of the flux in the surrounding or co-spatial active regions, or quiet Sun. Active regions might have a brightness of $F \approx 10^3-10^6$ DN/s in typical EUV images, while nanoflares exhibit only tiny brightness variations in the order of $F \approx 1-10^2$ DN/s \citep{Aschwanden00a,Aschwanden00b}. Nanoflares therefore have to be detected by their temporal variability, rather than by their absolute flux: consequently a time variability threshold between two consecutive images should be applied, 
\begin{equation}
	F(x,y; t_{i+1}) - F(x,y; t_i)
	\ge \Delta F_{thresh} = 3 \sigma_f \ ,
\end{equation}
rather than an absolute flux threshold. A possible threshold, \eg $F_{thresh}=3 \sigma_f$, can be specified by the photon Poisson noise in a time bin, with additional correction for spatial rebinning (to macropixels), exposure time, and other instrumental effects. An example of a solar EUV image is shown in \Fref{Fig_EUVNANO}, where the detected nanoflares are marked with ellipses.The location of detected nanoflares are not necessarily coincident with the locations of highest brightness, but their flux variability exceeds a  threshold $F > F_{thresh}$ in a difference image. Examples of variability maps are shown in \Fref{Fig_EUVTIME} which show the contours of EUV brightness, the pixels with significant variability (crosses), and pixels with significant variability that is cospatial in two subsequent images (diamonds). The automated detection criterion needs to include both spatial coherence and temporal contiguity. Those pixels that fulfill both criteria are marked with an elliptical area $A$ that  characterizes the Euclidean flare area, while the diamonds in \Fref{Fig_EUVTIME}  demarcate the instantaneous fractal flare area.

The numerical event detection code used for the examples shown in \Fref{Fig_EUVNANO} and \Fref{Fig_EUVTIME} was especially designed to detect solar microflares and nanoflares, which represent the faintest counterparts of solar flares, and thus are important to extend the dynamic range of frequency distributions of flare energies over nine orders of magnitude. Similar codes were also developed by \cite{KruckerBenz98} and \cite{ParnellJupp00}, which triggered controversial results on the index of the power law in the nanoflare regime. A number of assumptions were considered that contribute to the initially discrepant results of these indices, such as event definition, selection, and discrimination, sample completeness, observing cadence and exposure times, pattern recognition algorithms, threshold criteria, instrumental noise, wavelength coverage, fractal geometry, but also physical modeling issues of energy, temperature, electron density, line-of-sight integration, and fractal volume \citep[\eg][]{AschwandenParnell02, BenzKrucker02}.

\clearpage
\section{Summary And Conclusions}
In this review we have shown that the numerical detection of SOC is a research field onto itself. Although it remains difficult to state definitely that a system exists in a state of SOC based on feature detection alone, much progress has been made across all science fields that set out to attempt this feat. The basic studies of autocorrelations provides a powerful tool to determine if a system is in SOC. It can be used to determine if the particles in the system are spatially and temporally correlated in the appropriate manner, and is readily applicable to both simulations and experimental data. The structure function provides a complementary method using field increments, and provides an analytical connection to studies of SOC geometry. Future progress will surely consist of combining such methods with the more application-oriented methods such as marginal stability and statistical stationarity to high spatial resolution data. Even when such data is not available, block scaling provide a powerful technique to extract potential signatures of SOC.

The problems associated with working with less-than-optimum data are discussed in detail in Section 3. The scientist is reduced to applying some thresholds, and usually does not have all measurements in full 4 dimensions (3 space and 1 time). However, even with static 2D spatial slices, progress in this field has been made by adopting and adapting techniques of detecting power laws and fractals. Such features are undoubtedly ubiquitous in nature, and may well be a good signature of SOC systems. However we urge caution in adopting either of these as being a unique signature of SOC without further independent studies. In particular, the detection of power laws has undergone its own revolution in the past few years and powerful statistical tools are now freely and widely available for all scientists to use. Combined with a full understanding of instrumental effects of sub-sampling of the system, this opens up future studies in waiting time distributions as a signature of SOC, especially in those areas of study with long, homogenous, uninterrupted datasets. In terms of identifying features, it seems clear that the confidence in assigning the label of SOC to a system is much greater when we include as many datasets as possible, and as many dimensions as possible. In particular, if data can be used to move from units of DN (or counts per second) to units of energy (or energy per second) we will undoubtedly obtain a better measure of the energy release processes. It is these energy release processes that we then attempt to recognize. Probably the greatest untapped potential for the next 25 years lies in spatio-temporal studies. The concepts of spreading and avalanche exponents can be adopted for all future datasets. As hi-fidelity, multi-spectral data becomes more commonly available across all areas of science, perhaps the biggest obstacle to success is the risk of a lack of the interdisciplinary research avenues (such as the ISSI workshops), necessary to help us exploit each others' data. Numerical methods will play a key role in the advancement of clearly and unambiguously detect SOC in data. Scientists must continue to explore and understand these methods as applied to each others' data, as numerical methods will surely continue to provide a key link between simulations and experiments across all fields in scientific research. Advances in any field of research must spread across all of science. We must continue to seek to explore this interdisciplinary boundary over the next 25 years.
 
\acknowledgments
The author team acknowledges the hospitality and partial support for two workshops on {\em Self-Organized Criticality and Turbulence} at the {\sl International Space Science Institute (ISSI)} at Bern, Switzerland, during October 15-19, 2012, and September 16-20, 2013. One of us (JMA) was partially supported by a National Science Foundation Career award, NSF AGS-1255024, and NASA contracts NNH12CG10C and NNX13AE03G. One of us (MJA) was partially supported by NASA contract NNX11A099G and NASA contract NG04EA00C of the SDO/AIA instrument to LMSAL. One of us (MKG) was partially supported by EU FP7 grant PIRG07-GA-2010-268245. One of us (VIA) was partially supported by NASA LWS NNX11AO73G  grant and by the Program of the Presidium of Russian Academy of Sciences No. 21. The authors acknowledge the comprehensive and dedicated work of an anonymous referee.
\clearpage

\bibliography{ISSI_SOC_Numerics}

\end{document}